%% file: sir_agent_mcmc.tex
\documentclass[12pt]{article}

\usepackage{graphicx,psfrag,epsf}
\usepackage{enumerate}
\usepackage{color}
\usepackage{alltt}
\usepackage{framed}
\usepackage{listings}
\usepackage{longtable}
\usepackage[american]{babel}
\usepackage{cancel}
\usepackage{amsthm}
\usepackage{amsmath}
\usepackage{float}
\usepackage{mathtools}
\usepackage{amssymb}
\usepackage{tabularx}
\usepackage{multirow}
\usepackage{nicefrac}
\usepackage[font=small,margin=0.5in,labelfont=bf,tableposition=top]{caption}
\DeclareCaptionLabelFormat{andtable}{#1~#2  \&  \tablename~\thetable}
\usepackage{fancyhdr}
\usepackage[comma,authoryear]{natbib}
\usepackage{fullpage}
\usepackage{setspace}
\usepackage{subfigure}
\usepackage{kbordermatrix}
\usepackage{authblk}
\usepackage{booktabs}
\usepackage{parskip}
\usepackage{mathabx}
\usepackage{enumitem}
\usepackage{moresize}
\usepackage[final]{changes}

\colorlet{Changes@Color}{red}
\definechangesauthor[name={Jonathan Fintzi}, color={red}]{jf}
\definechangesauthor[name={Vladimir Minin}, color={blue}]{vm}
\definechangesauthor[name={Jon Wakefield}, color={red}]{jw}

\newcommand{\blind}{0}
\newcommand{\stkout}[1]{\ifmmode\text{\sout{\ensuremath{#1}}}\else\sout{#1}\fi}
\newcommand{\ind}[1]{\mathbf{\mathbb{I}} \left ( #1 \right )}

\newcommand{\bQ}{\boldsymbol{Q}}
\newcommand{\bP}{\mathbf{P}}
\newcommand{\bp}{\mathbf{p}}

\newcommand{\bw}{\mathbf{w}}

\newcommand{\mcS}{\mathcal{S}}
\newcommand{\mcI}{\mathcal{I}}

\newcommand{\mcC}{\mathcal{C}}
\newcommand{\btheta}{\boldsymbol{\theta}}
\newcommand{\btau}{\boldsymbol{\tau}}
\newcommand{\bLambda}{\boldsymbol{\Lambda}}

\newcommand{\diag}{\mathrm{diag}}
\newcommand{\bu}{\mathbf{u}}
\newcommand{\bX}{\mathbf{X}}
\newcommand{\rmd}{\mathrm{d}}
\newcommand{\e}{\mathrm{e}}
\newcommand{\bx}{\mathbf{x}}
\newcommand{\bY}{\mathbf{Y}}
\newcommand{\by}{\mathbf{y}}
\newcommand{\xnew}{\mathbf{x}^{\mathrm{new}}}
\newcommand{\xcur}{\mathbf{x}^{\mathrm{cur}}}

\newcommand{\beginsupplement}{%
		\setcounter{section}{0}
        \renewcommand{\thesection}{S\arabic{section}}
        \setcounter{page}{1}
        \renewcommand{\thepage}{S\arabic{page}}
        \setcounter{table}{0}
        \renewcommand{\thetable}{S\arabic{table}}%
        \setcounter{figure}{0}
        \renewcommand{\thefigure}{S\arabic{figure}}%
     }

\title{\textbf{}}

\date{}
\IfFileExists{upquote.sty}{\usepackage{upquote}}{}
\IfFileExists{upquote.sty}{\usepackage{upquote}}{}

\begin{document}

\def\spacingset#1{\renewcommand{\baselinestretch}%
{#1}\small\normalsize} \spacingset{1}

%%%%%%%%%%%%%%%%%%%%%%%%%%%%%%%%%%%%%%%%%%%%%%%%%%%%%%%%%%%%%%%%%%%%%%%%%%%%%%

\if0\blind
{
\title{Efficient Data Augmentation for Fitting Stochastic Epidemic Models to Prevalence Data}
  \author[1]{Jonathan Fintzi}
  \author[2]{Xiang Cui}
\author[1,2]{Jon Wakefield}
\author[2,3]{Vladimir N. Minin}
\affil[1]{Department of Biostatistics, University of Washington, Seattle}
\affil[2]{Department of Statistics, University of Washington, Seattle}
\affil[3]{Department of Biology, University of Washington, Seattle}
  \maketitle
} \fi

\if1\blind
{
  \bigskip
  \bigskip
  \bigskip
  \begin{center}
    {\LARGE\bf Title}
\end{center}
  \medskip
} \fi

\bigskip

\input{abstract_finalV2}

\noindent%

\spacingset{1.0}
\input{introduction_finalV2}
\input{methods_sir_finalV2}
\input{simulations_sir_finalV2}
\input{example_bbs_finalV2}
\input{conclusion_finalV2}

\bibliographystyle{plainnat}
\bibliography{agent_mcmc_refs}
\clearpage

\input{supplement_finalV2}

\end{document}

%% file: abstract_finalV2.tex
\renewcommand{\abstractname}{Abstract}
\begin{abstract}

Stochastic epidemic models describe the dynamics of an epidemic as a disease spreads through a population. Typically, only a fraction of cases are observed at a set of discrete times. The absence of complete information about the time evolution of an epidemic gives rise to a complicated latent variable problem in which the state space size of the \deleted{unobserved} epidemic grows large as the population size increases. This makes analytically integrating over the missing data infeasible for populations of even moderate size. We present a data augmentation Markov chain Monte Carlo (MCMC) framework for Bayesian estimation of stochastic epidemic model parameters, in which measurements are augmented with subject--level \replaced{disease histories}{trajectories}. In our MCMC algorithm, we propose each new subject--level path, conditional on the data, using a time--inhomogeneous continuous--time Markov process with rates determined by the infection histories of other individuals. The method is general, and may be applied, with minimal modifications, to a broad class of stochastic epidemic models. We present our algorithm in the context of \replaced{multiple}{a general} stochastic epidemic model\added{s} in which the data are binomially sampled prevalence counts, and apply our method to data from an outbreak of influenza in a British boarding school.
\end{abstract}

%% file: introduction_finalV2.tex
\section{Introduction}
\label{sec:intro}
Stochastic epidemic models (SEMs) are classic tools for modeling the spread of infectious diseases. A SEM represents the time evolution of an epidemic in terms of the disease histories of individuals as they transition through disease states. Incorporating stochasticity into epidemic models is important when the disease prevalence is low or when the population size is small. In both cases, the stochastic variability in the evolution of an epidemic greatly influences the probability and severity of an outbreak, along with the conclusions we draw about its dynamics \citep{keeling2008,allen2008introduction}. Moreover, many questions --- e.g., what is the \replaced{outbreak}{final} size distribution? What is the probability that a disease has been eradicated? --- cannot be answered using deterministic methods \citep{britton2010}.

The task of fitting a SEM is typically complicated by the limited extent of epidemiological data, which are recorded at discrete observation times\replaced{, }{and} commonly describe just one aspect of the disease process, e.g.\added{,} infections, and usually capture only a fraction of cases. \added{Complete }subject--level data, \replaced{which would consist of the exact times at which individuals transition through disease states}{such as infection and recovery times}, are \replaced{often unavailable}{rarely available} \citep{oneill2010}. Fitting SEMs in the absence of complete subject--level data \deleted{re}presents a complicated latent variable problem since it is usually impossible to analytically integrate over the missing data \deleted{when an epidemic is not fully observed}\citep{oneill2002}. This makes the \added{observed data} likelihood for a SEM intractable. 

Existing approaches to fitting SEMs with intractable likelihoods have largely fallen into four groups: martingale methods, approximation methods, simulation based methods, and data augmentation (DA) methods \citep{oneill2010}. Martingale methods estimate the parameters of interest using estimating equations based on martingales for the counting processes\deleted{that reflect events} within the SEM, e.g.\added{,} infections and recoveries \citep{becker1977general, watson1981application, sudbury1985proportion, andersson2000stochastic, lindenstrand2013estimation}. \replaced{These methods are not easily implemented for SEMs with complex models, and the resulting estimates are specific to the SEM dynamics.}{ The resulting estimates are specific to the SEM dynamics and are not easily implemented for more complex dynamics.} Approximation methods replace the epidemic model with a simpler model whose likelihood is \replaced{more}{analytically} tractable. For example, \citet{roberts2001} and \citet{cauchemez2008} use diffusion processes that approximate the SEM dynamics, while \citet{jandarov2014} use a Gaussian process approximation of a related gravity model. Another typical simplification is to discretize time \deleted{into ``generational units",} and to construct a transition model for the population flow \replaced{between model compartments at successive times}{at each generation time} \citep{longini1982, held2005, lekone2006, held2012}. These methods are computationally efficient and in many cases yield sensible estimates. However, the simplifying assumptions used in the various approximations are not always \replaced{appropriate}{realistic}. For instance, \added{the diffusion approximation may not be valid in small populations where the system is far from its deterministic limit \citep{andersson2000stochastic}}, while the discretization of time makes it awkward to approximate systems in which the observation times are not evenly spaced or the rates of events span several orders of magnitude \citep{glass2003, shelton2014}. \replaced{S}{Finally, s}imulation based methods use the underlying model to generate trajectories that serve as the basis for inference. This class of methods includes approximate Bayesian computation (ABC) methods \citep{mckinley2009,toni2009}, pseudo--marginal methods \citep{mckinley2014simulation}, and sequential Monte Carlo (or particle filter) methods \added{\citep{toni2009, andrieu2010particle, ionides2011iterated, dukic2012, koepke2016}}. \added{Within this class of methods, the particle marginal Metropolis--Hastings algorithm of \citet{andrieu2010particle} stands out in being a general method for Bayesian inference and is used as a benchmark method in this paper.} Although simulation--based methods have been used to fit complex models, they are computationally intensive and suffer from well known pitfalls. ABC methods are sensitive to the choice of summary statistic, rejection threshold, and prior  \citep{toni2009}. Sequential Monte Carlo methods, on which pseudo--marginal methods often rely, are prone to ``particle impoverishment" problems \citep{cappe2006inference, dukic2012}.

Traditional \added{agent--based} DA methods for fitting SEMs, first presented by \citet{oneill1999} and \citet{gibson1998}, target the joint posterior distribution of the missing data and model parameters to obtain a tractable complete data likelihood. \deleted{These methods have been used fruitfully in analyzing epidemics occurring in small to moderate sized populations in settings where some subject--level data \replaced{are}{is} available.}\added{That the augmentation is agent--based refers to the fact that subject--level disease histories, rather than population--level epidemic paths, are introduced as latent variables in the model. The advantage of this approach is} \deleted{A significant advantage} \replaced{in}{to agent--based DA is} that household structure and subject--level covariates may be incorporated into the model \citep{auranen2000,hohle2002,cauchemez2004bayesian, neal2004statistical, jewell2009bayesian, oneill2009}. However, existing DA methods suffer from convergence issues as the observed information becomes small relative to the missing data \citep{roberts2001, mckinley2014simulation, pooley2015}. The \textit{de facto} need for some subject--level data has precluded the use of classical DA machinery in many settings. Development of DA methods for SEMs is of continuing interest, and recent works by \citet{pooley2015}, \citet{QinShe15}\added{, and \citet{shestopaloff2016sampling}} have presented methods that do not rely on subject--level data\replaced{. However,}{, although} their algorithms forgo the flexibility of agent--based DA\added{, and in the case of the latter two papers have not been applied to SEMs}.

We present an agent--based DA Markov chain Monte Carlo (MCMC) framework for fitting SEMs to time series count data. We obtain a tractable complete data likelihood by augmenting the data with subject--level disease histories. Our MCMC targets the joint posterior distribution of the missing data and the model parameters as we alternate between updating subject--level paths and model parameters. We propose subject--paths, conditionally on the data, using a time--inhomogeneous continuous--time Markov chain (CTMC) with rates determined by the disease histories of the other individuals. These data--driven path proposals result in highly efficient perturbations to the latent epidemic path, and make our method practical for analyzing epidemic count data\replaced{ in the absence of any subject--level information.}{, even in moderately large population settings.} \replaced{Thus, o}{O}ur MCMC algorithm \deleted{requires no tuning and converges quickly. Furthermore, our algorithm does not require subject--level data, and thus} enables exact Bayesian inference for SEMs fit to datasets that would have been impossible to study with existing \added{agent--based} DA methods. Finally, our algorithm is not specific to any particular SEM dynamics or measurement process, and \deleted{thus} may be applied, with minimal modifications, to a broad class of SEMs.

%% file: methods_sir_finalV2.tex
\section{\replaced{The }{SIR Model and} Data Augmentation Algorithm \added{for an SIR Model}}
\label{sec:methods_sir}
For concreteness and clarity of exposition, we present our \added{Bayesian }DA algorithm \added{(BDA)} in the context of fitting a stochastic Susceptible--Infected--Recovered (SIR) model to \replaced{binomially distributed prevalence counts}{disease prevalence data}. \added{We outline in Section~\ref{sec:SEIR_SIRS_details} the minimal adaptations required for fitting Susceptible--Exposed--Infected--Recovered (SEIR) and Susceptible--Infected--Recovered--Susceptible (SIRS) models, which we \replaced{describe}{fit} in Sections \ref{sec:SIR_SEIR_SIRS_sim}, \ref{sec:SEIR_misspec_sim}, and \ref{sec:app_bbs}.}
	
The SIR model describes the time evolution of an epidemic in terms of the disease histories of \deleted{the} individuals as they transition through three \deleted{disease} states --- susceptible (S), infected/infectious (I), and recovered (R). For simplicity, we assume a closed, homogeneously mixing population in which each individual becomes \replaced{infectious immediately upon becoming infected}{infected, and hence infectious, immediately upon coming into contact with an infected person}. We also assume that recovery confers lifelong immunity and that there is no external force of infection. Therefore, the epidemic ceases once the pool of infectious individuals is depleted.

\subsection{Measurement process and data}
\label{sec:meas_proc}
Our data, $\bY = \lbrace Y_{1}, \dots, Y_L \rbrace$, are disease prevalence counts recorded at times $t_1,\dots,t_L \in [t_1,t_L]$. It should not beggar belief that the data could be subject to measurement error, for example, if asymptomatic individuals escape detection. Let $ S_\tau $, $ I_\tau $, and $ R_\tau $ denote the total susceptible, infected, and recovered people at time $ \tau $. We model the observed prevalence as a binomial sample, with \added{constant} detection probability $ \rho $, of the true prevalence at each observation time. Thus,
\begin{equation}\label{eqn:emit_dist}
Y_\ell | I_{t_\ell},\rho \sim \mathrm{Binomial}\left (I_{t_\ell}, \rho\right ).\end{equation}

\subsection{Latent epidemic process}
\label{subsec:pop_proc}
The data are sampled from a latent epidemic process, $ \bX = \lbrace \bX_1,\dots,\bX_N\rbrace $, that evolves \replaced{continuously in time}{in continuous--time} as individuals become infected and recover. The state space of this process is $ \mcS = \lbrace S,I,R\rbrace^N $, the Cartesian product of $ N $ state labels taking values in $ \lbrace S,I,R\rbrace $. The state space of a single subject, $ \bX_j $, is $\mcS_j = \lbrace S, I, R\rbrace $, and a realized subject--path is of the form \begin{equation} \bx_j(\tau) = \left \lbrace \begin{array}{ll}
S\ ,& \tau < \tau^{(j)}_{\mathrm{I}},\\
I\ ,& \tau^{(j)}_{\mathrm{I}} \leq \tau < \tau^{(j)}_{\mathrm{R}},\\
R\ ,& \tau^{(j)}_{\mathrm{R}} \leq \tau,
\end{array} \right . \end{equation} where $ \tau^{(j)}_{\mathrm{I}} $ and $ \tau^{(j)}_{\mathrm{R}} $ are the infection and recovery times for subject $ j $ (\replaced{though subject $ j $ may also never become infected or recover, or may become infected or recover outside of the observation period $ [t_1,t_L] $}{if they occur, possibly not in $ [t_1,t_L] $}). We write the configuration of $ \bX $ at time $ \tau $ as $ \bX(\tau) = \left (\bX_1(\tau),\dots,\bX_N(\tau)\right ) $, and adopt the convention that $ \bX(\tau) $ and derived quantities, e.g.\added{,} $ I_\tau $, depend on the configuration just before $ \tau $. We  use $ \tau^+ $ for quantities evaluated just after a particular time. The waiting times between transition events are taken to be exponentially distributed, and we denote by $ \beta $ and $ \mu $ the per--contact infectivity and recovery rates. Thus, the latent epidemic process evolves according to a time--homogeneous CTMC, with transition rate from configuration $ \bx $ to $ \bx^\prime $ given by
\begin{equation}
\lambda_{\bx,\bx^\prime} = \left \lbrace \begin{array}{rl}
\beta I,\ &\text{if } \bx\ \text{and } \bx^\prime\ \text{differ only in subject }j \text{, with }\bX_j=S\text{, and }\bX_j^\prime=I,\\
\mu,\ &\text{if } \bx\ \text{and } \bx^\prime\ \text{differ only in subject }j \text{, with }\bX_j=I\text{, and }\bX_j^\prime=R,\\
0,\ & \text{for all other configurations }\bx\ \text{and }\bx^\prime.
\end{array}\right.
\end{equation}
At the first observation time, we let $ \bX(t_1)|\bp_{t_1} \sim \mathrm{Categorical}\left (\lbrace S,I,R\rbrace, \bp_{t_1}\right ) $, where $ \bp_{t_1}=\left(p_{S}, p_{I},p_{R}\right) $ are the probabilities that an individual is susceptible, infected, or recovered. \replaced{Let $ \btau = \lbrace\tau_0,\dots,\tau_{K+1}\rbrace $, where $ t_1 \equiv \tau_0 $ and $ t_L \equiv \tau_{K+1} $, be the (ordered) set of $ K $ infection and recovery times of all individuals along with the endpoints of the observation period $ [t_1,t_L] $. Let $ \ind{\tau_k \corresponds I} $ and $ \ind{\tau_k \corresponds R} $ indicate whether $ \tau_k $ is an infection or recovery time\replaced{, and l}{.  L}et $ \btheta = (\beta, \mu, \rho, \bp_{t_1}) $ denote the vector of unknown parameters.}{Let $ \btau = \lbrace\tau_1,\dots,\tau_{K}\rbrace $ be the (ordered) infection and recovery times, and let $ \ind{\tau_k \corresponds I} $ and $ \ind{\tau_k \corresponds R} $ indicate if $ \tau_k $ is an infection or recovery time.  Let $ \btheta = (\beta, \mu, \rho, \bp_{t_1}) $.} The complete data likelihood is 
\begin{align} 
\label{eqn:comp_data_likelihood}
L(\bX, \bY | \btheta) &= \Pr(\bY|\bX, \rho)\times \Pr(\bX(t_1)|\bp_{t_1}) \times \pi(\bX |\bX(t_1),\beta, \mu) \nonumber \\
&=  \left [ \prod_{l = 1}^{L}\binom{I_{t_\ell}}{Y_\ell}  \rho^{Y_\ell}(1-\rho)^{I_{t_\ell} - Y_\ell}\right ] \times \left [p_{S}^{S_{t_1}} p_{I}^{I_{t_1}}p_{R}^{R_{t_1}}\right ]  \nonumber\\
&\hspace{0.2in} \times \prod_{k = 1}^{K}\left \lbrace \left [\beta I_{\tau_k}\times\ind{\tau_k \corresponds I} + \mu\times\ind{\tau_k \corresponds R}\right ] \exp{\left [-\left (\tau_k - \tau_{k-1}\right )\left (\beta I_{\tau_k} S_{\tau_k} + \mu I_{\tau_k}\right )\right ]}\right \rbrace \nonumber \\
& \hspace{0.2in} \times \exp \left [-\left (t_L - \tau_K\right )\left (\beta I_{\tau_K^+}S_{\tau_K^+} + \mu I_{\tau_K^+}\right )\right ]. 
\end{align}
We briefly reconcile what might seem like a discrepancy between our SIR model and the canonical construction of the model (see \citet{andersson2000stochastic}). Our model describes the time evolution of the subject--level collection of disease histories, and thus evolves on the state space of individual disease labels. The canonical SIR model describes the time evolution of the compartment counts, and thus evolves on the lumped state space of counts. The canonical construction would have been appropriate had we chosen to perform DA in terms of counts (for example, as in \citet{pooley2015}). However, the Markov process in the canonical model is a lumping of our process with respect to the partition induced by aggregating the individuals in each model compartment. Therefore, inference made on the full subject--level state space will exactly match inference based on the canonical model. We discuss this further in Section \ref{sec:lumpability} of the \replaced{s}{S}upplement.

\begin{figure}
	\centering
	\includegraphics[width=0.95\linewidth]{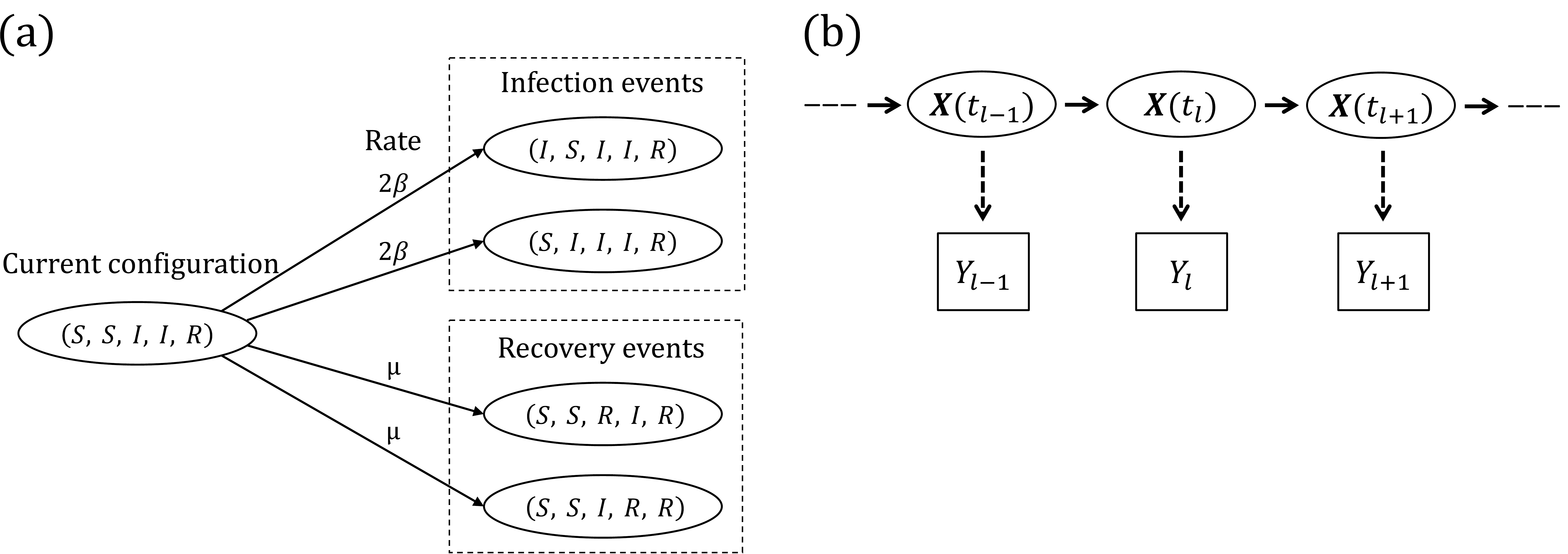}
	\caption{(a) SIR dynamics in a population of five subjects. The number of infecteds can increase from two to three via an infection of the first or second subject, reaching each of those configurations at rate $ 2\beta $. The number of recovered individuals can increase from one to two via a recovery of the third or fourth subject, reaching each of those configurations at rate $ \mu $. (b) Hidden Markov model for the joint distribution of the latent epidemic process and the data. The observations, $\mathbf{Y}_\ell,\ \ell=1,\dots,L$, are conditionally independent given $\bX(t)$, and $ \bY_\ell | I_{t_\ell}, \rho\sim\mathrm{Binomial}(I_{t_\ell}, \rho) $.}
	\label{fig:SIRdynamics_HMM.eps}
\end{figure}

\subsection{Subject--path proposal framework}
\label{subsec:subj_proc}
The observed data likelihood in the posterior  \added{$ \pi(\btheta|\bY) \propto \pi(\bY|\btheta)\pi(\btheta)=\int L(\bY|\bX, \btheta) \pi(\bX|\btheta) \pi(\theta) \mathrm{d}\pi(\bX)$} \deleted{$ \pi(\btheta|\bY) \propto \int L(\bY|\bX, \btheta) \pi(\bX|\btheta) \pi(\theta) \mathrm{d}\pi(\bX)$}
is analytically and numerically intractable for even moderately sized $ N $\added{, as it involves an extremely high dimensional integral over the collection of subject--paths, $ \bX $}. \replaced{The strategy employed in data augmentation methods is to introduce the \deleted{collection of}subject--paths, $ \bX $, as \replaced{latent variables}{additional parameters} in the model. This enables us to work with the tractable complete data likelihood in (\ref{eqn:comp_data_likelihood})\deleted{instead}.}{We can obtain the tractable complete data likelihood in (\ref{eqn:comp_data_likelihood}) by introducing the latent epidemic process, $ \bX $.} The joint posterior distribution is \begin{equation}
\label{eqn:jointpost}
\pi(\btheta, \bX| \bY) \propto \Pr(\bY|\bX,\rho) \times\pi(\bX|\bX(t_1),\beta,\mu) \times \Pr\left (\bX(t_1)|\bp_{t_1}\right) \times\pi(\beta)\pi(\mu) \pi(\rho) \pi(\bp_{t_1}),
\end{equation} where $ \pi(\beta)$, $\pi(\mu)$, $\pi(\rho)$, and $\pi(\bp_{t_1}) $ are prior densities. Our MCMC targets the joint posterior distribution (\ref{eqn:jointpost})  as we alternate between updating $ \bX|\btheta,\bY $ and $ \btheta|\bX,\bY $. 

Given the current collection of subject--paths, $ \xcur $, we propose $ \xnew $ by sampling \added{the} path of a single subject $\bX_j$, conditionally on the data, using a time--inhomogeneous CTMC \replaced{with}{on the} state space $ \mcS_j  $ \replaced{and}{with} rates \replaced{conditioned on the collection of}{ by} disease histories of \replaced{the}{all} other individuals, $ \bx_{(-j)}  = \lbrace \bx_1,\dots,\bx_{j-1},\bx_{j+1},\dots,\bx_N \rbrace$. The \replaced{proposed}{updated} collection of paths is accepted or rejected in a Metropolis--Hastings step. 

Let $ \btau^{(j)} = \lbrace \tau_\mathrm{I}^{(j)},\tau_\mathrm{R}^{(j)}\rbrace $ be the (possibly empty) set of infection and recovery times for subject $ j $, and define \replaced{$ \btau^{(-j)} = \lbrace t_1, t_L \rbrace \cup \lbrace\btau\setminus\btau^{(j)}\rbrace = \left \lbrace \tau_0^{(-j)} ,\tau_1^{(-j)}, \dots, \tau_{M}^{(-j)}, \tau_{M+1}^{(-j)}\right \rbrace$}{$ \btau^{(-j)} = \lbrace t_1, t_L \rbrace \cup \lbrace\btau\setminus\btau^{(j)}\rbrace = \left \lbrace \tau_0^{(-j)} ,\tau_1^{(-j)}, \dots, \tau_{M-1}^{(-j)}, \tau_M^{(-j)}\right \rbrace$}, where $ t_1 \equiv \tau_0^{(-j)} $ and \replaced{$ t_L\equiv\tau_{M+1}^{(-j)} $}{$ t_L=\tau_M^{(-j)} $}, to be the set of \added{$ M\leq K $} (ordered) times at which other subjects become infected or recover, along with $ t_1$ and $ t_L $. Let \replaced{$ \mcI = \lbrace \mcI_1, \dots,\mcI_{M+1}\rbrace $}{$ \mcI = \lbrace \mcI_1, \dots,\mcI_{M}\rbrace $} be the intervals that partition $ [t_1,t_L]$, i.e. \replaced{$ \mcI_1 = \left [\tau_0^{(-j)},\tau_1^{(-j)}\right ),\ \mcI_2=\left [\tau_1^{(-j)},\tau_2^{(-j)}\right ),\dots,\ \mcI_{M+1}=\left [\tau_{M}^{(-j)},\tau_{M+1}^{(-j)}\right )$}{$ \mcI_1 = \left [\tau_0^{(-j)},\tau_1^{(-j)}\right ),\ \mcI_2=\left [\tau_1^{(-j)},\tau_2^{(-j)}\right ),\dots,\ \mcI_M=\left [\tau_{M-1}^{(-j)},\tau_{M}^{(-j)}\right )$}. \deleted{Our Metropolis--Hastings proposal assumes that the CTMC for $ \bX_j $ is homogeneous within inter--event intervals.} Let $ I_\tau^{(-j)} = \sum_{i\neq j}\ind{\bX_i(\tau) = I} $ be the prevalence at time $ \tau $, excluding subject $ j $. \replaced{Let }{Define the rate matrices }\replaced{$ \bLambda^{(-j)} = \left \lbrace\bLambda_1^{(-j)}(\btheta),\dots,\bLambda_{M+1}^{(-j)}(\btheta) \right \rbrace$}{$ \bLambda^{(-j)} = \left \lbrace\bLambda_1^{(-j)}(\btheta),\dots,\bLambda_{M}^{(-j)}(\btheta) \right \rbrace$} \added{be the sequence of rate matrices} corresponding to each interval in $ \mcI $, where for \replaced{$ m=1,\dots,M+1 $. The}{$ m=1,\dots,M $, the} rate matrix for subject $ j $ is
\begin{equation} \bLambda_m^{(-j)}(\btheta) = \bordermatrix{ & S & I & R \cr
	S & -\beta I_{\tau_m}^{(-j)} & \beta I_{\tau_m}^{(-j)} & 0 \cr 
	I & 0 & -\mu & \mu \cr
	R & 0 & 0 & 0 }.
\end{equation}

We can construct the transition probability matrix for \deleted{each} interval\deleted{,} \added{$ m $ as} $$ \bP^{(j)}(\tau_{m-1},\tau_m) = \left (
p_{a,b}^{(j)}(\tau_{m-1},\tau_m)\right )_{a,b\in \mcS_j}, $$ where $ p_{a,b}^{(j)}(\tau_{m-1},\tau_m) = \Pr(\bX_j(\tau_m)=b|\bX_j(\tau_{m-1})=a, \btheta) $, using the matrix exponential \added{$$
	\bP^{(j)} (\tau_{m-1},\tau_m)= \exp\left [(\tau_m - \tau_{m-1})\bLambda^{(-j)}_m(\btheta)\right ].
	$$} \deleted{$\bP^{(j)} (\tau_{m-1},\tau_m)= \exp\left [(\tau_m - \tau_{m-1})\bLambda^{(-j)}_s(\btheta)\right ].$} 
This computation requires an eigen--decomposition of each rate matrix\replaced{, which may be carried out efficiently by computing the decomposition analytically. We may further lessen the total computational burden by caching the eigen decompositions to avoid duplicate computations.}{, the parts of which are cached.} \added{One additional point to note is the eigen--values of any SIR rate matrix are always real. However, this is not generally true, e.g.\added{,} it is possible for the rate matrix of an SIRS model to have complex eigenvalues. In this case, we obtain a real valued transition probability matrix by first applying a rotation to each rate matrix with complex eigenvalues in order to obtain its real canonical form \citep{hirsch2013differential}. This is discussed in Section \ref{sec:mtx_exp}.} 

By the Markov property, the time--inhomogeneous CTMC density over the observation period $ [t_1,t_L] $, denoted $ \pi(\bX_j | \bx_{(-j)}, \btheta) = \pi\left (\bX_j | \bLambda^{(-j)}; \mathcal{I}\right ) $, can be written as \replaced{a}{the} product of time--homogeneous CTMC densities over the inter--event intervals $ \mcI_1,\dots,\mcI_{M} $. Thus,
\begin{equation}
\label{eqn:subj_level_dens}
\pi\left (\bX_j | \bLambda^{(-j)};\mathcal{I}\right ) = \Pr(\bX_j(t_1) | \bp_{t_1}) \prod_{m=1}^{M}\pi\left (\bX_j  | \bX_j(\tau_{m-1}), \bLambda^{(-j)}_m(\btheta);\mcI_m\right ).
\end{equation} 
Similarly, the transition probability matrix over an interval $ \mathcal{I}_\ell = [t_{\ell-1},t_\ell] $ can be written as the product of transition probability matrices over the sub--intervals in $ \mathcal{I}_\ell $, within which the \added{subject--level} CTMC is time--homogeneous. Thus, the transition probability matrix over an inter--observation interval, $ \mcI_\ell = [t_{\ell-1}, t_\ell] $, partitioned by $ S $ transition events that define inter--event intervals with endpoints given by times \replaced{$ t_{\ell-1} = \tau_{\ell,0}^{(-j)} < \tau_{\ell,1}^{(-j)}<\dots<\tau_{\ell,S-1}^{(-j)}  < \tau_{\ell,S}^{(-j)} \equiv t_\ell $}{$ t_{\ell-1} = \tau_{\ell,0}^{(-j)} < \tau_{\ell,1}^{(-j)}<\dots<\tau_{\ell,S}^{(-j)}  < _{\ell,S+1}^{(-j)} = t_\ell $}, is constructed as
\begin{equation*}\label{eqn:inhomog_tpmprod} \bP^{(j)}(t_{\ell - 1},t_\ell) = \prod_{s=1}^{S}\bP^{(j)}\left(\tau_{\ell,s-1}^{(-j)},\tau_{\ell,s}^{(-j)}\right) .\end{equation*}

\begin{figure}[ht!]
\centering
\includegraphics[width=0.95\linewidth]{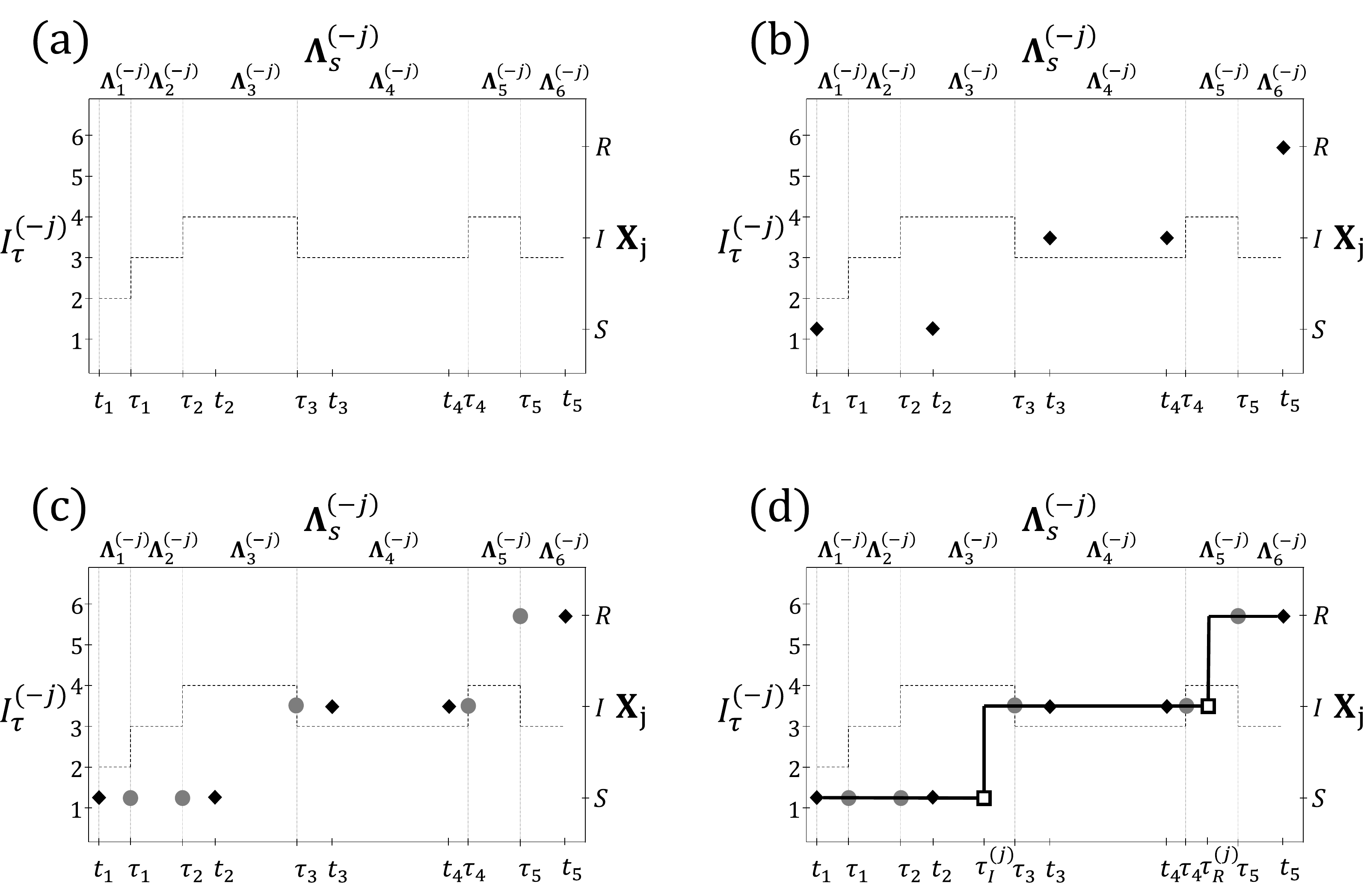}
\caption{Procedure for constructing a subject--path proposal \added{with SIR dynamics}. (a) The dashed line depicts the number of infected individuals, excluding $ \bX_j $, the subject whose path is being sampled. The observation times, $ t_1,\dots,t_5 $, and times at which other subjects change disease states, $ \tau_1,\dots,\tau_5 $, are shown on the bottom axis. Rate matrices of the time--inhomogeneous CTMC (top axis) are constant within inter--event intervals (vertical lines). The state space of the subject--level process, $ \bX_j $, is shown on the right axis. (b) \added{\textit{HMM step}: }Sample the state of $ \bX_j $ at $ t_1,\dots,t_5 $, conditional on the data and on the disease histories of other subjects. (c) \added{\textit{Discrete time skeleton step}:} Sample the infection status at  $ \tau_1,\dots,\tau_5 $, conditional on the sequence of states sampled in the HMM step. (d) \added{\textit{Event time step}:} Sample the infection and recovery times from endpoint-conditioned time--homogeneous CTMC distributions, conditional on the sequence of disease states sampled in the HMM and discrete time skeleton steps.}
\label{fig:sampling_diagram}
\end{figure}

The MCMC algorithm for constructing a subject--path proposal proceeds in three steps (Figure \ref{fig:sampling_diagram}):  
\begin{enumerate}[nolistsep]
	\item \textit{HMM step}: sample the disease state \added{of the subject under consideration} at \added{the} observation times, conditional on the data and disease histories of other subjects.
	\item \textit{Discrete time skeleton step}: sample the state at times when the time--inhomogeneous CTMC rates change, conditional on the states sampled in the HMM step. 
	\item \textit{Event time step}: sample the exact transition times conditional on the discrete sequence of states drawn in the previous steps. 
\end{enumerate}

\subsubsection{HMM step}
The key to sampling a sequence of disease states at \added{the} observation times is to rewrite the emission probability given by (\ref{eqn:emit_dist}) as
\begin{equation}\label{eqn:hmm_emit}
Y_\ell| X_j(t_\ell), I_{t_\ell}^{(-j)},\rho \sim \mathrm{Binomial}\left (\ind{X_j(t_\ell)=I} + I_{t_\ell}^{(-j)}, \rho\right ).
\end{equation}
The emission probability in (\ref{eqn:hmm_emit}) only depends on whether subject $ j $ is infected at time $ t_\ell $, since we treat the parameters and other subject\replaced{ paths}{s} as fixed. Furthermore, the observations are conditionally independent of one another, given $ \bx$ and $ \btheta $, which induces a hidden Markov model (HMM) over the joint distribution $ \bX $ and $ \bY $ (Figure \ref{fig:SIRdynamics_HMM.eps}b). 

We sample the discrete path of $ \bX_j $ at times $ t_1,\dots,t_L $ from the conditional distribution of $ \bX_j $, denoted $ \pi(\bX_j | \bY,\bx_{(-j)}, \btheta; t_1,\dots,t_L) $, using the standard stochastic forward--backward algorithm \citep{scott2002}. The algorithm efficiently computes the conditional probabilities of the paths that $ \bX_j $ can take through $ \mcS_j $ in the forward recursion. A discrete path is then sampled in the backward recursion. We provide details about the HMM sampling step in Section \ref{sec:fb_algo}\deleted{ of the Supplement}.

\subsubsection{Discrete-time skeleton step}
It would be straightforward to sample the exact infection and recovery times of subject $ j $, conditional on the sequence of states at times $ t_1,\dots, t_L $, if the subject--level CTMC rates did not possibly vary over each inter--observation interval. We may reduce our problem to the time--homogeneous case by first sampling the disease state at the intermediate event times when the CTMC rates change, and then sampling the full path within each inter--event interval. Consider an inter--observation interval, $ \mcI_\ell = [t_{\ell-1}, t_\ell] $, containing inter--event intervals whose endpoints are given by times \replaced{$ t_{\ell-1} = \tau_{\ell,0}^{(-j)} < \tau_{\ell,1}^{(-j)}<\dots<\tau_{\ell,n-1}^{(-j)}  < \tau_{\ell,n}^{(-j)} = t_\ell $}{$ t_{\ell-1} = \tau_{\ell,0}^{(-j)} < \tau_{\ell,1}^{(-j)}<\dots<\tau_{\ell,n-1}^{(-j)}  < _{\ell,n}^{(-j)} = t_\ell $}, and let $ x_i = \bx_j(\tau^\ell_i) $. We recursively sample $ \bX_j $ at each intermediate event time, beginning at $ \tau_1^\ell $, from the discrete distribution with masses 
\begin{align}
&\added{\Pr\left (\bX_j(\tau_{l,i}^{(-j)}) = x_{l,i} |  \bX_j(\tau^{(-j)}_{i-1}) = x_{i-1}, \bX_j(\tau^{(-j)}_n) = x_n\right )} \nonumber \\  &\hspace{1in}\added{= \frac{\Pr\left (\bX_j(\tau^{(-j)}_{l,i}) = x_{l,i}, \bX_j(\tau^{(-j)}_{l,i-1}) = x_{l,i-1}, \bX_j(\tau^{(-j)}_n) = x_n\right )}{\Pr\left (\bX_j(\tau^{(-j)}_{l,i-1}) = x_{l,i-1}, \bX_j(\tau^{(-j)}_n) = x_n\right )}} \nonumber\\
&\hspace{1in}\added{= \frac{\Pr\left (\bX_j(\tau^{(-j)}_{l,i}) = x_{l,i} | \bX_j(\tau^{(-j)}_{l,i-1}) = x_{l,i-1}\right )\Pr\left (\bX_j(\tau^{(-j)}_n) = x_n | \bX_j(\tau^{(-j)}_{l,i}) = x_{l,i}\right )}{\Pr\left (\bX_j(\tau^{(-j)}_n) = x_n | \bX_j(\tau^{(-j)}_{l,i-1}) = x_{l,i-1}\right)}} \nonumber\\
&\hspace{1in} \added{= \frac{\left [\bP^{(j)}(\tau^{(-j)}_{l,i-1},\tau^{(-j)}_{l,i})\right ]_{x_{l,i-1},x_{l,i}}\left [\prod_{k=i}^{n-1}\bP^{(j)}(\tau^{(-j)}_k, \tau^{(-j)}_{k+1})\right ]_{x_{l,i},x_n}}{\left [\prod_{k=i-1}^{n-1}\bP^{(j)}(\tau^{(-j)}_k, \tau_{k+1}^{(-j)})\right]_{x_{l,i-1},x_n}}.}
\label{eqn:dt_skel}
\end{align}
%\footnotesize
%\begin{align}
%\begin{split}
%&\textcolor{red}{\stkout{\Pr\left (\bX_j(\tau_i^\ell) = x_i |  \bX_j(\tau^\ell_{i-1}) = x_{i-1}, \bX_j(\tau^\ell_n) = x_n\right ) = \frac{\Pr\left (\bX_j(\tau^\ell_i) = x_i, \bX_j(\tau^\ell_{i-1}) = x_{i-1}, \bX_j(\tau^\ell_n) = x_n\right )}{\Pr\left (\bX_j(\tau^\ell_{i-1}) = x_{i-1}, \bX_j(\tau^\ell_n) = x_n\right )}}} \nonumber\\
%& \textcolor{red}{\stkout{= \frac{\Pr\left (\bX_j(\tau^\ell_i) = x_i | \bX_j(\tau^\ell_{i-1}) = x_{i-1}\right )\Pr\left (\bX_j(\tau^\ell_n) = x_n | \bX_j(\tau^\ell_i) = x_i\right )}{\Pr\left (\bX_j(\tau^\ell_n) = x_n | \bX_j(\tau^\ell_{i-1}) = x_{i-1}\right)}}} \nonumber\\
%& \textcolor{red}{\stkout{= \frac{\left [\bP^{(j)}(\tau^\ell_{i-1},\tau^\ell_i)\right ]_{x_{i-1},x_i}\left [\prod_{k=i}^{n-1}\bP^{(j)}(\tau^\ell_k, \tau^\ell_{k+1})\right ]_{x_i,x_n}}{\left [\prod_{k=i-1}^{n-1}\bP^{(j)}(\tau^\ell_k, \tau_{k+1}^\ell)\right]_{x_{i-1},x_n}}.}}
%\end{split}
%\end{align}

%\added[id=vm]{I am confused by the indices above. Should we be using $\tau^{(-j)}_{l,i}$ inside the equation?}.

\subsubsection{Event time step}
The final step in constructing a subject--path is to sample the exact infection and recovery times given the discrete sequence of states obtained in the previous two steps. This amounts to simulating the path of an endpoint--conditioned time--homogeneous CTMC, a task for which there exist a variety of efficient methods \citep{hobolth2009}. \replaced{When fitting the SIR model, we}{We} chose to use modified rejection sampling, a modification of Gillespie's direct algorithm \citep{gillespie1976} that explicitly avoids simulating constant paths. This method is known to be efficient when the states differ at the endpoints of small time intervals. 
\added{We used  uniformization{--based sampling \citep{hobolth2009}} when fitting SEIR and SIRS models, which was more robust when sampling paths in intervals with multiple transitions.} \replaced{Fast implementations of these methods are }{A fast implementation is} available \replaced{in}{through} the \texttt{ECctmc} package in \texttt{R} \citep{ECctmc}. We briefly summarize the \replaced{algorithms}{modified rejection sampling algorithm} in Section \ref{sec:ecctmc}\replaced{.}{, and refer the reader to (Hobolth and Stone, 2009) for a thorough discussion of methods foran excellent discussion of methods for simulating paths of endpoint--conditioned time--homogeneous CTMCs.} 

\subsubsection{Metropolis--Hastings step}
 Having constructed a complete subject--path proposal, we decide whether to accept or reject the proposal via a Metropolis--Hastings step. It is important to understand that the true distribution of $ \bX_j | \bx_{(-j)},\btheta $ is neither Markovian nor analytically tractable, and therefore, does not match the time--inhomogeneous CTMC in our proposal. The target distribution of the subject--path proposal is $ \pi(\bX | \bY) \propto\pi(\bY | \bX)\pi(\bX) $. 
 Thus, we accept a path proposal with probability   
 \begin{align}
 a_{\xcur \longrightarrow \xnew} &= \min \left \lbrace \frac{\pi(\xnew | \by)}{\pi(\xcur | \by)}\frac{q(\xcur | \xnew,\by)}{q(\xnew | \xcur,\by)},\ 1\right \rbrace \nonumber \\
 &=\min \left \lbrace  \frac{\pi(\xnew)}{\pi(\xcur)}\frac{\pi\left(\xcur_j| \bLambda^{(-j)}; \mcI\right)}{\pi\left(\xnew_j| \bLambda^{(-j)}; \mcI\right)}, 1 \right \rbrace,
 \end{align}
 where we have suppressed the dependence on $ \btheta $. Hence, the Metropolis--Hastings ratio is equal to the ratio of population-level time--homogeneous CTMC densities, multiplied by the ratio of time--inhomogeneous CTMC proposal densities (see Section \ref{sec:MH_ratio} for the derivation). 
 
\subsubsection{Initializing the collection of subject--paths}
We initialize the collection of subject paths at the start of our MCMC by simulating paths using Gillespie's direct algorithm \citep{gillespie1976} until we have found one under which the data have non--zero probability. 
A sufficient condition for this \added{under the binomial sampling model} is that the number of infected individuals is greater than the observed prevalence at each observation time. \deleted{For the sake of efficiency, we simulate paths using the canonical SIR model on the lumped state space of compartment counts (Andersson and Britton, 2000). A valid population--level path may then be mapped to a collection of subject--level paths by selecting the subject for each infection or recovery uniformly at random from among the subjects who are at risk for that event. Thus, the individual associated with a particular infection event is sampled uniformly at random from among the subjects who are susceptible just prior to that infection time, while the subject associated with a given recovery event is sampled uniformly at random from among the individuals who are infected just prior to that recovery time.}

 \subsection{Parameter updates}
One MCMC iteration includes a number of subject--path updates, followed by a set of parameter updates. \added{The optimal number of subject--path updates per MCMC iteration is specific to the dynamics of the SEM and the epidemic setting (e.g., endemic vs. epidemic, high vs. low escape probability), but ultimately boils down to the cost of subject--path updates vis--a--vis parameter updates. We discuss this further in Section \ref{sec:num_subj_per_iter}.}\deleted{We will discuss the choice of how many subject--path updates to undertake per \replaced{MCMC iteration in \added{Section} \ref{sec:num_subj_per_iter}}{set of parameter updates in a subsequent section}.} \replaced{In the case of the SIR model, as well as the other models we will fit in subsequent sections, c}{C}onjugate priors are available for all our model parameters. Thus, we use Gibbs sampling to draw new parameter values from their univariate full conditional distributions \replaced{(see Section \ref{sec:priors}).}{(shown in Table \ref{tab:param_gibbs}). 
	We refer the reader to for a derivation of the full conditional distributions.}
 
 \subsection{Implementation}
We provide the \texttt{R} and \texttt{C++} code base for this paper, along with examples \added{and the code for reproducing the results we present in the following sections}, in the form of an \texttt{R} package in a stable GitHub repository (\texttt{https://github.com/fintzij/BDAepimodel}). Future implementations, including extensions to the algorithm presented here along with improvements to the implementation, will be incorporated into the \texttt{stemr} package (\texttt{https://github.com/fintzij/stemr}).

%% file: simulations_sir_finalV2.tex
\section{Simulation Results}
\subsection{\added{Inference under various epidemic dynamics}}
\label{sec:SIR_SEIR_SIRS_sim}
\begin{figure}[!h]
	\centering
	\includegraphics[width=\linewidth]{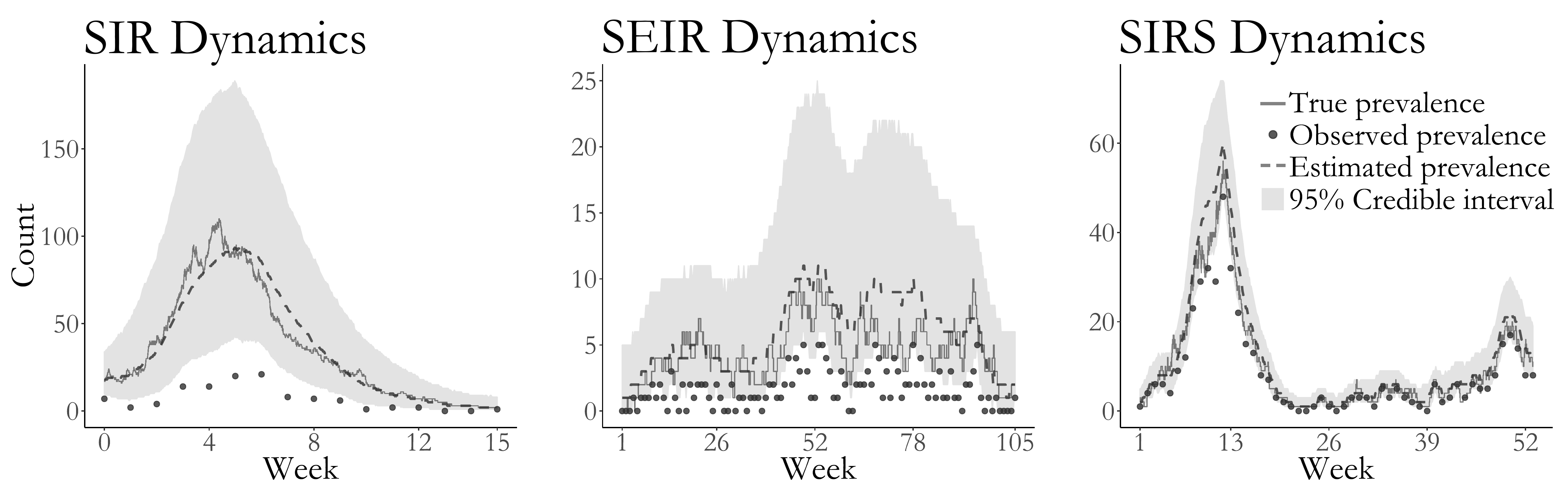}
	\caption{\added{Estimated latent posterior distributions of disease prevalence in outbreaks simulated under SIR (left), SEIR (middle), and SIRS (right) dynamics. Depicted are the true unobserved prevalence (solid line), observed data (dots), pointwise posterior median prevalence (dashed line), and pointwise 95\% credible intervals (shaded region). Latent posterior estimates are based on a thinned sample, with every $250^{th}$ sample retained.}}
	\label{fig:sim1_latent_posts}
\end{figure}

\added{We fit SIR, SEIR, and SIRS dynamics to binomially distributed prevalence counts sampled from epidemics simulated under corresponding dynamics in populations of 750, 500, and 200 individuals (details provided in Section \ref{sec:sim1_details}). Priors for the rate parameters and binomial sampling probability were scaled so that the priors spanned reasonable ranges of values (e.g. recovery durations ranging from days to weeks/months rather than seconds to eons under extremely diffuse priors), but were otherwise only mildly informative, while the initial distribution parameters were assigned informative priors (see tables \ref{tab:sim1_sir_priors}, \ref{tab:sim1_seir_priors}, and \ref{tab:sim1_sirs_priors}). The three datasets, depicted in Figure \ref{fig:sim1_latent_posts} along with the estimated pointwise posterior prevalence, presented a range of challenges. The SIR example was arguably the most ``standard'' as the observation period captured the exponential growth and decline of the epidemic. Thus, much of the curvature in the latent path was reflected in the data. In contrast, data from the outbreak simulated under near--endemic SEIR dynamics contained very little information about the shape of the epidemic curve. The task of disentangling whether the data were sampled with low probability from a high--prevalence outbreak, or visa--versa, was further complicated by the inclusion of an additional disease state --- the exposed state --- that was not directly observed. Finally, the SIRS model was more computationally challenging for two reasons. First, the recurrent nature of the disease process demanded that the disease state at each event time, and the path within each inter--event interval, be sampled in the subject--path proposal. Second, it was possible for CTMC rate matrices to have complex eigen--decompositions, which made computing transition probability matrices more expensive. This affected the optimal number of subject--path updates per MCMC iteration (see Section \ref{sec:num_subj_per_iter} for further discussion on this point). Simulation details, along with minor adaptations to our algorithm for fitting the SEIR and SIRS models, are presented in Section \ref{sec:SEIR_SIRS_details}.}

\begin{figure}[!h]
	\centering
	\includegraphics[width=\linewidth]{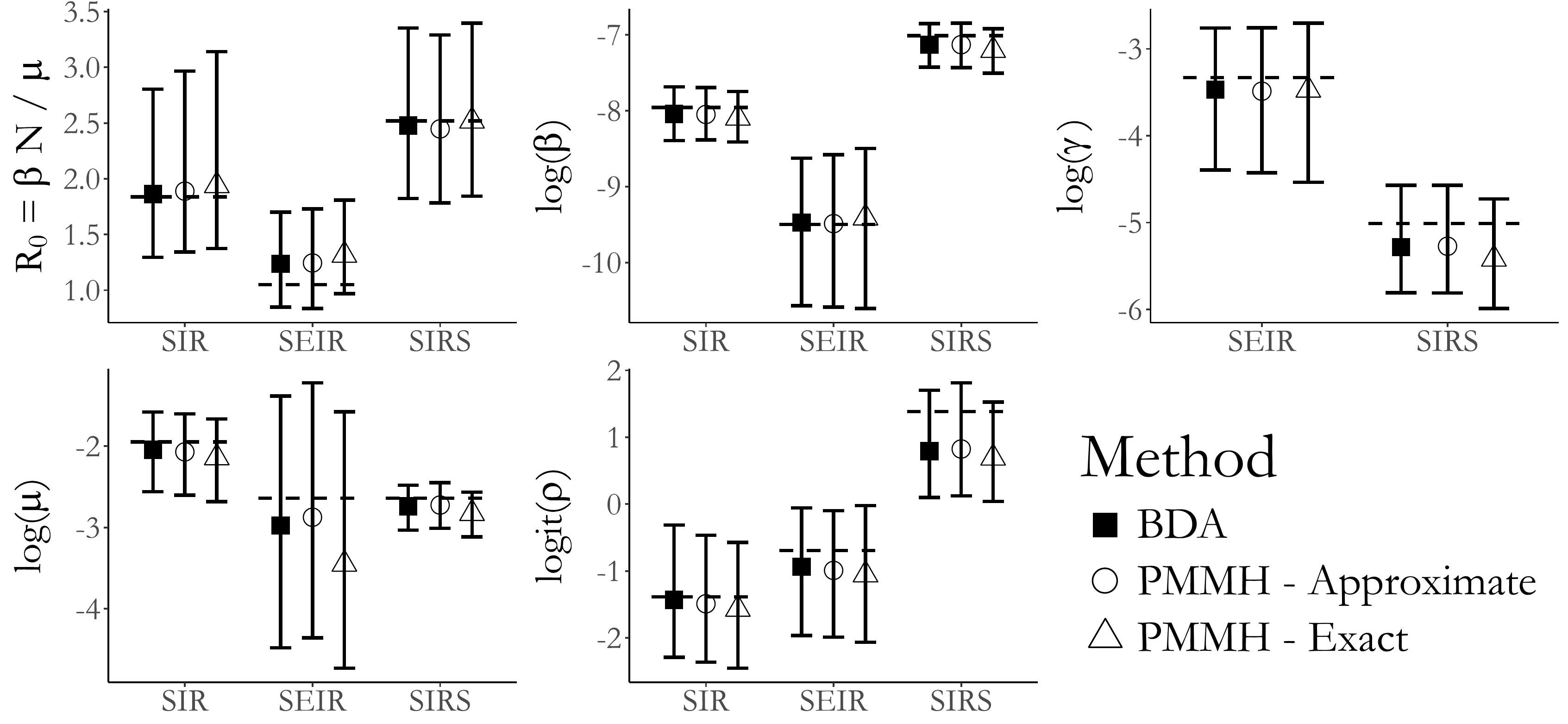}
	\caption{\added{Posterior medians and 95\% credible intervals of parameters in the SIR, SEIR, and SIRS models fit with Bayesian data augmentation (BDA) and particle marginal Metropolis--Hastings (PMMH) with particle paths simulated approximately (using $ \tau $--leaping) and exactly (using Gillespie's direct algorithm). Displayed are estimates of the basic reproductive number, $ R_0 $, the rate parameters, and the binomial sampling probability. In all models, $ \beta $ is the per--contact infectivity rate, $ \mu $ is the recovery rate, and $ \rho $ is the binomial sampling probability. In the SEIR model, $ \gamma $ denotes the rate at which an exposed individual becomes infectious, while in the SIRS model $ \gamma $ denotes the rate at which immunity is lost.}}
	\label{fig:sim1_credint}
\end{figure}

\added{The true epidemic paths and parameter values fell well within the 95\% Bayesian credible intervals in all three simulations (Figure \ref{fig:sim1_latent_posts} presents the estimated latent posterior prevalence; Figure \ref{fig:sim1_credint} presents posterior estimates of model parameters; Figure \ref{fig:sim1_latent_post_all} presents estimated latent posterior distributions and true epidemic paths for all model compartments). The acceptance rates for subject--path proposals were roughly 92\% for the SIR model, 91\% for the SEIR model, and 77\% for the SIRS model. Our posterior estimates of the model parameters also closely match estimates obtained using the particle marginal Metropolis--Hastings (PMMH) algorithm of \citet{andrieu2010particle}, implemented using the \texttt{pomp} package in \texttt{R} \citep{pomp}. We simulated particle paths in the PMMH algorithm in two ways; exactly using Gillespie's direct algorithm \citep{gillespie1976}, and approximately using a multinomial modification of $ \tau $--leaping \citep{breto2011compound}. In these small population examples, the exact algorithm is arguably more appropriate, as the leap conditions for $ \tau $--leaping may not be met in small populations, but it is also substantially slower. In these simple settings, PMMH tended to outperform our algorithm in terms of log--posterior effective sample size (ESS) per CPU time. When PMMH particle paths were simulated by $ \tau $--leaping, the average ESS per CPU compared to BDA was roughly $ 350\times $ greater for the SIR model, $ 4.4\times $ greater for the SEIR model, and $ 13\times $ greater for the SIRS model. Exact simulation of PMMH particle paths reduced the computational advantage of PMMH substantially. In this case, the average log--posterior ESS per CPU time was $ 10.5\times $ greater for PMMH in fitting the SIR model, $ 2\times $ for the SEIR model, and $ 0.7\times $ for the SIRS model. These comparisons did not include the time required to tune the MCMC for PMMH, which was nontrivial. In contrast, our algorithm required no tuning beyond selecting the number of subject--paths to update per MCMC iteration. We also note that in fitting the models using PMMH, we were required to make several implementation decisions to prevent particle degeneracy and to balance speed with precision. These included selecting the number of particles and the time--step in the approximate $ \tau $--leaping algorithm. For example, when using $ \tau $--leaping to simulate particle paths, the number of particles required to obtain good mixing for the SIRS model fit with PMMH was much higher than for the other two models. Details of the PMMH implementations and further results are discussed in Section \ref{sec:sim1_details}.}

\subsection{\added{Inference under model misspecification}}
\label{sec:SEIR_misspec_sim}
\added{In practice, every stochastic epidemic model is misspecified with respect to the real world epidemic process from which the data arise, and the malignancy of the model misspecification is often imposible to diagnose a priori. We can build up an understanding of an epidemic's dynamics by fitting SEMS under a range of dynamics, beginning with simple, easily interpretable models. The results of each model are interpretted counterfactually --- e.g. ``If the true epidemic followed SIR dynamics, our best guess of the dynamics that gave rise to the data would be...''. The iterative nature of epidemic modeling suggests that some minimal criteria for the usefulness of any computational algorithm would be that MCMC converges to some reasonable estimate of the model dynamics, and that the estimated latent posterior distribution under the hypothetical dynamics should reflect the true epidemic.}

\begin{figure}[!ht]
	\centering
	\includegraphics[width=0.5\linewidth]{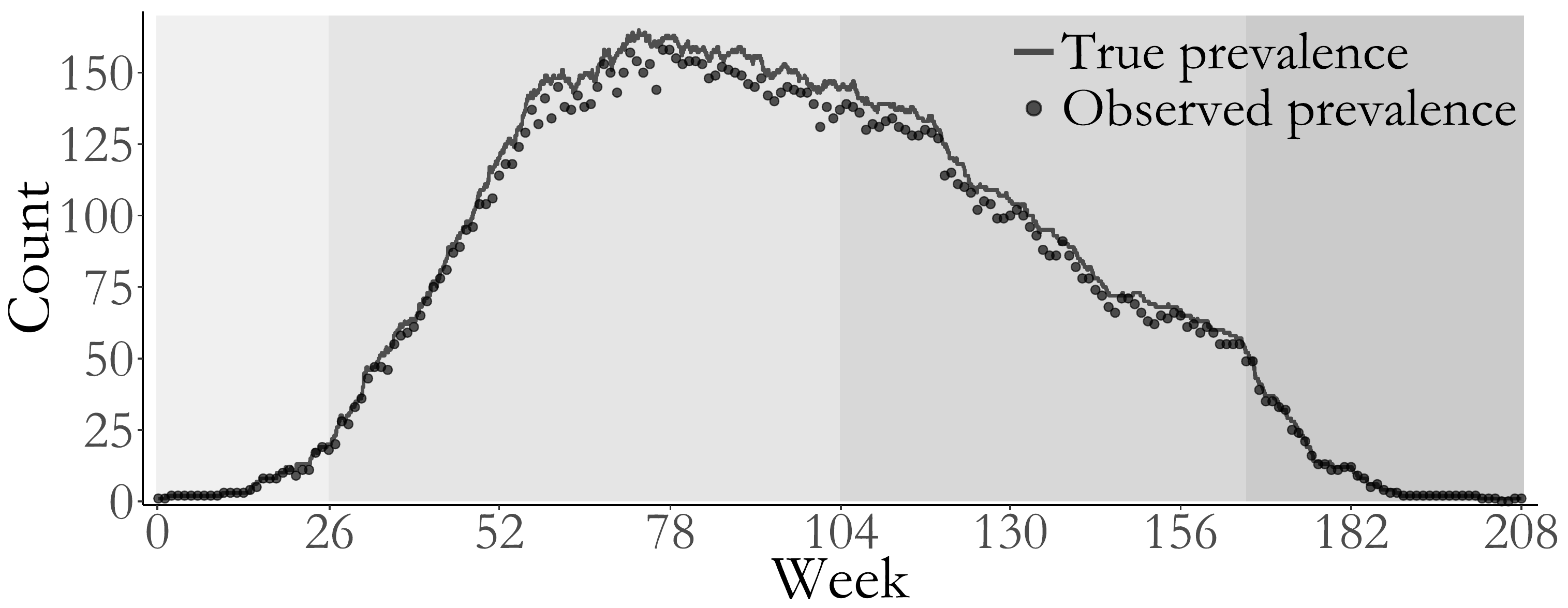}
	\qquad
	\begin{tabular}[b]{lllll}
		& \multicolumn{4}{c}{Epoch}\\ \cmidrule{2-5}
		Parameter & 1 & 2 & 3 & 4 \\ \hline	
		$ R_0^{\mathrm{Eff}} $ & 14.9& 9.2& 0.1& 0\\
		$ 1/\gamma $ (days) & 210&210&90& 180\\
		$ 1/\mu $ (days) & 150 &330&300&70	\\
		&&&&
	\end{tabular}
	\captionsetup{labelformat=andtable}
	\caption{\added{Simulated outbreak with SEIR dynamics that varied over four epochs (shaded regions). Weekly prevalence counts (points) were binomially sampled with sampling probability $ \rho = 0.95 $ from the true unobserved prevalence (solid line). The table presents the effective reproductive number computed based on the number of susceptibles at the beginning of each epoch, $ R_0^{\mathrm{Eff}} = \beta(\tau) S(\tau) / \mu(\tau) $, the mean latent period, $ 1/\gamma $, and the mean infectious period, $ 1/\mu $.}}
	\label{fig:misspec_data}
\end{figure}

\added{However, it is precisely the inherent misspecification of SEMs that leads simulation--based methods struggle in many instances, and it is here that we highlight a critical advantage of our data augmentation algorithm. Our subject--path proposals are driven, not just by the SEM dynamics, but also by the data. This enables us to overcome model misspecification in situations in which simulation--based methods degenerate due to their reliance on an adequately accurate model for simulating epidemic paths. We demonstrate this in a simple example in which we fit SIR and SEIR models to four years of weekly prevalence data sampled from an epidemic simulated under time--varying SEIR dynamics, where the latent period, infectious period, and per--contact infectivity rate were modulated over four discrete epochs (depicted in Figure \ref{fig:misspec_data}, details presented in Section \ref{sec:SEM_misspec_details}).}

\setcounter{table}{1}

\added{We fit SIR and SEIR models to the data using our DA algorithm, and using PMMH with 2,500 particles, the paths for which were simulated approximately via $ \tau $--leaping with a time--step of 1 day. We assigned weakly informative priors for the rate parameters governing the epidemic dynamics in both models, and informative priors for the binomial sampling probability and the initial state probabilities (Table \ref{tab:misspec_priors}). The MCMC chains for models fit with PMMH suffered from severe particle degeneracy and did not converge (see Figures \ref{fig:misspec_sir_bda_traceplots} and \ref{fig:misspec_seir_bda_traceplots}).}

\begin{figure}[!h]
	\centering
	\includegraphics[width=0.9\linewidth]{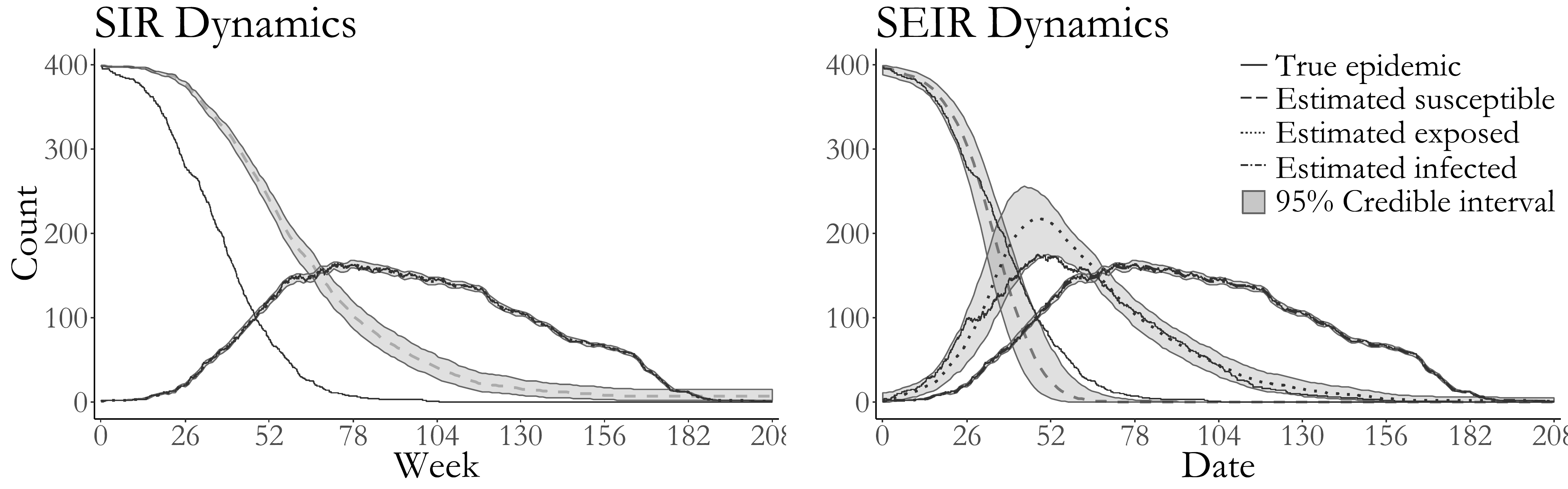}
	\caption{\added{True epidemic path (solid lines), pointwise posterior median estimate of the numbers of susceptibles (dashed line), exposed (dotted line), and infected individuals (dash--dotted line) and pointwise 95\% credible intervals (shaded regions) under SIR and SEIR dynamics.}}
	\label{fig:misspec_latent_posts}
\end{figure}

\added{Both models fit via DA yield reasonable estimates for the within--subject disease dynamics (i.e. the infectious period, as well as the latent period in the case of the SEIR model). The posterior median average infectious period duration was estimated to be 292 days (95\% BCI: 263 days, 323 days) under SIR dynamics, and 287 days (95\% BCI: 260 days, 318 days) under SEIR dynamics. The posterior median average latent period under SEIR dynamics was 211 days (95\% BCI: 165 days, 260 days). The posterior median estimate of $ R_0 $ under SIR dynamics was 4.05 (95\% BCI: 3.40, 4.81), while under SEIR dynamics, the posterior median estimate of $ R_0 $ was 23.8 (95\% BCI: 15.1, 37.0). While the true prevalence fell well within the pointwise 95\% credible interval for both models (Figure \ref{fig:misspec_latent_posts}), we notice that the degree of model misspecification drastically affected our ability to estimate the history of the numbers of noninfectious people over the course of the epidemic. Under SIR dynamics, we drastically overestimate the number of susceptible individuals. The SEIR model much more closely resembles the time--varying SEIR model used to simulate the epidemic. Although the true path for the number of susceptible still falls outsize the 95\% credible interval at times, we are still able to reconstruct a reasonabe range of paths for the number of exposed individuals. This contrasts with the models fit in Section \ref{sec:SIR_SEIR_SIRS_sim}, which were not misspecified with respect to the true epidemic dynamics. In that case, the complete path of the epidemic fell well within the estimated credible intervals for all disease states for all three models (Figure \ref{fig:sim1_latent_post_all}). Therefore, we advise caution in  reconstructing the epidemic history for disease states that were not measured, particularly when severe model misspecification is suspected.}

\subsection{\added{Inference under population size misspecification}} 
\added{Model misspecficiation often extends not only to the SEM  dynamics, but also to the assumed population size. This is most often the case in settings where subject--level data is unvailable, e.g. surveillance settings, and may result in biased estimates of the SEM dynamics. This bias is the result of a missmatch between the intensive dynamics of the epidemic process, which are a function of the fractions of individuals in the population who in each disease state, and the extensive scale of prevalence counts, which are not normalized by the population size. Without knowing the true population size, it is difficult to know whether the scale of the counts reflects a high prevalence/low detection rate setting, or visa--versa. Moreover, wrongly assuming too large, or too small, of a population size could bias posterior inference of the epidemic dynamics.}

\added{We simulated weekly prevalence counts under a binomial measurement process with detection probability $ \rho = 0.3 $ from an epidemic with SIR dynamics in a population of $ N=1,250 $ individuals. We then fit SIR models using a series of assumed population sizes under a flat prior for the binomial sampling probability and diffuse priors for the epidemic dynamics (see Section \ref{sec:popsize_misspec_details} for complete simulation details and prior specifications), and compared the resulting scaled parameter estimates. The per--contact infectivity rate, $ \beta $, was rescaled by the population size, $ N $, so that it could be interpreted as the rate of disease transmission. We computed $ R_0 $ using the assumed population size. Finally, we scaled the binomial sampling probability by the assumed population size to give the expected number of observed infections in a completely infected population.}

\begin{figure}
	\centering
	\includegraphics[width=0.9\linewidth]{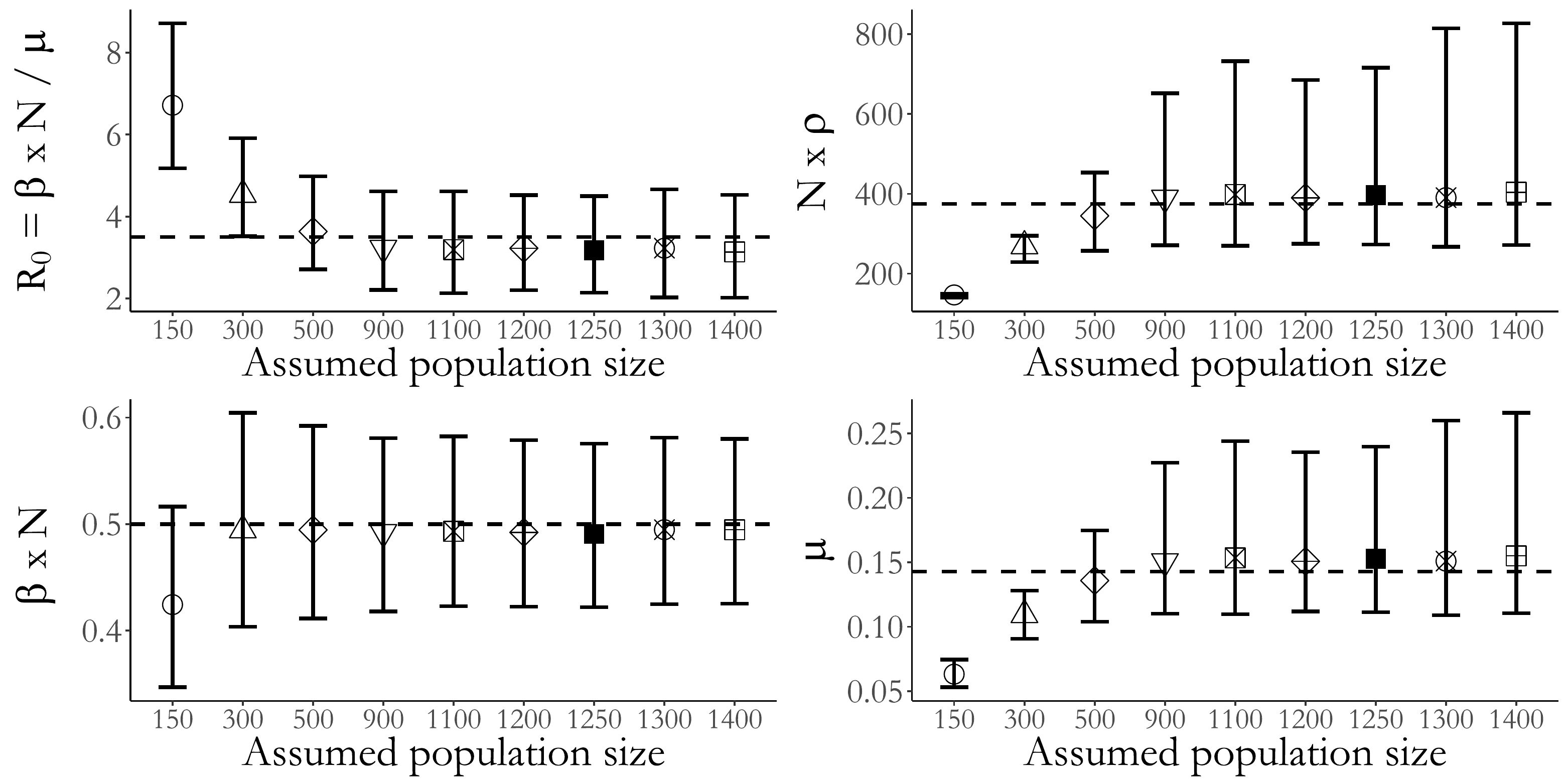}
	\caption{\added{Posterior medians and 95\% credible intervals for the basic reproductive number, $R_0$, infectivity rate, recovery rate, and binomial sampling probability scaled by the assumed population size. The dashed lines indicate the true values in the population of size 1,250. The population size, N, indicates the assumed population size used in fitting the model.}}
	\label{fig:popsize_misspec_credints}
\end{figure}

\begin{figure}
	\centering
	\includegraphics[width=\linewidth]{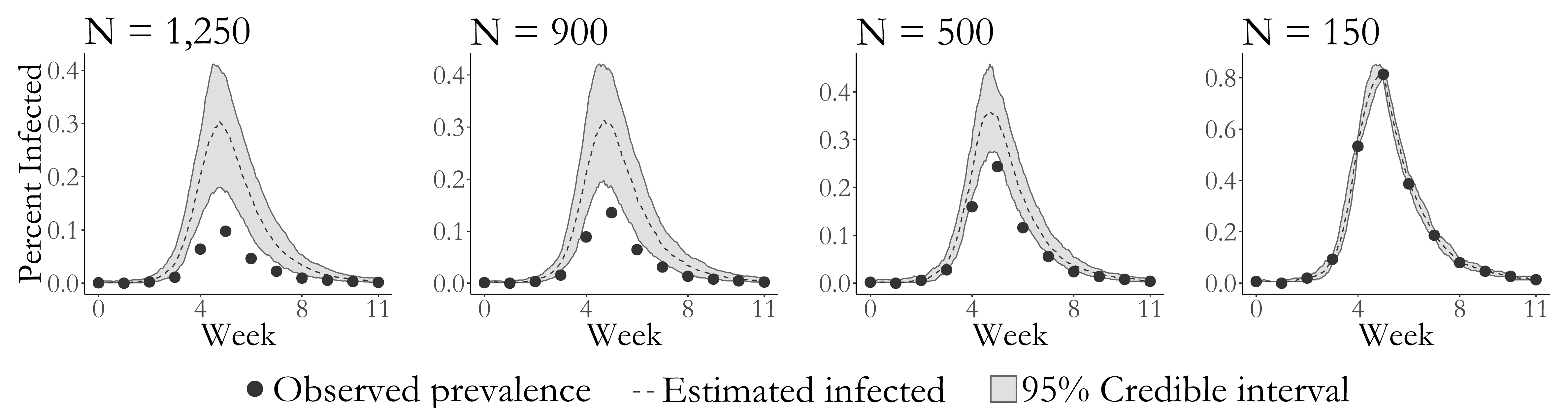}
	\caption{\added{Estimated latent posterior distributions of disease prevalence under SIR dyamics. The true population size is 1,250. Depicted are the observed prevalence (dots), pointwise posterior median prevalence (dashed line), and pointwise 95\% credible intervals (shaded region) all scaled by the assumed population size. Latent posterior estimates are based on a thinned sample, with every $250^{th}$ sample retained.}}
	\label{fig:popsize_misspec_latent_posts}
\end{figure}

\added{We are able to obtain approximately valid inference under moderate misspecification of the population size. However, estimates of the epidemic dynamics and the case detection probability become severely biased as the magnitude of the misspecification increases. Furthermore, the widths of the credible intervals for the model parameters shrink as misspecification of the population size becomes more severe. The constrained ranges of model dynamics also manifest in a narrowing of the widths of the pointwise credible intervals for disease prevalence (Figure \ref{fig:popsize_misspec_latent_posts}). Under severe misspecification of the population size (N = 150), the latent posterior distribution has 95\% of its mass within only a narrow band of epidemic paths. In contrast, under moderate misspecification of the population size, the widths of the latent posterior credible intervals are quite similar to the estimated range using the true population size.}

\added{There are two final points that we wish to make based on this simulation. The first is that it might be possible to deliberately misspecify the true population size in order to speed up computation time and still obtain approximately valid inference. The average run time using the true population size of 1250 individuals was roughly $ 2\times $ and $ 7\times $ the average run times in populations of 900 and 500 individuals. Yet, posterior inferences about the epidemic dynamics were not substantially affected. Longer run times in large populations result from having to sample more subject--paths per MCMC iteration at a relatively higher cost per subject--path. The second point is that in situations where the true population size is unknown, 
SEM likelihood--based inference has some robustness to misspecification of the population size, at least in a neighborhood of population sizes around the true number of individuals. Thus, comparing posterior inferences under a range of population sizes could be a useful heuristic diagnostic for population size misspecification.}

\subsection{\replaced{Effect of prior specification on posterior inference}{Inference for the posterior distribution of the latent process}}
\label{sec:prior_effect_sim}
\replaced{Given the relatively limited extent of aggregated prevalence counts compared to a  setting in which subject--level data is available, we must be concerned with how our choices of prior distributions influence our posterior inferences. We simulated an outbreak with SIR dynamics in a population of 750 individuals for which $ R_0 = \beta \times 763 / \mu = 1.8375 $ and the mean infectious period was $ 1/\mu = 7 $ days. We fit four SIR models to binomially distributed weekly prevalence data, sampled with detection probability $ \rho = 0.2$, under the following four prior regimes: Regime 1 --- informative priors for all model parameters; Regime 2 --- vague priors for the rate parameters and an informative prior for the sampling probability; Regime 3 --- informative priors for the rate parameters and a flat prior for the sampling probability; Regime 4 --- vague priors for the rate parameters and a flat prior for the sampling probability. Complete simulation details and convergence diagnostics are supplied in Section \ref{sec:prior_effect_details}.}{We simulated an epidemic in a population of size $N=750$ with proportions of initially susceptible, infected, and recovered individuals of 0.9, 0.03, and 0.07. The per--contact infectivity rate was $\beta=0.00185$ and the recovery rate was $\mu=0.5$, which together correspond to a basic reproduction number of $R_0 = \beta N/\mu = 2.78$ and a mean infectious duration of two days. Binomially sampled prevalence counts were drawn at observation times 0, 1,..., 15 with sampling probability $\rho = 0.2$. We ran three MCMC chains, each for 250,000 iterations, updating the paths of ten subjects, chosen uniformly at random, per MCMC iteration. Traceplots of the log--likelihood and parameters were monitored for convergence and are shown in Section \ref{sec:sim1_supp}.}

\begin{figure}[!h]
	\centering
	\includegraphics[width=\linewidth]{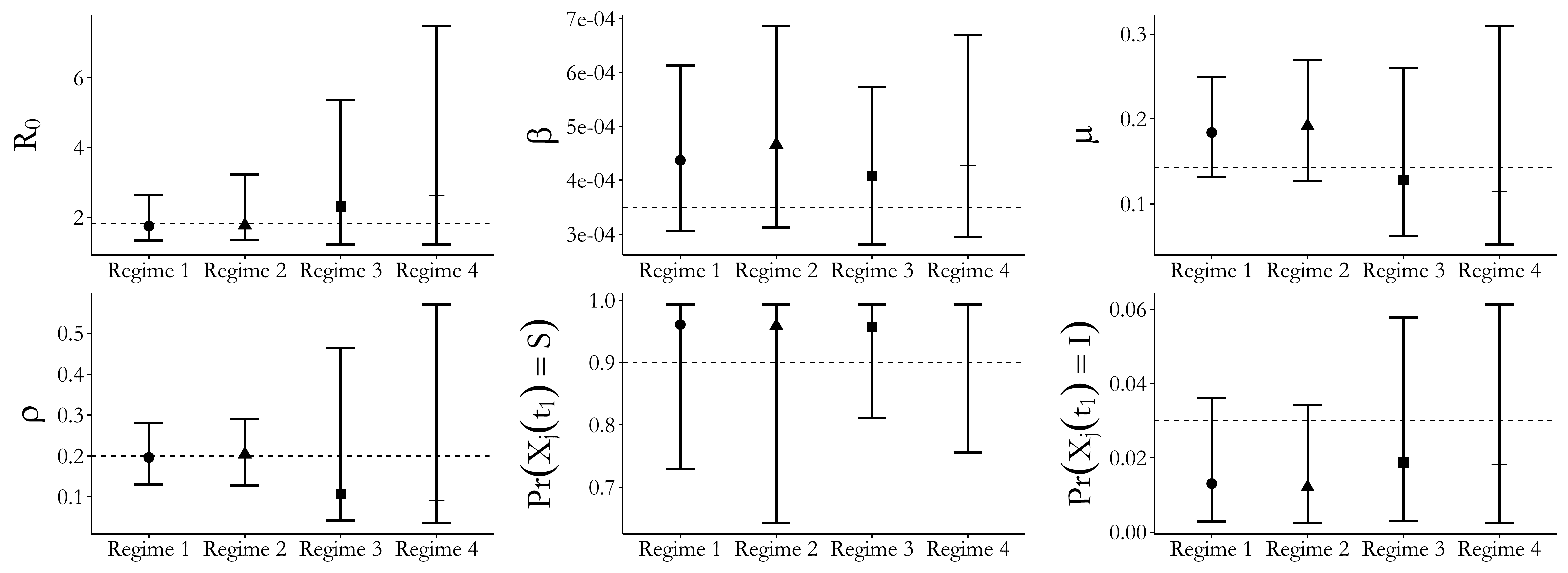}
	\caption{\added{Posterior median estimates and 95\% credible intervals for all SIR model parameters under four different prior regimes (Table \ref{tab:prior_effect_priors}). Regimes 1 and 3 set informative priors for the per--contact infectivity and recovery rates. Regimes 1 and 2 set informative priors for the binomial sampling probability. The same mildly informative prior for the initial state probabilities was used in all four regimes.}}
	\label{fig:prior_credints}
\end{figure}

\begin{figure}[!h]
	\centering
	\includegraphics[width=0.6\linewidth]{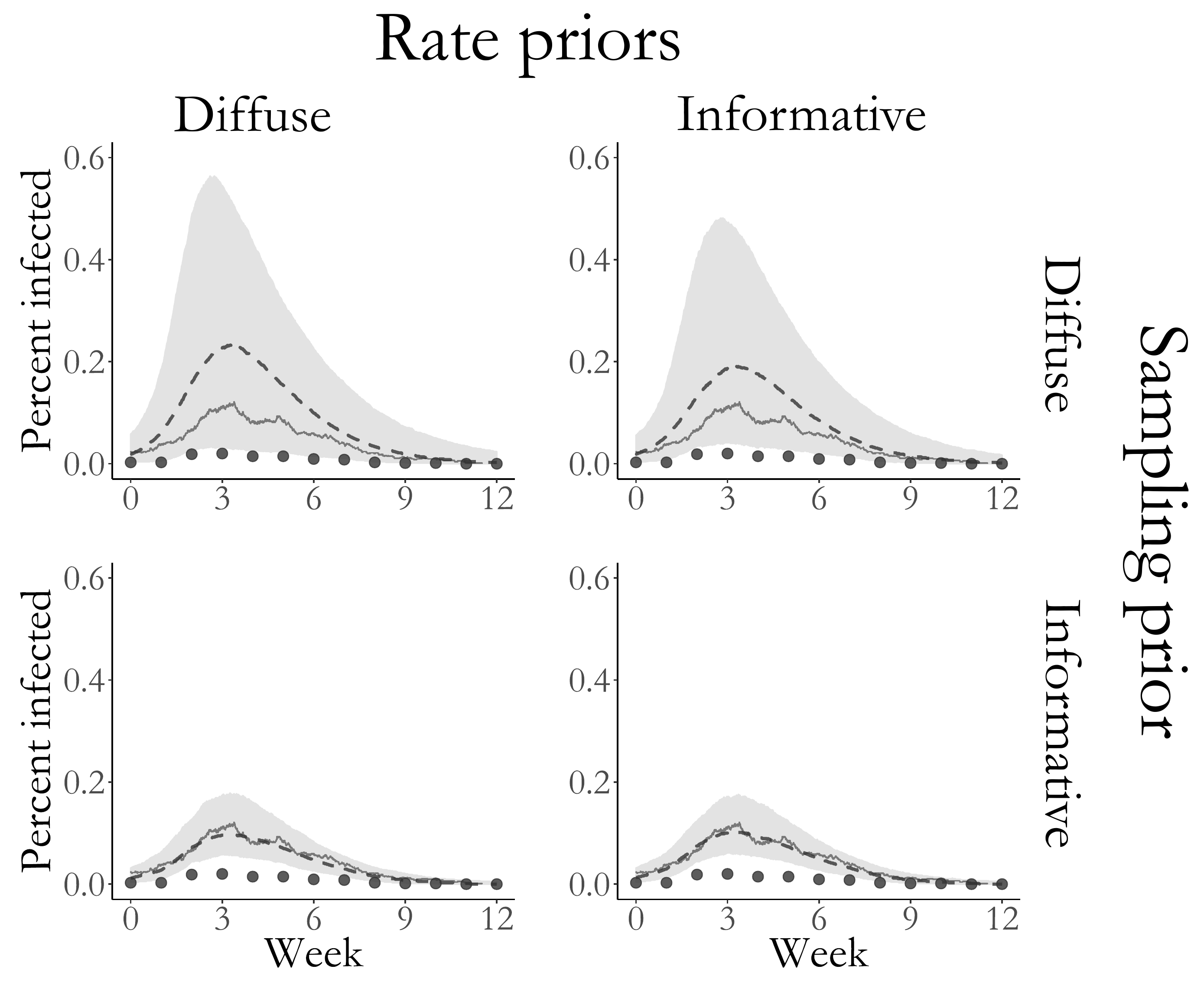}
	\caption{\added{Estimated latent posterior distributions of disease prevalence in outbreaks simulated under four prior regimes for SIR model rate parameters and the binomial sampling probability. Depicted are the true unobserved prevalence (solid line), observed data (dots), pointwise posterior median prevalence (dashed line), and pointwise 95\% credible intervals (shaded region). Latent posterior estimates are based on a thinned sample, with every $250^{th}$ sample retained.}}
	\label{fig:prior_latent_posts}
\end{figure}

The true values for all model parameters fell within the 95\% credible intervals under all \replaced{four}{three} prior regimes. \replaced{Informative priors tended to result in narrower credible intervals for the parameters (Figure \ref{fig:prior_credints}) as well as for the latent process (Figure \ref{fig:prior_latent_posts}), though the effects of the prior for the detection probability were particularly pronounced. The strength of prior information about the sampling probability affected the widths of credible intervals to a much greater extent than the priors for the rate parameters. Strong prior information about the sampling probability also resulted in substantially narrower credible intervals for disease prevalence under each of the prior regimes for the rate parameters. In contrast, informative priors for the rate parameters yielded only slightly narrower credible intervals for disease prevalence when holding constant the strength of the sampling probability prior. The effects on the initial state probability parameters seem to reverse this pattern, although we caution against overinterpretion given the paucity of data available for estimating those parameters. MCMC chains with strong priors for the binomial sampling probability also appeared mix somewhat better than chains with diffuse priors for the sampling probabilty (see traceplots in Section \ref{sec:prior_effect_details}).}{The choice of prior regime did not affect the acceptance rate for subject--path proposals, and was roughly 93\% acceptance under all three prior regimes. However, the widths of Bayesian credible intervals (Figure \ref{fig:sim_credint}), and the widths of pointwise posterior distributions for the latent process (Figure \ref{fig:latent_post_all}) were sensitive to the prior regime.  In particular, additional information about the detection probability improved inference for both the model parameters and the latent process. Finally, posterior samples of the recovery rate and binomial sampling probability were highly correlated (supplementary Figures \ref{fig:sim1pairsdiffuse}, \ref{fig:sim1pairsinformsamp}, and \ref{fig:sim1pairsinformative}).}

%% file: example_bbs_finalV2.tex
\section{Influenza in a British boarding school}
\label{sec:app_bbs}

\begin{figure}[ht!]
	\centering
	\includegraphics[width=\linewidth]{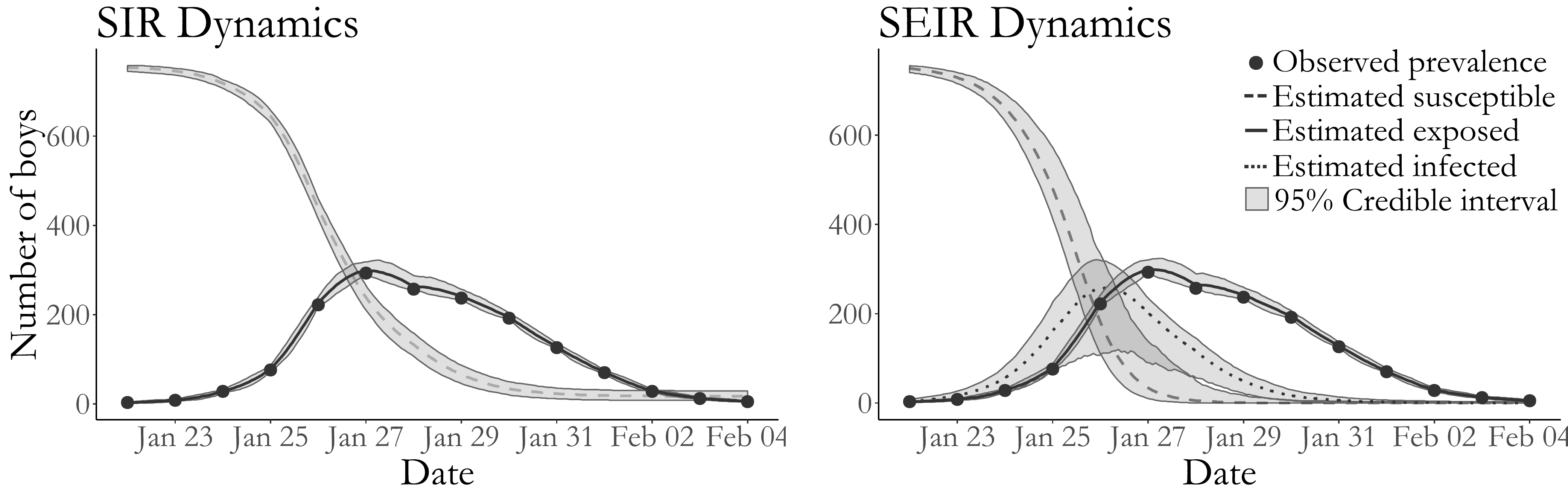}
	\caption{\added{Boarding school data, pointwise posterior median estimates and pointwise 95\% credible intervals (grey shaded areas) under SIR and SEIR dynamics of the numbers of susceptible boys (dashed line), exposed boys (dotted line), and infected boys (solid line). Posterior estimates based on a thinned sample, with every 250$ ^{th} $ configuration retained.}}
	\label{fig:bbs_dat}
\end{figure}

As an application, we analyze data from an outbreak of influenza in a British boarding school \citep{anon1978, davies1982}. This outbreak took place shortly after the Easter term began in January 1978, and was estimated to eventually infect roughly 90\% of the 763 boys aged 10--18. Daily counts of the boys who were confined to the infirmary from January $22^{nd}$ through February $4^{th}$ were accessed via the \texttt{pomp} package in \texttt{R} \citep{pomp}, and are displayed in Figure \ref{fig:bbs_dat}. 

\begin{figure}
\centering
\includegraphics[width=0.95\linewidth]{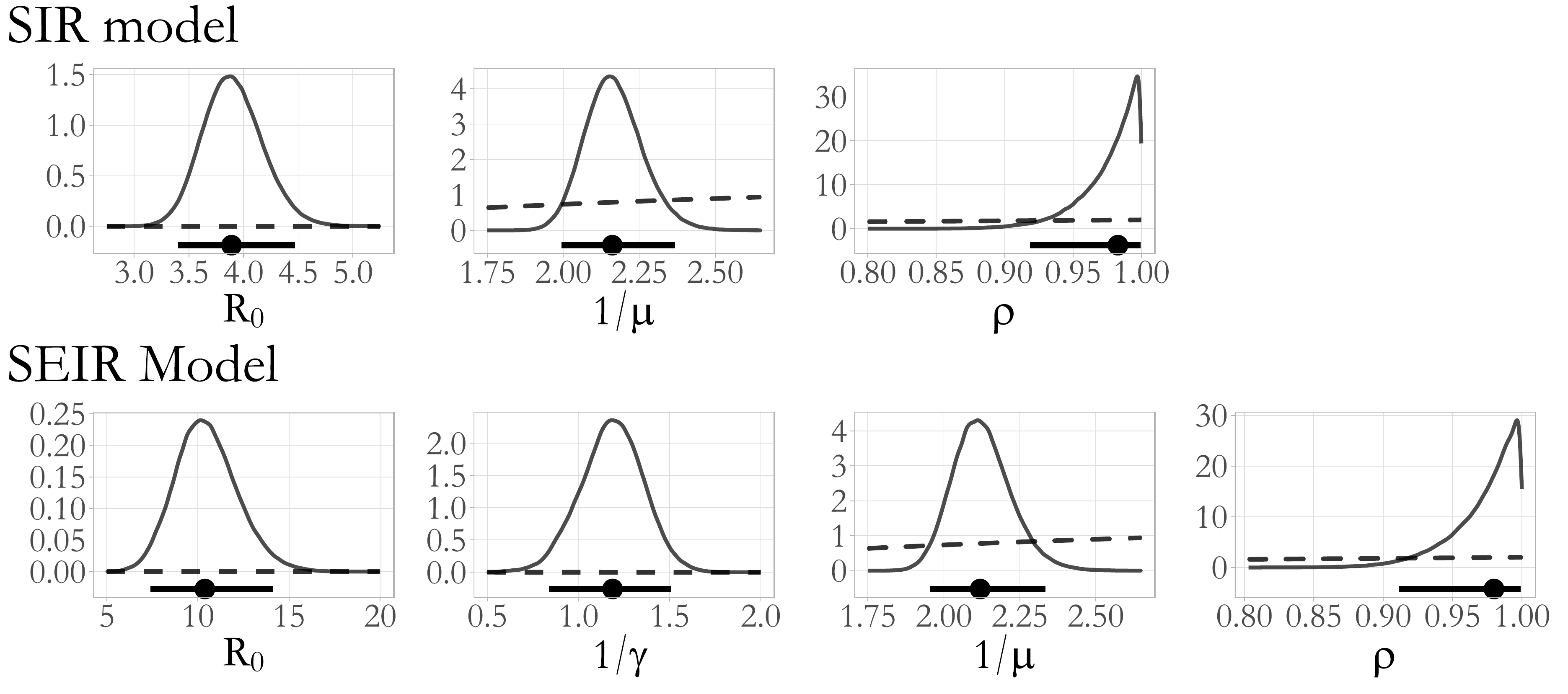}
\caption{\added{Posterior density estimates for $ R_0 = \beta N /\mu $, the mean latent and infectious periods, $ 1/\gamma $ and $ 1/mu $, and the binomial sampling probability, $ \rho $, from SIR and SEIR model parameters fit to the British boarding school data (solid lines). The posterior median and 95\% Bayesian credible intervals are drawn below the density plots (solid lines with circles). The implied prior densities (dashed lines) for $ R_0 $ and the latent and infectious periods, and the prior density for the binomial sampling probability, are plotted over the posterior ranges.}}
\label{fig:bbs_densities}
\end{figure}

\replaced{We used our DA algorithm and PMMH to fit SIR and SEIR models with a binomial emission distribution to the data (see Section \ref{sec:bbs_supp} of the supplement for complete details). All of the parameters were assigned diffuse priors, which are plotted over the posterior ranges in Figure \ref{fig:bbs_densities}. The PMMH algorithm failed to converge for both models, which we suspect was due to a combination of model misspecification and the constrained state space of the binomial measurement process. We also fit a set of supplementary SIR and SEIR models in section \ref{sec:bbs_neg_binom}, in which we assumed a negative--binomial emission distribution. This was done in order to facilitate comparison with PMMH, although we feel that a negative binomial emission distribution is not appropriate in such a closely monitored outbreak setting since it does not rule out over--reporting of cases.}{We fit an SIR model to the data, running MCMC five chains in parallel, each for 250,000 iterations. We discarded the first 1,000 iterations as burn-in, and sampled ten subject--paths per MCMC iteration. We also ran another set of three chains in parallel for 250,000 iterations each, in which we sampled 75 subject paths per MCMC iteration. The total run time for this second set of chains was over seven times longer than for the set of five chains and yielded virtually identical estimates. Convergence of the chains was assessed visually (see Section \ref{sec:bbs_supp} of the Supplement), and the retained parameter samples and latent posterior paths following the burn--in period were combined to form the final posterior sample.}

\added{Together, the SIR and SEIR models suggest that cases were detected with high probability and that the outbreak, though aggressive, was not atypical given the closed environment in which it occurred. The posterior median estimates of the detection probability, roughly 0.98 for both models (SIR 95\% BCI: 0.92, 1.00; SEIR 95\% BCI: 0.91, 1.00), suggested that while almost all of the infectious boys were detected, a handful of cases likely went unnoticed. The posterior median recovery rate under SIR dynamics corresponds to an average period of 2.16 days (95\% BCI: 1.99, 2.37) during which an infectious boy could transmit an infection to other boys before being confined to the infirmary. Under SEIR dynamics, the posterior median average infectious period was 2.12 days (95\% BCI: 1.95, 2.33), and the posterior median average latent period was 1.19 days (95\% BCI: 0.84, 1.51). These results are consistent with the typical progression of influenza, in which individuals typically incubate for between one to four days before symptoms manifest, and are typically infectious for one day before, and up to a week after, symptom onset \citep{cdcFlu}. The posterior median estimates of $ R_0 $ were 3.89 (95\% BCI: 3.40, 4.47) under SIR dynamics, and 10.38 (95\% BCI: 7.40, 14.11) under SEIR dynamics. Previous analyses of this dataset with trajectory matching estimate $ R_0 $ to be roughly 3.7 for the SIR model and 35.9 for the SEIR model \citep{wearing2005, keeling2008}, though we note that these estimates are based on deterministic models that do not properly account for distributional properties of the data. Our results for both models are also in agreement with estimates of SIR and SEIR model dynamics under a negative binomial emission distribution (see Section \ref{sec:bbs_neg_binom}).}

\deleted{The posterior median recovery rate corresponds to an average period of 2.17 days (95\% BCI: 2.0, 2.38) during which an infectious individual could contact susceptible boys before being confined to the infirmary. Our posterior median estimate of $R_0$ was 3.90 (95\% BCI: 3.41, 4.48). Previous analyses of this dataset, using trajectory matching, estimated the mean infectious duration to be roughly 2.2 days, and $R_0$ to be roughly 3.7 (Wearing et al., 2005, Keeling and Rohani, 2008). These estimates of $R_0$ are similar to those from other influenza outbreaks occurring in closed environments (Biggerstaff et al., 2014). Our posterior median estimates of the probabilities that an individual was susceptible, infected, or recovered at the start of the outbreak were, respectively, 0.99 (95\% BCI: 0.98, 0.99), 0.003 (95\% BCI: 0.001, 0.007), and 0.009 (0.004, 0.017). Finally, our posterior median estimate of $ \rho $ was 0.98 (95\% BCI: 0.92, 1.00), suggesting that, while almost all of the infections were observed on each day, a handful of cases likely remained undetected. This is consistent with the typical progression of influenza, in which individuals are infectious for about a day before becoming symptomatic (Centers for Disease Control and Prevention).}

\deleted{We specified a Beta(2,1) prior, which density linearly increases from on its support between 0 to 1, for $ \rho $, since the majority, though perhaps not all, of the infections on any given day would likely have been detected in such a closely monitored environment. 
We selected diffuse priors for the infectivity and recovery rates, taking  $ \beta\sim $Gamma(0.001,1) and $ \mu \sim$Gamma(1,2), the former being highly diffuse and the latter chosen by assuming that on average individuals developed symptoms and got isolated 2 days after the infection. Since roughly 90\% of the students were eventually infected, it is likely that only a few students were either infected or immune prior to the start of the outbreak. Therefore, we set a Dirichlet(900, 3, 9) prior for the initial disease state probabilities.}

%% file: conclusion_finalV2.tex
\section{Conclusion}
\label{sec:conc}

We have presented an agent--based Bayesian DA algorithm for fitting SEMs to disease prevalence time series counts. This was previously difficult, if not impossible, to carry out using traditional \added{agent--based} DA methods in the absence of subject--level data. \replaced{Although we outlined the BDA algorithm in the context of fitting an SIR model to binomially distributed prevalence data, our algorithm represents a general solution for fitting SEMs to prevalence counts. In simulations and the applied example, we fit SEIR and SIRS models to prevalence data, and in the supplement also fit SIR and SEIR models with a negative binomial emission distribution to the British boarding school data. We have demonstrated that our algorithm yields approximately valid inference when the population size is misspecified. Moreover, our algorithm is usable in settings in which simulation--based methods, such as PMMH, break down due to misspecification of the SEM.}{Although we have presented our DA algorithm in the context of fitting an SIR model to prevalence data, the machinery in our algorithm relied neither on the SEM dynamics, nor on the binomial distribution used to model the data. Therefore, our algorithm represents a general solution to the problem at hand, and may be applied, with minor modifications, to a broad class of SEMs. We have demonstrated that updating a small fraction of the subject--paths per MCMC iteration optimizes the MCMC efficiency. Furthermore, moderate misspecification of the assumed population size still yields approximately valid inference. Therefore, our DA algorithm is likely to be of practical use, even in analyzing epidemics in moderately large populations. However, we do not view this algorithm as a solution for analyzing epidemics in very large populations as run length times and MCMC mixing will eventually deteriorate. Still, in many scenarios, this DA algorithm will mitigate the need for extremely computationally expensive simulation--based methods, or for approximate methods.} Finally, our DA algorithm is carried out entirely at the subject level, making it possible to also incorporate subject--level covariates and fit models to subject--level data. 

\added{There are two fundamental limitations of agent-based DA methods from which our algorithm is not excepted. First, the bookkeeping required to track the collection of subject--paths increases in size and complexity as the number of events grows large. Attempts to fit stochastic epidemic models in large populations using agent-based DA may be thwarted by prohibitive computational overhead. MCMC run times using our implementation, which was coded for reliability rather than speed, substantially degraded once the assumed population size was greater than a few thousand people. Second, we suspect that MCMC mixing in large populations could eventually become too slow for agent--based DA to be of practical use, even if solutions could be found for the computational bottlenecks. As the population size gets large, perturbations to the likelihood from re-sampling one subject at a time become relatively less significant. For this reason, we view extensions for jointly sampling multiple subject--paths as a critical step in mitigating slow MCMC mixing in large populations.}

\replaced{Finally}{To conclude}, we would like to comment on directions for future work that we intend to pursue. \replaced{T}{First, t}he DA algorithm in this paper addresses the problem of fitting SEMs to prevalence data. This type of data summarizes total number of infections in the population at a particular time. However, \replaced{outbreak}{the epidemiological} data often consist of incidence counts, which are the number of new cases accumulated in each inter-observation interval. Extending our DA algorithm to accommodate incidence data is an important next step and should be straightforward in situations where the state space for the subject level process is finite --- for instance, if a subject cannot become reinfected more than once or twice in a given inter-observation interval. \replaced{We also believe it is important to investigate whether there is a way to make our DA algorithm more efficient by selecting the subjects whose paths are resampled in each iteration in a way that maximizes the perturbation to the population--level path and does not invalidate the MCMC. Designing an optimal schedule of subject--path updates could be critical to the application of our algorithm to more complex models fit to epidemics in large, structured populations.}{Second, although we have shown that it suffices to sample only a small fraction of the subject-level paths, it is likely that our already efficient DA algorithm could be made even more so if the schedule of subject--paths to update could be chosen to maximally perturb the population level path. Finally, an obvious next step in assessing the usefulness of the algorithm is to fit SEMs with more complex dynamics to a variety of datasets.}

\section{Acknowledgements}
J.F., J.W., and V.N.M. were supported by the NIH grant U54 GM111274. J.W. was supported by the NIH grant R01 CA095994. V.N.M. was supported by the NIH grant R01 AI107034. \added{We would also like to thank Aaron King and the rest of the authors of the \texttt{pomp} package for their help with the PMMH algorithm that served as a benchmark for the methods presented in this paper.}

%% file: supplement_finalV2.tex
\beginsupplement   
\section{SIR Model Construction and Lumpability of CTMCs}
\label{sec:lumpability}
In this section, we outline why the SIR model of Section \ref{subsec:pop_proc} is equivalent to the canonical SIR model \citep{kermack1927, andersson2000stochastic} via a property called \textit{lumpability}. The following discussion is not meant to be a comprehensive presentation of the theoretical details behind the connection between the two models. We refer readers seeking a more thorough presentation to \citet{tian2006lumpability}.

Given a Markov process, $ \bX $ with state space $ \mcS = \lbrace s_1,\dots,s_P\rbrace $ and initial probability vector $ \pi $, we define a new process, $ \overline{\bX} $ on state space $ \overline{\mcS} = \lbrace S_1,\dots,S_\mathcal{L}\rbrace $, a partition of $ \mcS $. The jump chain of this new chain is obtained by taking the sequence of subsets of $ \overline{\mcS} $ that contain the corresponding states of the original jump chain. The initial probability distribution of $ \overline{\bX}(t) $ is \begin{equation*}
\Pr(\overline{\bX}(t_0) = S_i) = \mathrm{Pr}_\pi(\bX(t_0) \in S_i)
\end{equation*}
and its transition probabilities are given by
\begin{equation*}
\Pr(\overline{\bX}(t+\Delta t) = S_j | \overline{\bX}(t)=\overline{\bx}(t^\prime), t^\prime \leq t) = \Pr(\overline{\bX}(t+\Delta t) \in S_j | \bX(t)=\bx(t^\prime), t^\prime \leq t),
\end{equation*}
where $ \overline{\bx}(t^\prime) $  and $ \bx(t^\prime) $ denote the paths of the original process and the new process. The new process is called the \textit{lumped process}. We say that the original process is \textit{lumpable} with respect to a partition $ \overline{\mcS} $ of $ \mcS $, and that $ \overline{\bX}(t) $ is the \textit{lumped Markov process} corresponding to $ \bX(t) $, if for every choice of $ \pi $ we have that $ \overline{\bX}(t) $ is Markov and the transition probabilities do not depend on $ \pi $. A necessary and sufficient condition for a CTMC to be lumpable is that its rate matrix, $ \bLambda = (\lambda_{a,b}) $, where $ \lambda_{a,b} $ being the rate of transition from $ s_a $ to $ s_b $, satisfies
\begin{equation*}
\sum_{s_b \in S_B}\lambda_{a,b} = \sum_{s_b \in S_B}\lambda_{c,b}
\end{equation*}
for any pair of sets $S_A$ and $ S_B$ and for any pair of states  $(s_a, s_c)$  in $S_A \in \overline{\mcS} $. 
 
\begin{figure}
	\centering
	\includegraphics[width=0.9\linewidth]{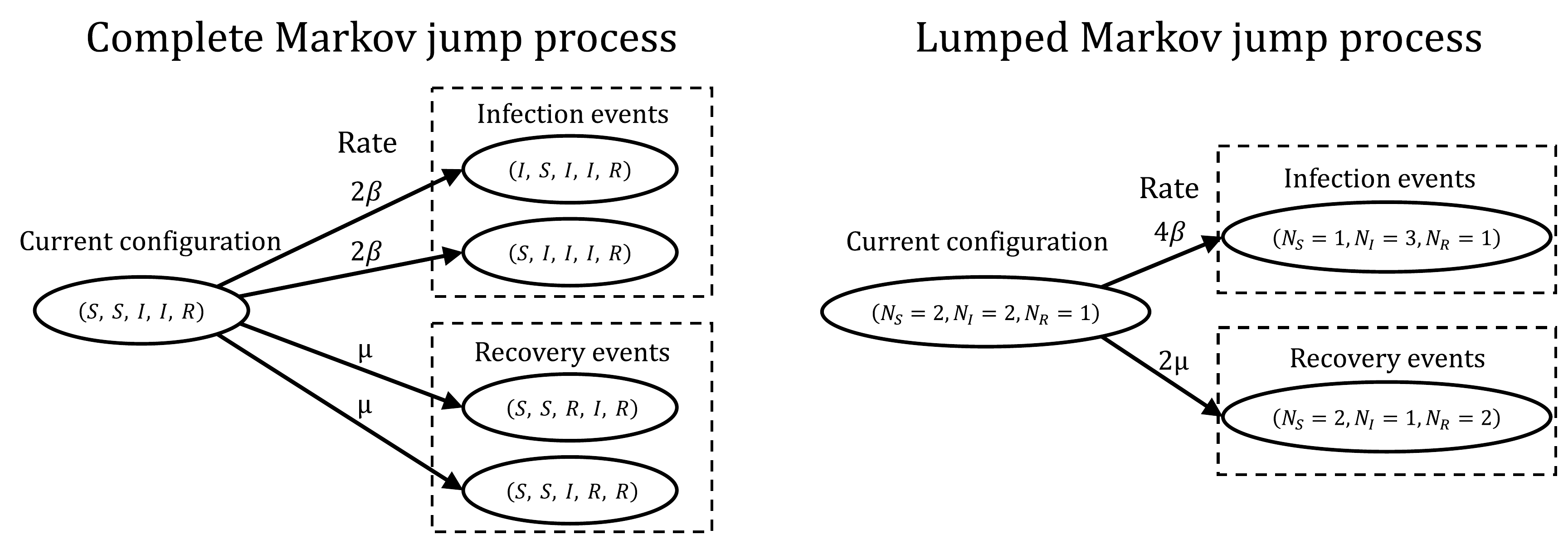}
	\caption{\added{Complete and lumped representations of SIR dynamics in a population of five individuals. The per--contact infectivity rate, $ \beta $, and the recovery rate, $ \mu $, parameterize exponential waiting time distributions between transition events. The complete Markov jump process evolves on the state space of subject state labels, $ \mcS = \lbrace S,I,R\rbrace^N $, with dynamics determined by the subject--level transition rates. Each susceptible may contact two infected individuals, while each infected individual recovers independently. The lumped process evolves on the state space of compartment counts, $ \widetilde{\mcS} = \lbrace N_S,N_I,N_R:\ N_S + N_I+N_R=N\rbrace $, with dynamics determined by lumped transition rates. The waiting time distributions between transitions are derived by noting that if $ \tau_1\sim \exp(\lambda_1) $ and $ \tau_2\sim\exp(\lambda_2) $, then $ \tau_{\min} = \min(\tau_1,\tau_2)\sim\exp(\lambda_1+\lambda_2) $.}}
	\label{fig:sirrepresentations}
\end{figure}

In Section \ref{sec:methods_sir}, we defined the latent process, $ \bX(\tau) = (\bX_1,\dots,\bX_N)$, with state space $ \mcS = \lbrace S, I, R\rbrace^N $. Let $ c_u = (x_1,\dots,x_N) $ denote a configuration of the state labels (e.g. $ c_u = (S, I, S, R, I) $), and denote the set of configurations that correspond to a vector of compartment counts by $$ \mcC_{lmn} = \left \lbrace c_u:l=\sum_{i=1}^N\ind{x_i = S},m=\sum_{i=1}^N\ind{x_i=I},n = \sum_{i=1}^N\ind{x_i=R},\ l+m+n=N \right \rbrace. $$ 
The state space of count vectors, 
$$ \overline{\mcS} = \left \lbrace \mcC_{lmn}: l,m,n \in \lbrace1,\dots,N\rbrace, l+m+n = N \right \rbrace, $$ 
defines a partition of $ \mcS $ that is obtained by stripping away the subject labels and summing the number of individuals in each disease state. 

Given the partition $ \overline{\mcS} $ of $ \mcS $, we may define the CTMC for the canonical SIR model, $ \overline{\bX} = (S_\tau, I_\tau, R_\tau) $, on the state space of compartment counts, depicted in Figure \ref{fig:sirrepresentations}. This construction is usually presented for computational reasons since discarding the subject labels for infection and recovery events substantially reduces the computational overhead. When the sojourn times are exponentially distributed, the transition rates for the time-homogeneous CTMC are 
\begin{equation*}
\begin{array}{cc}
\underline{\text{Transition}} & \underline{\text{Rate}} \\
(S,I,R) \longrightarrow (S-1,I+1,R) & \beta S I ,\\
(S,I,R) \longrightarrow (S,I-1,R+1) & \mu I .
\end{array}
\end{equation*}

The state space $ \overline{\mathcal{S}} $ partitions the state space $ \mathcal{S} $ into groups of configurations for which the triple of compartment counts are the same. The CTMC $\overline{\bX}$ trivially satisfies the condition for lumpability, and thus is the lumped Markov chain of $ \bX $ with respect to this partition.

\newpage

\section{\added{Computing the matrix exponential}}
\label{sec:mtx_exp}
\added{The transition probability matrix (TPM), $ \bP(t) $, for a time--homogeneous CTMC over an interval of length $ t $, solves the matrix differential equation \[ \frac{\rmd}{\rmd t}\bP(t) = \bLambda\bP(t),\hspace{0.2in} \mathrm{s.t.\ }\mathbf{P}(0) = \mathbf{I}, \] where $ \bLambda $ is the transition rate matrix for the CTMC and $ \mathbf{I} $ is an identity matrix of the same size as $ \bLambda $ \citep{wilkinson2011stochastic}.  Therefore, $ \bP $ is computed using the matrix exponential solution of the above differential equation, $ \bP = \exp(t\bLambda) $. This is the most intensive step in our algorithm. However, we may lessen the computational burden to a large extent by leveraging the fact that we are computing the matrix exponential for the same rate matrix for possibly many values of $ t $. Therefore, computing the matrix exponential using the eigen decomposition of $ \bLambda $ and caching the resulting eigenvalues and eigenvectors will be relatively efficient \citep{moler2003nineteen}. We outline this computation in the following two cases: when the eigenvalues of $ \bLambda $ are all real (e.g. as with the SIR and SEIR models), and when $ \bLambda $ has complex eigenvalues (e.g. as is possibly the case with the SIRS model).}

\subsection{\added{Case 1: $ \bLambda $ has real eigenvalues}}
\added{Suppose that $ \bLambda_{n\times n} =  \mathbf{UVU}^{-1}$, where $ \mathbf{V} $ is a diagonal matrix of eigenvalues, $ v_1,\dots,v_n $, and $ \mathbf{U} $ is the matrix whose columns are the corresponding right eigenvectors. Then, \[ \e^{t\mathbf{V}} = \diag(\e^{v_1t},\dots,\e^{v_nt}). \] That $ \mathbf{U} $ is nonsingular yields \[ \e^{t\bLambda} = \mathbf{U} \e^{t\mathbf{V}}\mathbf{U}^{-1}. \] }

\subsection{\added{Case 2: $ \bLambda $ has complex eigenvalues}}
\added{In the event that $ \bLambda $ has complex eigenvalues, we may obtain a real-valued TPM by transforming $ \bLambda $ into its real canonical form \citep{hirsch2013differential}. Suppose that $ \bLambda $ has $ r $ real eigenvalues, $ v_1,\dots,v_r $, with corresponding real eigenvectors, $ \bu_1,\dots,\bu_r $, and $ n-r $ pairs of complex conjugate eigenvalues. Let $ (\bu_j|\bw_j) $ denote the real and imaginary parts of the eigenvector corresponding to the $ j^{th} $ eigenvalue, $ \alpha_j + i\beta_j $, for $ j = r+1,\dots,n $, and define the matrix $ \mathbf{T} = \left (\bu_1|\dots|\bu_r|\bu_{r+1}|\bw_{r+1}|\dots|\bu_n|\bw_n\right ) $. 
The real canonical form for a rate matrix with complex eigenvalues can now be written as $ \mathbf{V} = \mathbf{T}^{-1}\bLambda \mathbf{T} $, where $ \mathbf{V} = \diag(v_1,\dots,v_r,\mathbf{B}_{r+1},\dots, \mathbf{B}_n) $, and each $ \mathbf{B}_j,\ j=r+1,\dots,n $ is given by } \[ \mathbf{B}_j = \left (\begin{array}{cc}
	\alpha_j & \beta_j \\
	-\beta_j & \alpha_j
	\end{array}\right ), \]
\added{which implies that }
\[ \e^{t\mathbf{B}_j} = \e^{\alpha_jt}\left (\begin{array}{cc}
	\cos(\beta_jt) & \sin(\beta_jt) \\
	-\sin(\beta_jt) & \cos(\alpha_j)
	\end{array}\right ), \]
\added{and hence $ \e^{t\mathbf{V}} = \diag(\e^{v_1t},\dots,\e^{v_rt},\e^{t\mathbf{B}_{r+1}},\dots, \e^{t\mathbf{B}_n}) $. Therefore, we can compute the matrix exponential of $ t\bLambda $ as $ \e^{t\bLambda} = \mathbf{T}\e^{t\mathbf{V}}\mathbf{T}^{-1}. $}

\newpage 

\section{Forward-Backward Algorithm for Sampling the Disease State at Observation Times}
\label{sec:fb_algo}

The stochastic forward-backward algorithm \citep{scott2002} enables us to efficiently sample from $ \pi\left (\bX \mid \bY, \bX_{(-j)}, \btheta\right ) $ by recursively accumulating, in a ``forward'' pass, information about the probability of various paths through $ \mcS $, conditional on the data, and then recursively sampling a trajectory in a ``backwards'' pass. Let $ \bY_{t_1}^{t_\ell} = (Y_{1},\dots,Y_{\ell}) $ denote the observations made at times $ t_1,\dots,t_\ell $, and similarly, let $ \bX_{j,t_{L-\ell+1}}^{t_L} = (\bX_{j}(t_{L-\ell + 1}),\dots, \bX_j(t_L)) $ denote the state of $ \bX_j $ at times $ t_{L-\ell+1},\dots,t_L $. In the forward recursion, we construct a sequence of matrices $ \bQ^{(t_2)}_j,\dots,\bQ^{(t_L)}_j $, where $ \bQ^{(t_\ell)}_j = \left (q_{j,r,s}^{(t_\ell)}\right )$, and $ q_{j,r,s}^{(t_\ell)} = \Pr\left (\bX_j(t_\ell) = s, \bX_j(t_{\ell - 1}) = r \mid \bY_{t_1}^{t_{\ell}},\bX_{(-j)},\btheta \right )$. Let $ \bP^{(j)}_{r,s}(t_{\ell - 1},t_\ell) = \Pr\left (\bX_j(t_\ell) = s \mid \bX(t_{\ell-1})=r, \btheta; \bX_{(-j)}\right ) $. If there are changes in the numbers of infected individuals in interval $ \mathcal{I}_\ell $, we construct the transition probability matrix for that interval as in (\ref{eqn:inhomog_tpmprod}). Then, 
\begin{equation}
q_{j,r,s}^{(t_\ell)} \propto \pi_{j}^{(t_\ell)}\left (r \mid \bX_{(-j)}, \btheta\right ) \times \bP^{(j)}_{r,s}\left (t_{\ell-1},t_\ell\right ) \times f\left (Y_{t_\ell} \mid \bX_j(t_\ell),\bX_{(-j)}(t_\ell), \rho, \bp_{t_1}\right ),	
\end{equation}
where $  \pi_{j}^{(t_\ell)}\left (r \mid \bX_{(-j)}, \btheta, \rho\right ) = \sum_r q_{j,r,s}^{(t_j)}$ and with proportionality reconciled via $ \sum_r\sum_s q_{j,r,s}^{(t_j)}=1 $.

In the backwards pass, we sample the sequence of states at times $ t_1,\dots,t_L $ from the distribution $ \pi\left (\bX \mid \bY, \bX_{(-j)},\btheta, \rho,\bp_{t_1}\right )$. To do this, we first note that\footnotesize \begin{align*}
	\pi\left (\bX \mid \bY, \bX_{(-j)},\btheta, \rho,\bp_{t_1}\right ) &= \pi\left (\bX_j(t_L) \mid \bY_{t_1}^{t_L}, \bX_{(-j)},\btheta,\rho, \bp_{t_1}\right ) \prod_{\ell = 1}^{L-1}\pi \left (\bX_j(t_{L-\ell}) \mid \bX_{j,t_{L-\ell+1}}^{t_L}, \bX_{(-j)}, \bY_{t_1}^{t_L}, \btheta, \rho, \bp_{t_1}\right )\\
	&= \pi\left (\bX_j(t_L) \mid \bY_{t_1}^{t_L}, \bX_{(-j)},\btheta,\rho, \bp_{t_1}\right ) \prod_{\ell = 1}^{L-1}\pi \left (\bX_j(t_{L-\ell}) \mid \bX_{j,t_{L-\ell+1}}, \bX_{(-j)}, \bY_{t_1}^{t_{L-\ell+1}}, \btheta, \rho, \bp_{t_1}\right ),
\end{align*}  \normalsize
where the second equality follows from the conditional independence of the HMM. We proceed by first drawing $ \bX_j(t_L) $ from $ \pi_j^{(t_L)}\left (\cdot \mid \bX_{(-j)}, \btheta,\rho\right  ) $, and then drawing $ \bX_j(t_\ell),\ \ell = L-1,\dots,1, $ each in turn from the categorical distribution with masses proportional to column $ \bx_j(t_{\ell+1}) $ of $ \bQ_j^{(t_{\ell+1})} $. 

\newpage
	
\section{\replaced{Simulating endpoint Conditioned Time--Homogeneous CTMC Paths}{Simulating Endpoint Conditioned Time--Homogeneous CTMC Paths via Modified Rejection Sampling}}
\label{sec:ecctmc}
In this section, we briefly \replaced{summarize}{outline} the modified rejection sampling \added{and uniformization} algorithm\added{s} for simulating a path from an endpoint-conditioned time-homogeneous CTMC. The following discussion is not meant to be comprehensive, and we refer the reader to \added{the excellent paper by} \citet{hobolth2009} for a more thorough discussion. We also refer the reader to the \texttt{ECctmc R} package for a fast implementation \added{of these algorithms} which we relied upon in implementing our data augmentation algorithm \citep{ECctmc}.  

Our goal is to simulate a path for a time--homogeneous CTMC, $ \bX $, in the interval $ [0,T] $, conditional on $ \bX(0) = a $ and $ \bX(T) = b $. Let $ \bLambda $ be the rate matrix for the process. Let $ \Lambda_{a} $ denote the $ a,a $ diagonal element of $ \bLambda $, and similarly let $ \Lambda_{a,b} $ denote the rate given by the $ a,b $ element. \deleted{Let $ \tau $ denote the hitting time for the first state transition.} \added{We also denote by $ \mathbf{P}(T) $ the transition probability matrix for the CTMC over $ [0,T] $, and $ P_{ab}(T) $ the probability of beginning in state $ a $ and ending in state $ b $. }

\subsection{\added{Modified rejection sampling}}
\added{The modified rejection algorithm proposes paths by explicitly sampling the first transition time when it is known that at least one transition occurred (i.e. when $ a  \neq b $). The remainder of the path is proposed by forward sampling, for instance, via Gillespie's direct algorithm. The proposed path is then accepted if $ \bX(T) = b $. When it is not known whether a transition occurred (i.e. when $ a = b $), a path is proposed via ordinary forward simulation and accepted if $ \bX(T) = b $.}

\deleted{The probability that at least one state change occurs in the interval $ [0,T] $ given that the chain begins in state $ a $ is $
\Pr(0\leq\tau\leq T | \bX(0) = a) = 1 - \e^{-T\Lambda_a}.$
The density of the first transition time given that the chain begins in state $ a $ and that at least one transition occurs in the interval $ [0,T] $ is $
f(\tau | 0 \leq \tau \leq T,\ \bX(0) = a) = \frac{\Lambda_a \e^{-\tau\Lambda_a}}{1 - \e^{-T\Lambda_a}}.$ 
Thus, the CDF of $ \tau $ given that at least one transition occurs in $ [0, T] $ and that $ \bX(0) = a $ is
$
F(\tau | 0\leq\tau\leq T,\ \bX(0) = a) = \int_{0}^{\tau}\frac{\Lambda_a \e^{-t\Lambda_a}}{1 - \e^{-T\Lambda_a}}dt
= \frac{1 - \e^{-\tau\Lambda_a}}{1 - \e^{-T\Lambda_a}}.
$}

\replaced{We sample the first transition time via the inverse--CDF method}{We can now sample the first transition time by the inverse-CDF method}, sampling $ u\sim \mathrm{Unif}(0,1) $ and applying the inverse-CDF function 
\begin{equation} F^{-1}(u) = \frac{-log\left [1 - u \times \left (1 - \e^{-T\Lambda_a}\right )\right ]}{\Lambda_a}. 
\end{equation}

\added{We found that the modified rejection algorithm worked well in fitting the SIR and SIRS models. In the examples we studied in which these models were fit, subject--paths over intervals where the endpoints required multiple jumps ($ S \rightarrow R $, or $ I \rightarrow S $) were almost never considered. Therefore, usually only a single transition time was required to be sampled in a given interval, and accomplishing this using the inverse--CDF method was quite fast.}

\deleted{The modified rejection algorithm proposes paths by explicitly sampling the first transition time when it is known that at least one transition occurred (i.e. when $ a  \neq b $). The remainder of the path is proposed by forward sampling, for instance, via Gillespie's direct algorithm. The proposed path is then accepted if $ \bX(T) = b $. When it is not known whether a transition occurred (i.e. when $ a = b $), a path is proposed via ordinary forward simulation and accepted if the proposed path is valid, i.e. if $ \bX(T) = b $.} 

\subsection{\added{Uniformization}}
\added{The uniformization algorithm samples the path for a time--homogeneous CTMC conditional on the state at the interval endpoints by coupling the original process to a Markov chain determined by an auxilliary Poisson point process. State transitions, including virtual transition where the state does not change, occur at points of this auxilliary process, and the sequence of state labels is drawn from the corresponding Markov chain.}

\added{We construct the transition rate matrix of the auxilliary Markov chain, $ \bY $, as $ R = I + \frac{1}{\mu}\bLambda $, where $ \mu = \max_a \bLambda_a$. Then number of state transitions, $ N $, conditional on $ \bX(0) = a,\ \bX(T) = b $, can be shown to be }
\begin{equation}
\added{P(N=n|\bX(0) = a, \bX(T) = b) = \e^{-\mu T}\frac{(\mu T)^n}{n!}R_{ab}^n / P_{ab}(T).}
\end{equation}
\added{The algorithm proceeds by first sampling the number of state transitions from this distribution. If there are no transitions, or if there is one transition and the states at the endpoints are the same, the algorithm terminates. Otherwise, we drawn $ n $ independent uniform values in $ [0,T] $ and sort them to obtain the times of state transitions. The state labels at the sorted sequence of times, $ \tau_i $, $ i = 1,\dots,n-1 $, is then drawn from the discrete distribution with masses given by}
\begin{equation}
\added{P(\bX(\tau_i) | \bX(\tau_{i-1}, \bX(T) = b) = \frac{R_{x_{i-1},x_i}(R^{n-i})_{x_ib}}{(R^{n-i+1})_{x_{i-1}b})}.}
\end{equation}
\added{We found that uniformization was preferable to modified rejection sampling when fitting the SEIR model. In this case, modified rejection sampling tended to get hung up when sampling paths in intervals where the endpoints suggested that at least two state transitions occured (which though it seldom occured, significantly slowed down the MCMC). We also note that the transition probability, $ P_{ab}(T) $, is computed and cached in carrying out the HMM step of our algorithm. Therefore, there are no additional eigen--decompositions or matrix exponentiations required in using the uniformization algorithm to sample the exact times of state transition.}

\newpage
\section{Metropolis-Hastings Ratio Details}
\label{sec:MH_ratio}
 Our target distribution is $ \pi(\bX | \bY) \propto\pi(\bY|\bX)\pi(\bX) $. Note that $ \xnew $ and $ \xcur $ differ only in the path of the $ j^{th} $ subject, so $ \Lambda^{(-j)}(\xcur) = \Lambda^{(-j)}(\xnew)=\bLambda^{(-j)} $. Suppressing the dependence on $ \btheta$ for clarity, the acceptance ratio is
 \begin{equation*}
 a_{\xcur \longrightarrow \xnew} = \min \left \lbrace \frac{\pi(\xnew|\by)}{\pi(\xcur|\by)}\frac{q(\xcur|\xnew)}{q(\xnew|\xcur)},\ 1\right \rbrace
 \end{equation*}
 
 Now, \begin{eqnarray*}
 	\pi(\xnew|\by) &\propto& \Pr(\by|\xnew)\pi(\xnew),\\
 	\pi(\xcur|\by) &\propto& \Pr(\by|\xcur)\pi(\xcur),
 	\end{eqnarray*}
 	
where $ \Pr(\by|\xnew) $ and $ \Pr(\by|\xcur )$ are binomial probabilities for the measurement process, and $ \pi(\xnew) $ and $ \pi(\xcur) $ are the time--homogenous CTMC densities of the current and the proposed population--level paths that appear in Equation (\ref{eqn:comp_data_likelihood}). Let $ \pi(\xnew_j|\bLambda^{(-j)}; \mcI) $ and $ \pi(\xcur_j| \bLambda^{(-j)}; \mcI) $ denote the time--inhomogeneous subject--level CTMC proposal densities given by (\ref{eqn:subj_level_dens}). Then,
 	\begin{eqnarray*}
	q(\xnew|\xcur) &=& \Pr(\xnew|\by; \bLambda^{(-j)}(\xcur), \mcI)\\
	&=& \frac{\pi(\xnew, \by; \bLambda^{(-j)}(\xcur), \mcI)}{\Pr(\by; \bLambda^{(-j)}, \mcI)}\\
	&=& \frac{\Pr(\by|\xnew)\pi(\xnew_j| \bLambda^{(-j)}; \mcI)}{\Pr(\by; \bLambda^{(-j)}(\xnew), \mcI)}\\
	\shortintertext{and similarly, } q(\xcur|\xnew) &=& \frac{\Pr(\by|\xcur)\pi(\xcur_j| \bLambda^{(-j)}; \mcI)}{\Pr(\by; \bLambda^{(-j)}(\xcur), \mcI)}.
	\end{eqnarray*}
Therefore, 
	\begin{eqnarray*}
	\frac{\pi(\xnew|\by)}{\pi(\xcur|\by)}\frac{q(\xcur|\xnew)}{q(\xnew|\xcur)} &=&  \frac{\Pr(\by|\xnew)\pi(\xnew)}{\Pr(\by|\xcur)\pi(\xcur)}\frac{\Pr(\by|\xcur)\pi(\xcur_j; \bLambda^{(-j)})}{\Pr(\by|\xnew)\pi(\xnew_j; \bLambda^{(-j)})}\\
	&=& \frac{\pi(\xnew)}{\pi(\xcur)}\frac{\pi(\xcur_j| \bLambda^{(-j)}; \mcI)}{\pi(\xnew_j| \bLambda^{(-j)}; \mcI)}.
	\end{eqnarray*}
Hence,
	\begin{equation*}
	a_{\xcur \longrightarrow\xnew}=\min \left \lbrace  \frac{\pi(\xnew)}{\pi(\xcur)}\frac{\pi(\xcur_j| \bLambda^{(-j)}; \mcI)}{\pi(\xnew_j| \bLambda^{(-j)}; \mcI)} , 1 \right \rbrace .
	\end{equation*}

\newpage
\section{\added{Fitting SEIR and SIRS models via Bayesian data augmentation}}
\label{sec:SEIR_SIRS_details}
\subsection{SEIR model formulation}
\added{The Susceptible--Exposed--Infectious--Recovered (SEIR) model adds an additional latent state to the SIR model in which subjects who are exposed to an infected individual incubate before becoming infectious. As with the SIR model, recovery is assumed to confer lifelong immunity. The structure of this model does not affect any of the machinery involved in the subject--path proposal mechanism, but rather merely redefines the population--level time--homogeneous CMTC for the epidemic process, and the subject--level time--inhomogeneous CTMC used in the subject--path proposals.}

\added{Under this model, we suppose that the data are sampled from a latent epidemic process, $ \bX = \lbrace \bX_1,\dots,\bX_N\rbrace $, that evolves in continuous--time as individuals become exposed, infectious, and recover. The state space of this process is $ \mcS = \lbrace S,E,I,R\rbrace^N $, the Cartesian product of $ N $ state labels taking values in $ \lbrace S,E,I,R\rbrace $. The state space of a single subject, $ \bX_j $, is $\mcS_j = \lbrace S, E, I, R\rbrace $, and a realized subject--path is of the form \begin{footnotesize}$ \bx_j(\tau) = \left (S,\ \tau < \tau^{(j)}_{\mathrm{E}};E,\  \tau^{(j)}_{\mathrm{E}} \leq \tau < \tau^{(j)}_{\mathrm{I}};I,\ \tau^{(j)}_{\mathrm{I}} \leq \tau < \tau^{(j)}_{\mathrm{R}};R\ ,\ \tau^{(j)}_{\mathrm{R}} \leq \tau
	\right ) $\end{footnotesize}
where $ \tau^{(j)}_{\mathrm{E}} $, $ \tau^{(j)}_{\mathrm{I}} $, and $ \tau^{(j)}_{\mathrm{R}} $ are the times at which subject $ J $ becomes exposed, infectious, and recovers. As with the SIR model, some or all of these events may not transpire in the observation period $ [t_1,t_L] $, or at all. We let $ \beta $ be the per--contact infectivity rate, $ \gamma $ be the rate at which an exposed individual becomes infectious, and $ \mu $ be the rate at which an infectious individual recovers. Furthermore, we write the vector of disease state probabilities as $ \bp_{t_1} = (p_S,p_E,p_I,p_R) $. The latent epidemic process evolves according to a time--homogeneous CTMC, with transition rate from configuration $ \bx $ to $ \bx^\prime $ that differ only in the state of one subject $ j $ is given by $ \bLambda = \beta I $ if $ \bX_j = S $ and $ \bX^\prime_j = E$, $ \gamma $ if $ \bX_j = E $ and $ \bX^\prime_j = I$, and $ \mu $ if $ \bX_j = I $ and $ \bX^\prime_j = R$. Finally, the time--inhomogeneous CTMC rate matrices used in the subject--path proposal distribution have the form}
\begin{equation} \bLambda_m^{(-j)}(\btheta) = \bordermatrix{ & S & E & I & R \cr
	S & -\beta I_{\tau_m}^{(-j)} & \beta I_{\tau_m}^{(-j)} & 0 & 0 \cr 
	E & 0 & -\gamma & \gamma & 0 \cr
	I & 0 & 0 & -\mu & \mu \cr
	R & 0 & 0 & 0 & 0 }.
\end{equation}
\added{As is the case with the SIR model, the eigen--values of the CTMC rate matrices for the SEIR model are always real valued. The only computational modification, relative to the SIR model, that we suggest is that times of state transition in inter--event intervals be sampled conditional on the state at the endpoints via uniformization (see Section \ref{sec:ecctmc} of the supplement).}

\subsection{SIRS model formulation}
\added{The Susceptible--Infected--Recovered--Susceptible (SIRS) model modifies the SIR model to allow for loss of immunity. Again, fitting this model using our Bayesian data augmentation algorithm does not affect any of the machinery involved in the subject--path proposal mechanism, although the recurrent nature of the disease dynamics increase the computational burden of the algorithm since the disease state at the interval endpoints does absolve us of sampling the path within each inter--event interval where the states at the endpoints are the same.}

\added{Under the SIRS model, we suppose that the data are sampled from a latent epidemic process, $ \bX = \lbrace \bX_1,\dots,\bX_N\rbrace $, that evolves in continuous--time as individuals become exposed, infectious, and recover. The state space of this process is $ \mcS = \lbrace S,I,R\rbrace^N $, the Cartesian product of $ N $ state labels taking values in $ \lbrace S,I,R\rbrace $. The state space of a single subject, $ \bX_j $, is $\mcS_j = \lbrace S, I, R\rbrace $, and a realized subject--path is of the form\\ $ \bx_j(\tau) = \left (S,\ \tau < \tau^{(j)}_{\mathrm{I}_1}; 
	I,\ \tau^{(j)}_{\mathrm{I}_1} \leq \tau < \tau^{(j)}_{\mathrm{R}_1};
	R\ ,\ \tau^{(j)}_{\mathrm{R}_1} \leq \tau < \tau^{(j)}_{\mathrm{L}_1}; 
	S\ ,\ \tau^{(j)}_{\mathrm{L}_1} \leq \tau < \tau^{(j)}_{\mathrm{I}_2};\dots
	\right ) $,\\
	where $ \tau^{(j)}_{\mathrm{I}_k} $, $ \tau^{(j)}_{\mathrm{R}_k} $, and $ \tau^{(j)}_{\mathrm{L}_k} $ are times at which subject $ J $ becomes infected, recovers, and loses immunity, and are ennumerated by the subscript $ k $ as the process may revisit each state multiple time. As with the SIR and SEIR models, some or all of these events may not transpire in the observation period $ [t_1,t_L] $, or at all. We let $ \beta $ be the per--contact infectivity rate, $ \mu $ be the rate at which an infectious individual recovers, and $ \gamma $ be the rate at which immunity is lost. Furthermore, we write the vector of disease state probabilities as $ \bp_{t_1} = (p_S,p_I,p_R) $. The latent epidemic process evolves according to a time--homogeneous CTMC, with transition rate from configuration $ \bx $ to $ \bx^\prime $ that differ only in the state of one subject $ j $ is given by $ \bLambda = \beta I $ if $ \bX_j = S $ and $ \bX^\prime_j = E$, $ \mu $ if $ \bX_j = I $ and $ \bX^\prime_j = R$, and $ \gamma $ if $ \bX_j = R $ and $ \bX^\prime_j = S$. Finally, the time--inhomogeneous CTMC rate matrices used in the subject--path proposal distribution have the form}
\begin{equation} \bLambda_m^{(-j)}(\btheta) = \bordermatrix{ & S & I & R \cr
		S & -\beta I_{\tau_m}^{(-j)} & \beta I_{\tau_m}^{(-j)} & 0  \cr 
		I & 0 & -\mu & \mu \cr
		R & \gamma & 0 & -\gamma }.
\end{equation}
\added{Unlike the SIR and SEIR models, eigenvalues of each CTMC rate matrix may be complex. In order to obtain a real valued transition probability matrix over an interval for which eigen--values of the rate matrix are complex, we must rotate that rate matrix to obtain its real canonical form. This is further discussed in Section \ref{sec:mtx_exp} of the Supplement.}

\newpage
\section{\added{Selecting the Number of Subject--Paths to Update per MCMC Iteration}}
\label{sec:num_subj_per_iter}
	\added{There is no need to re--sample the path of every subject within each MCMC iteration. Indeed, we might suspect that the efficiency of our MCMC could be improved by sampling only a few subject--paths between parameter updates. Subject--path proposals could result in high autocorrelation, as is the case for traditional DA methods \citep{roberts2001}, and frequently updating model parameters may help to break this correlation. Parameter updates also tend to produce high autocorrelation. However, subject--path proposals are costly compared to updates of model parameters. Therefore, it is reasonable to suspect that the effective sample size (ESS) per CPU time might be improved by sampling only a handful of subject--paths per MCMC iteration.}
	
	\added{Many factors, including the SEM dynamics, population size, efficiency of the implementation, and the degree of model misspecification could affect the optimal number subject--path updates per MCMC iteration. It is clearly impossible to disentangle the effects of all of the possible factors that could affect the optimal number of subject--path updates per iteration. In the main paper, we set the number of subject--paths per iteration on the basis of log--posterior effective sample size (ESS) per CPU time in an initial run of 5,000--10,000 iterations (depending on the simulation).}

\newpage
\section{\added{Prior and Full--Conditional Distributions of SEM parameters}}
\label{sec:priors}
 \begin{table}[ht!]
	\begin{center}
		\small
		\begin{tabular}{clcc}
			\hline \rule[-2ex]{0pt}{5.5ex} Parameter & \shortstack{\\Conjugate \\ Prior Dist.} & \shortstack{\\Prior\\ Hyperparameters} & Full Conditional Hyperparameters \\ 
			
			\hline \hline
			
			\rule[-2ex]{0pt}{5.5ex} $R_0$ & Beta$ ^\prime $ & $ a_\beta,\ a_\mu,\ 1,\ \frac{b_\mu N}{b_\beta} $& ---\\
			
			\hline \rule[-2ex]{0pt}{5.5ex} $\beta$ & Gamma & $a_\beta,\  b_\beta$ & $a_\beta + \sum_{j=1}^{M}\ind{\tau_j\corresponds I},\  b_\beta + \sum_{j=1}^{M}S_{\tau_{j-1}}I_{\tau_{j-1}}(\tau_j - \tau_{j-1})$\\ 
			
			\hline \rule[-2ex]{0pt}{5.5ex} $\mu$ & Gamma & $a_\mu,\ b_\mu$ & $a_\mu + \sum_{j=1}^{M}\ind{\tau_j \corresponds R},\  b_\mu + \sum_{j=1}^{M}I_{\tau_{j-1}}(\tau_j - \tau_{j-1})$\\ 
			
			\hline \rule[-2ex]{0pt}{5.5ex} $\rho$ & Beta & $a_\rho,\ b_\rho$ & $a_\rho + \sum_{j=1}^{L}Y_{t_j},\ b_\rho + \sum_{j=1}^{L}(I_{t_j} - Y_{t_j})$\\ 
			
			\hline \rule[-2ex]{0pt}{5.5ex} $\mathbf{p}_{t_1}$ & Dirichlet & $a_{S},\ b_{I},\ c_{R}$ & $a_{S} + S_{t_1},\ b_{I} + I_{t_1},\ c_{R} + R_{t_1}$\\ 
			\hline 
		\end{tabular}
		\caption{\added{Prior and full conditional distributions for SIR model parameters. $ \beta $ is the per--contact infectivity rate, $ \mu $ is the recovery rate, $ \rho $ is the binomial sampling probability, and $ \bp_{t_1} $ is the vector of initial state probabilities. Gamma priors are parameterized with rates, so a Gamma($ a,b $) distribution has mean $ a/b $. The Beta prime prior for $ R_0 = \beta N / \mu $ is the implied prior induced by the prior distributions for $ \beta $ and $ \mu $. The indicators $ \ind{\tau_j \corresponds I} $ and $\ind{\tau_j \corresponds R} $ equal 1 if $ \tau_j $ corresponds to a time when an individual becomes infected or recovers.}}
		\label{tab:SIR_priors}
	\end{center}
\end{table}

\begin{table}[ht!]
	\begin{center}
		\small
		\begin{tabular}{clcc}
			\hline \rule[-2ex]{0pt}{5.5ex} Parameter & \shortstack{\\Conjugate \\ Prior Dist.} & \shortstack{\\Prior\\ Hyperparameters} & Full Conditional Hyperparameters \\ 
			
			\hline \hline
			
			\rule[-2ex]{0pt}{5.5ex} $R_0$ & Beta$ ^\prime $ & $ a_\beta,\ a_\mu,\ 1,\ \frac{b_\mu N}{b_\beta} $& ---\\
			
			\hline \rule[-2ex]{0pt}{5.5ex} $\beta$ & Gamma & $a_\beta,\  b_\beta$ & $a_\beta + \sum_{j=1}^{M}\ind{\tau_j \corresponds E},\  b_\beta + \sum_{j=1}^{M}S_{\tau_{j-1}}I_{\tau_{j-1}}(\tau_j - \tau_{j-1})$\\
			
			\hline \rule[-2ex]{0pt}{5.5ex} $\gamma$ & Gamma & $a_\gamma,\ b_\gamma$ & $a_\gamma + \sum_{j=1}^{M}\ind{\tau_j\corresponds I},\  b_\gamma + \sum_{j=1}^{M}E_{\tau_{j-1}}(\tau_j - \tau_{j-1})$\\ 
			
			\hline \rule[-2ex]{0pt}{5.5ex} $\mu$ & Gamma & $a_\mu,\ b_\mu$ & $a_\mu + \sum_{j=1}^{M}\ind{\tau_j \corresponds R},\  b_\mu + \sum_{j=1}^{M}I_{\tau_{j-1}}(\tau_j - \tau_{j-1})$\\ 
			
			\hline \rule[-2ex]{0pt}{5.5ex} $\rho$ & Beta & $a_\rho,\ b_\rho$ & $a_\rho + \sum_{j=1}^{L}Y_{t_j},\ b_\rho + \sum_{j=1}^{L}(I_{t_j} - Y_{t_j})$\\ 
			
			\hline \rule[-2ex]{0pt}{5.5ex} $\mathbf{p}_{t_1}$ & Dirichlet & $a_{S},\ b_{I},\ c_{R}$ & $a_{S} + S_{t_1},\ b_{I} + I_{t_1},\ c_{R} + R_{t_1}$\\ 
			\hline 
		\end{tabular}
		\caption{\added{Prior and full conditional distributions for SEIR model parameters. $ \beta $ is the per--contact infectivity rate, $ \gamma $ is the rate at which an exposed individual becomes infectious, $ \mu $ is the recovery rate, $ \rho $ is the binomial sampling probability, and $ \bp_{t_1} $ is the vector of initial state probabilities. Gamma priors are parameterized with rates, so a Gamma($ a,b $) distribution has mean $ a/b $. The Beta prime prior for $ R_0 = \beta N / \mu $ is the implied prior induced by the prior distributions for $ \beta $ and $ \mu $. The indicators $ \ind{\tau_j \corresponds E} $, $ \ind{\tau_j \corresponds I} $ and $\ind{\tau_j \corresponds R} $ equal 1 if $ \tau_j $ corresponds to a time when an individual becomes exposed, becomes infectious, or recovers.}}
		\label{tab:SEIR_priors}
	\end{center}
\end{table}

\begin{table}[ht!]
	\begin{center}
		\small
		\begin{tabular}{clcc}
			\hline \rule[-2ex]{0pt}{5.5ex} Parameter & \shortstack{\\Conjugate \\ Prior Dist.} & \shortstack{\\Prior\\ Hyperparameters} & Full Conditional Hyperparameters \\ 
			
			\hline \hline
			
			\rule[-2ex]{0pt}{5.5ex} $R_0$ & Beta$ ^\prime $ & $ a_\beta,\ a_\mu,\ 1,\ \frac{b_\mu N}{b_\beta} $& ---\\
			
			\hline \rule[-2ex]{0pt}{5.5ex} $\beta$ & Gamma & $a_\beta,\  b_\beta$ & $a_\beta + \sum_{j=1}^{M}\ind{\tau_j \corresponds E},\  b_\beta + \sum_{j=1}^{M}S_{\tau_{j-1}}I_{\tau_{j-1}}(\tau_j - \tau_{j-1})$\\
			
			\hline \rule[-2ex]{0pt}{5.5ex} $\mu$ & Gamma & $a_\mu,\ b_\mu$ & $a_\mu + \sum_{j=1}^{M}\ind{\tau_j \corresponds R},\  b_\mu + \sum_{j=1}^{M}I_{\tau_{j-1}}(\tau_j - \tau_{j-1})$\\ 
			
			\hline \rule[-2ex]{0pt}{5.5ex} $\gamma$ & Gamma & $a_\gamma,\ b_\gamma$ & $a_\gamma + \sum_{j=1}^{M}\ind{\tau_j\corresponds L},\  b_\gamma + \sum_{j=1}^{M}R_{\tau_{j-1}}(\tau_j - \tau_{j-1})$\\ 
			
			\hline \rule[-2ex]{0pt}{5.5ex} $\rho$ & Beta & $a_\rho,\ b_\rho$ & $a_\rho + \sum_{j=1}^{L}Y_{t_j},\ b_\rho + \sum_{j=1}^{L}(I_{t_j} - Y_{t_j})$\\ 
			
			\hline \rule[-2ex]{0pt}{5.5ex} $\mathbf{p}_{t_1}$ & Dirichlet & $a_{S},\ b_{I},\ c_{R}$ & $a_{S} + S_{t_1},\ b_{I} + I_{t_1},\ c_{R} + R_{t_1}$\\ 
			\hline 
		\end{tabular}
		\caption{\added{Prior and full conditional distributions for SIRS model parameters. $ \beta $ is the per--contact infectivity rate, $ \mu $ is the recovery rate, $ \gamma $ is the rate at which a recovered individual loses immunity, $ \rho $ is the binomial sampling probability, and $ \bp_{t_1} $ is the vector of initial state probabilities. Gamma priors are parameterized with rates, so a Gamma($ a,b $) distribution has mean $ a/b $. The Beta prime prior for $ R_0 = \beta N / \mu $ is the implied prior induced by the prior distributions for $ \beta $ and $ \mu $. The indicators $ \ind{\tau_j \corresponds I} $, $\ind{\tau_j \corresponds R} $, and $ \ind{\tau_j \corresponds L} $  equal 1 if $ \tau_j $ corresponds to a time when an individual becomes infected, recovers, or loses immunity.}}
		\label{tab:SIRS_priors}
	\end{center}
\end{table}

\newpage

\section{\added{Simulation 1 --- Inference Under Various Epidemic\\ Dynamics --- Setup and Additional Results}}
\label{sec:sim1_details}
\subsection{\added{Simulation details for the SIR model}}
\added{We simulated an epidemic in a population of 750 individuals, 90\% of whom were initially susceptible and 3\% of whom were initially infected. Prevalence was observed with detection probability $ \rho=0.2 $ at weekly intervals over a four month period which captured both the exponential growth and decline of the epidemic. The mean infectious period was $ 1/\mu = 7 $ days and the per-contact infectivity rate was 0.00035, which combined to give a basic reproductive number was $ R_0 = \beta N /\mu \approx 1.8$.}

\added{We ran three chains for 100,000 iterations each, sampling the paths for 75 subjects, chosen uniformly at random, per MCMC iteration. We discarded the first 10 iterations from each chain as burn-in. Priors for the rate parameters (summarized in Table \ref{tab:sim1_sir_priors}) were scaled so that the prior mass spanned a reasonable range of values, but were otherwise mild. Similarly, the prior for the binomial sampling probability reflected a general prior belief that fewer than 40\% of cases were detected. The prior for the initial distribution parameters was informative, and was chosen as such because of the paucity of data available for estimation of the initial distribution parameters.}

\begin{table}[ht]
	\centering
	\begin{tabular}{lll}
		\hline
		Param. & True Value & Prior distribution \\ 
		\hline
		$ R_0 $ & 1.8 & Beta$ ^\prime $(0.3, 1, 1, 6) \\
		$ \beta $ & 0.00035 & Gamma$ (0.3, 1000) $ \\ 
		$ \mu $ & 0.14 & Gamma(1, 8)  \\ 
		$ \bp_{t_1} $ & (0.9, 0.03, 0.07) & Dirichlet(90, 2, 5)  \\ 
		$ \rho $ & 0.2 & Beta(2, 7) \\
		\hline
	\end{tabular}
	\caption{\added{Prior distributions for SIR model and measurement process parameters. The prior for $ R_0 $ is the induced prior implied by $ \beta $ and $ \mu $. The per--contact infectivity rate is $ \beta $, the recovery rate is $ \mu $, the binomial sampling probability is $ \rho $, and the initial state probabilities are $ \bp_{t_1} $.}}
	\label{tab:sim1_sir_priors}
\end{table}

\added{We also fit the SIR model to the data using PMMH. We ran two sets of three MCMC chains with the PMMH algorithm for 50,000 iterations each with 100 particles per chain, and discarded the first 100 iterations as burn-in. The first set of chains simulated particle paths approximately using $ \tau $--leaping with a time step of two hours, while the second chain simulated paths exactly via Gillespie's direct algorithm. Parameters were updated using random walk Metropolis--Hastings (RWMH) with a proposal covariance matrix estimated from an initial run of 5,000 iterations using an adaptive RWMH algorithm with a target acceptance rate of 23.4\%. We updated parameters on transformed scales in  order to remove restrictions on the parameter space, applying a log transformation to $ \beta $ and $ \mu $, a logit transformation to $ \rho $, and a generalized logit transformation to $ \bp_{t_1} $.}

\subsection{Additional results and MCMC diagnostics for the SIR model}
\begin{table}[ht]
	\centering
	\begin{tabular}{lrrrr}
		\hline
		Method & Chain & Hours & ESS & ESS per CPU time \\ 
		\hline
		BDA &  1 & 9.9 & 87.7 & 8.8 \\ 
		BDA &  2 & 8.7 & 67.9 & 7.8 \\ 
		BDA &  3 & 8.5 & 63.8 & 7.5 \\ 
		PMMH--A &  1 & 0.6 & 1847.4 & 2871.6 \\ 
		PMMH--A &  2 & 0.6 & 1942.2 & 2995.7 \\ 
		PMMH--A &  3 & 0.7 & 1876.6 & 2615.9 \\ 
		PMMH--E &  1 & 26.1 & 1568.3 & 60.1 \\ 
		PMMH--E &  2 & 20.4 & 2123.7 & 104.0 \\ 
		PMMH--E &  3 & 20.5 & 1849.4 & 90.2 \\ 
		\hline
	\end{tabular}
	\caption{\added{Log--posterior run times, effective sample sizes (ESSs), and effective sample sizes per CPU time measure in hours (ESS.per.CPU.time). BDA indicates our Bayesian data augmentation algorithm, PMMH--A indicates PMMH with paths simulated approximately via $ \tau $--leaping algorithm, and PMMH--E indicates PMMH with paths simulated exactly using Gillespie's direct algorithm. The BDA chains were run for 100,000 iterations each, while the PMMH chains were run for 50,000 iterations following a tuning run of 5,000 iterations.}}
	\label{tab:sim1_sir_ess}
\end{table}

\begin{figure}
	\centering
	\includegraphics[width=0.9\linewidth]{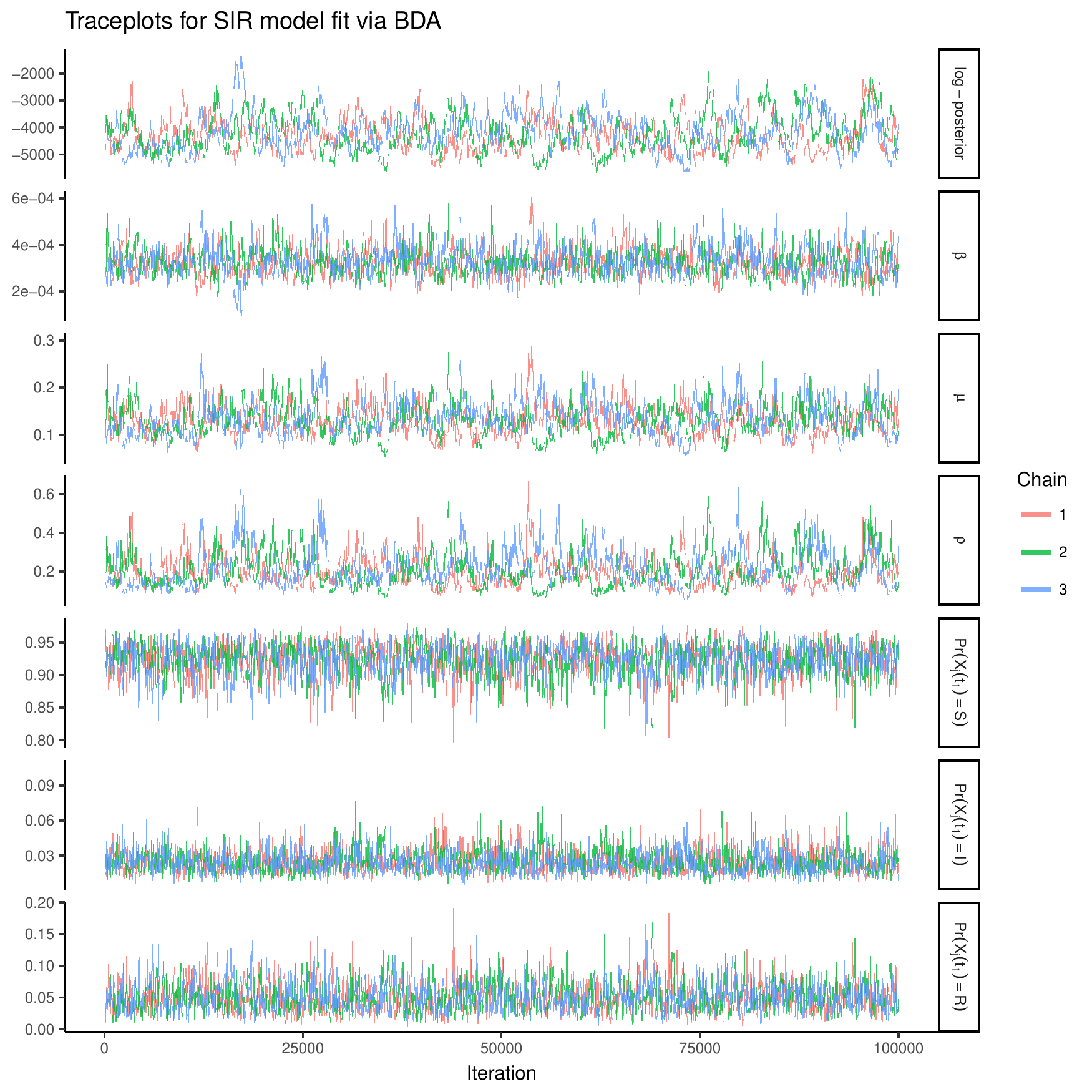}
	\caption{\added{Traceplots of the log--posterior and model parameters for the SIR model fit using Bayesian data augmentation following an initial burn--in of 10 iterations. $ \beta $ denotes the per--contact infectivity rate, $ \mu $ is the recovery rate, $ \rho $ is the binomial sampling probability. Traceplots are thinned to display every 50\textsuperscript{th} iteration.}}
	\label{fig:sirbdatraceplots}
\end{figure}

\begin{figure}
	\centering
	\includegraphics[width=0.9\linewidth]{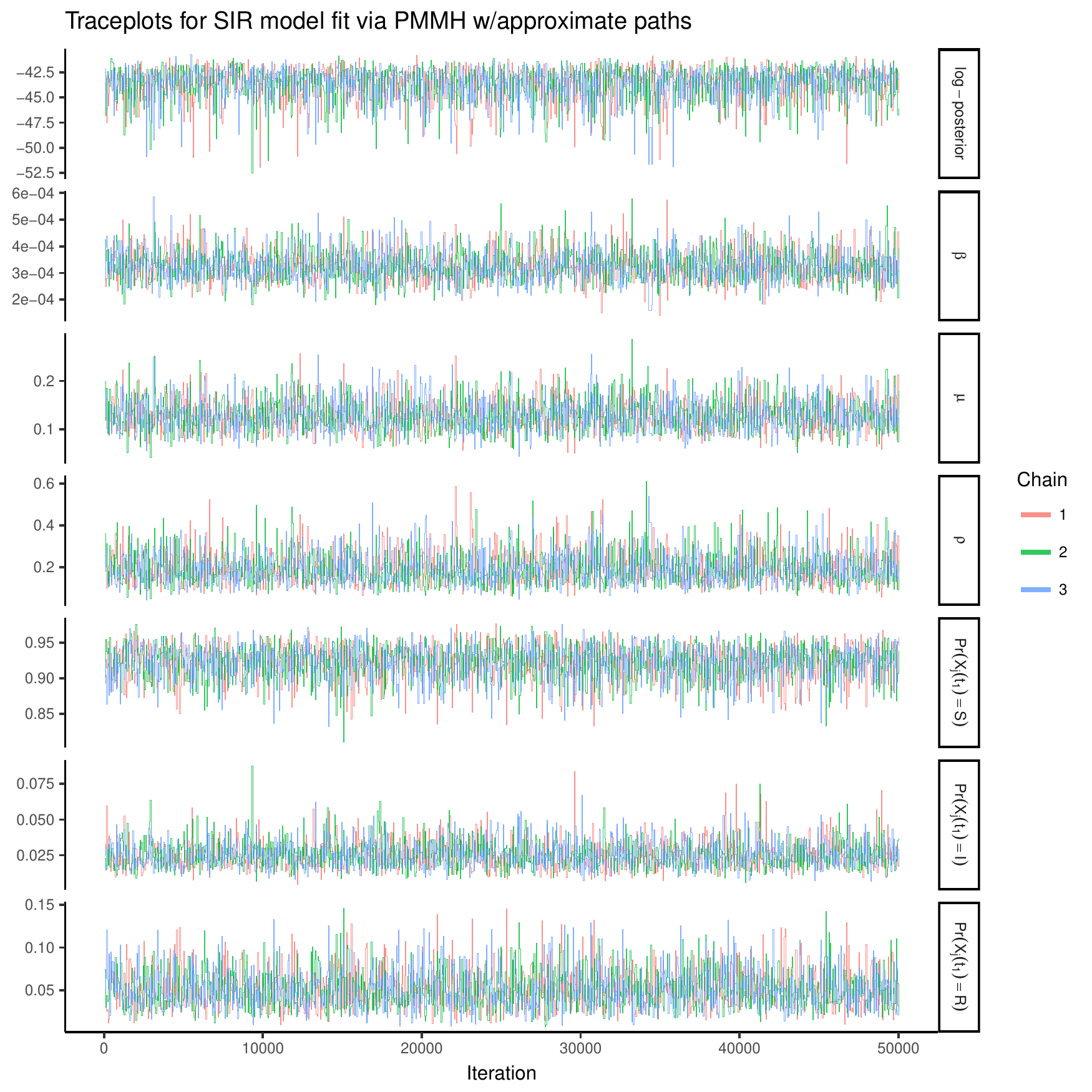}
	\caption{\added{Traceplots of the log--posterior and model parameters for the SIR model fit using PMMH with 100 particles and a time step of 8 hours, following a tuning run of 5,000 iterations used to estimate the covariance matrix for the RWMH and an initial burn--in of 100 iterations. $ \beta $ denotes the per--contact infectivity rate, $ \mu $ is the recovery rate, $ \rho $ is the binomial sampling probability. Traceplots are thinned to display every 50\textsuperscript{th} iteration.}}
	\label{fig:sirpompapproxtraceplots}
\end{figure}

\begin{figure}
	\centering
	\includegraphics[width=0.9\linewidth]{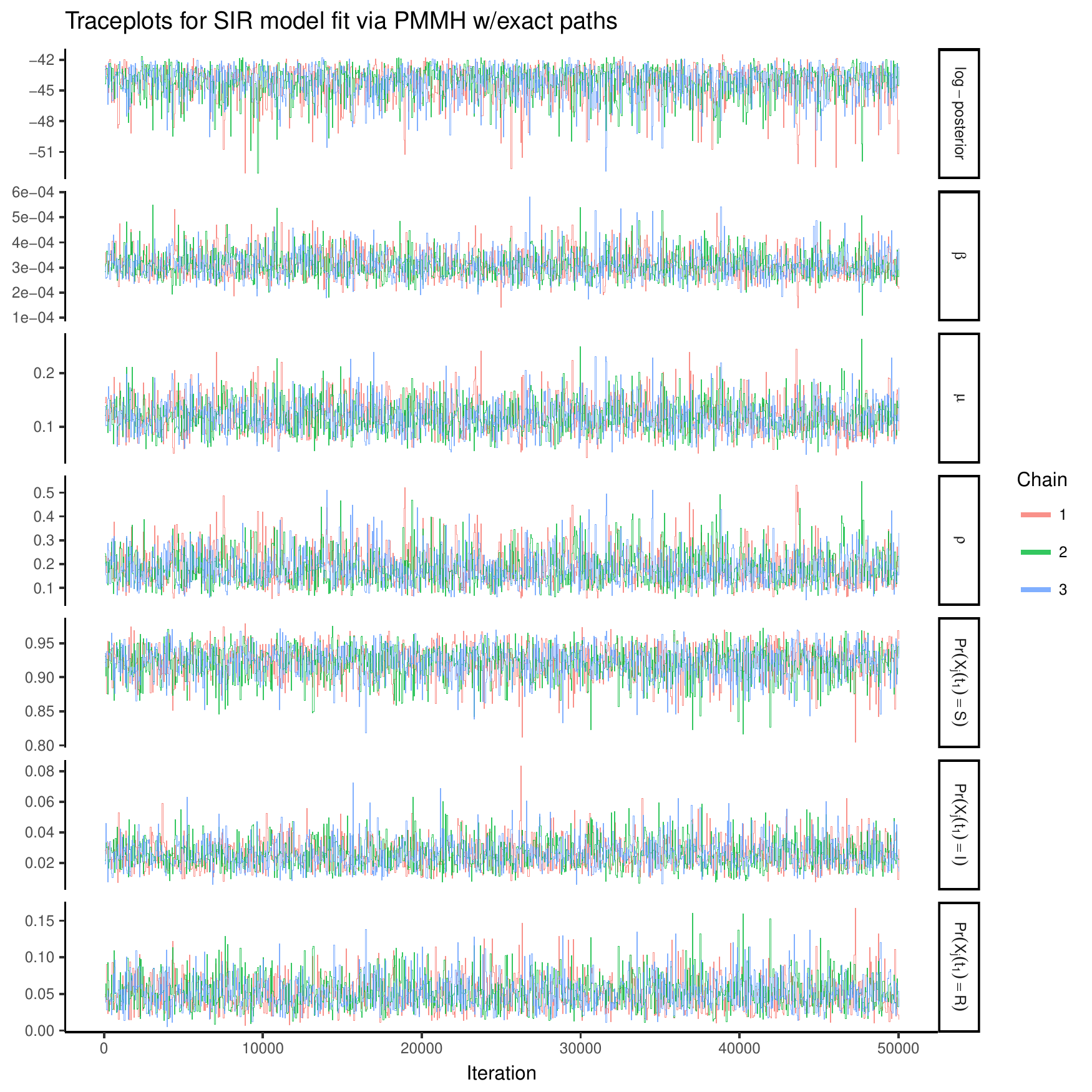}
	\caption{\added{Traceplots of the log--posterior and model parameters for the SIR model fit using PMMH with 100 particles, following a tuning run of 5,000 iterations used to estimate the covariance matrix for the RWMH and an initial burn--in of 100 iterations. $ \beta $ denotes the per--contact infectivity rate, $ \mu $ is the recovery rate, $ \rho $ is the binomial sampling probability. Traceplots are thinned to display every 50\textsuperscript{th} iteration.}}
	\label{fig:sirpompexacttraceplots}
\end{figure}

\newpage
\subsection{\added{Simulation details for the SEIR model}}
\added{We simulated an outbreak under near-endemic SEIR dynamics, with $ R_0 = \beta N / \mu = 1.05 $, in a population of 500 individuals. The outbreak was initiated by a single infected individual in an otherwise susceptible population, 121 of whom eventually became infected. The mean sojourn time in the exposed state was $ 1/\gamma = 14 $ days, while the mean infectious period duration was $ 1/\mu = 28$ days. Prevalence was observed at weekly intervals, with detection probability $ \rho = 0.3 $, over a two year period.} 

\added{We ran three chains for 100,000 iterations each, sampling the paths for 100 subjects, chosen uniformly at random, per MCMC iteration. We discarded the first 10 iterations from each chain as burn-in. Priors for the rate parameters (summarized in Table \ref{tab:sim1_seir_priors}) were scaled so that the prior mass spanned a reasonable range of values, but were otherwise mild. The prior for the binomial sampling probability was chosen so that 80\% of the mass was between roughly 15 and 55 percent. The prior for the initial distribution parameters was informative.}

\begin{table}[ht]
	\centering
	\begin{tabular}{lll}
		\hline
		Param. & True Value & Prior distribution \\ 
		\hline
		$ R_0 $ & 1.05 & Beta$ ^\prime $(1, 3.2, 1, 5) \\
		$ \beta $ & 0.000075 & Gamma$ (1, 10000) $ \\
		$ \gamma $ & 0.071 & Gamma$ (1, 11) $\\ 
		$ \mu $ & 0.036 & Gamma(3.2, 100)  \\ 
		$ \bp_{t_1} $ & (0.998, 0.006, 0.002, 0, 0) & Dirichlet(100, 0.1, 0.4, 0.01)  \\ 
		$ \rho $ & 0.3 & Beta(3.5, 6.5) \\
		\hline
	\end{tabular}
	\caption{\added{Prior distributions for SEIR model and measurement process parameters. The prior for $ R_0 $ is the induced prior implied by $ \beta $ and $ \mu $. The per--contact infectivity rate is $ \beta $, the rate at which an exposed individual becomes infectious is $ \gamma $, the recovery rate is $ \mu $, the binomial sampling probability is $ \rho $, and the initial state probabilities are $ \bp_{t_1} $.}}
	\label{tab:sim1_seir_priors}
\end{table}

\added{We also fit the SEIR model to the data using PMMH. We ran two sets of three MCMC chains with the PMMH algorithm for 50,000 iterations each with 200 particles per chain, and discarded the first 100 iterations as burn-in. The first set of chains simulated particle paths approximately using $ \tau $--leaping with a time step of 8 hours, while the second chain simulated paths exactly via Gillespie's direct algorithm. Parameters were updated using random walk Metropolis--Hastings (RWMH) with a proposal covariance matrix estimated from an initial run of 5,000 iterations using an adaptive RWMH algorithm with a target acceptance rate of 23.4\%. We updated parameters on transformed scales in order to remove restrictions on the parameter space, applying a log transformation to $ \beta $, $ \gamma $, and $ \mu $, a logit transformation to $ \rho $, and a generalized logit transformation to $ \bp_{t_1} $.}

\subsection{Additional results and MCMC diagnostics for the SEIR model}

\begin{table}[ht]
	\centering
	\begin{tabular}{llrrrr}
		\hline
		Model & Method & Chain & Time & ESS & ESS per CPU time \\ 
		\hline
		SEIR & BDA &  1 & 9.2 & 149.9 & 16.2 \\ 
		SEIR & BDA &  2 & 9.2 & 146.0 & 15.9 \\ 
		SEIR & BDA &  3 & 9.0 & 143.9 & 16.0 \\ 
		SEIR & PMMH - A &  1 & 8.1 & 483.6 & 59.5 \\ 
		SEIR & PMMH - A &  2 & 8.3 & 684.8 & 82.2 \\ 
		SEIR & PMMH - A &  3 & 8.4 & 570.5 & 67.9 \\ 
		SEIR & PMMH - E &  1 & 15.8 & 411.9 & 26.1 \\ 
		SEIR & PMMH - E &  2 & 15.9 & 589.8 & 37.1 \\ 
		SEIR & PMMH - E &  3 & 14.1 & 466.3 & 33.1 \\ 
		\hline
	\end{tabular}
	\caption{\added{Log--posterior run times, effective sample sizes (ESSs), and effective sample sizes per CPU time measure in hours (ESS.per.CPU.time). BDA indicates our Bayesian data augmentation algorithm, PMMH--A indicates PMMH with paths simulated approximately via $ \tau $--leaping algorithm, and PMMH--E indicates PMMH with paths simulated exactly using Gillespie's direct algorithm. The BDA chains were run for 100,000 iterations each, while the PMMH chains were run for 50,000 iterations following a tuning run of 5,000 iterations.}}
	\label{tab:sim1_seir_ess}
\end{table}

\begin{figure}
	\centering
	\includegraphics[width=0.9\linewidth]{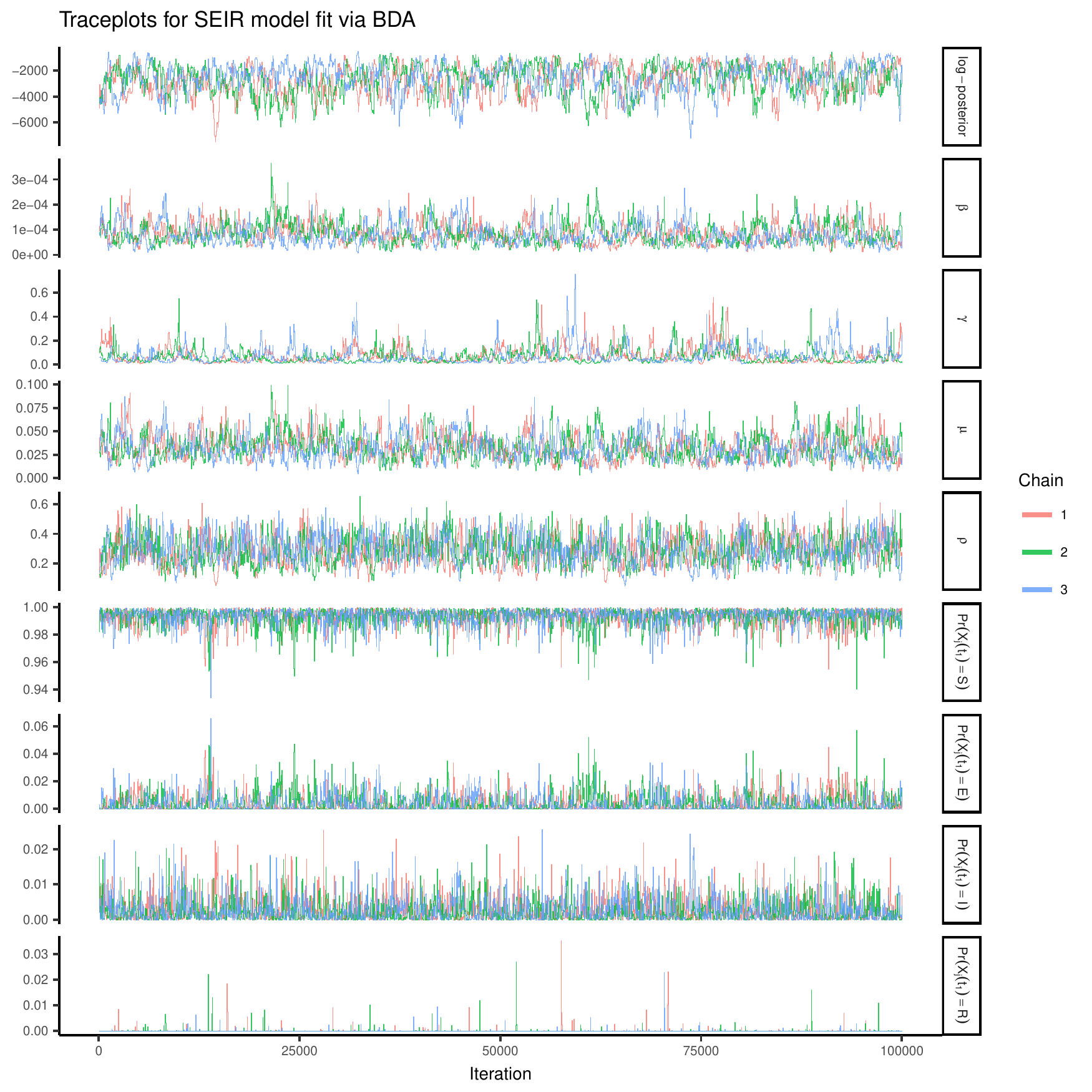}
	\caption{\added{Traceplots of the log--posterior and model parameters for the SIR model fit using Bayesian data augmentation following an initial burn--in of 10 iterations. $ \beta $ denotes the per--contact infectivity rate, $ \gamma $ is the rate at which exposed individuals become infectious, $ \mu $ is the recovery rate, $ \rho $ is the binomial sampling probability. Traceplots are thinned to display every 50\textsuperscript{th} iteration.}}
	\label{fig:seirbdatraceplots}
\end{figure}

\begin{figure}
	\centering
	\includegraphics[width=0.9\linewidth]{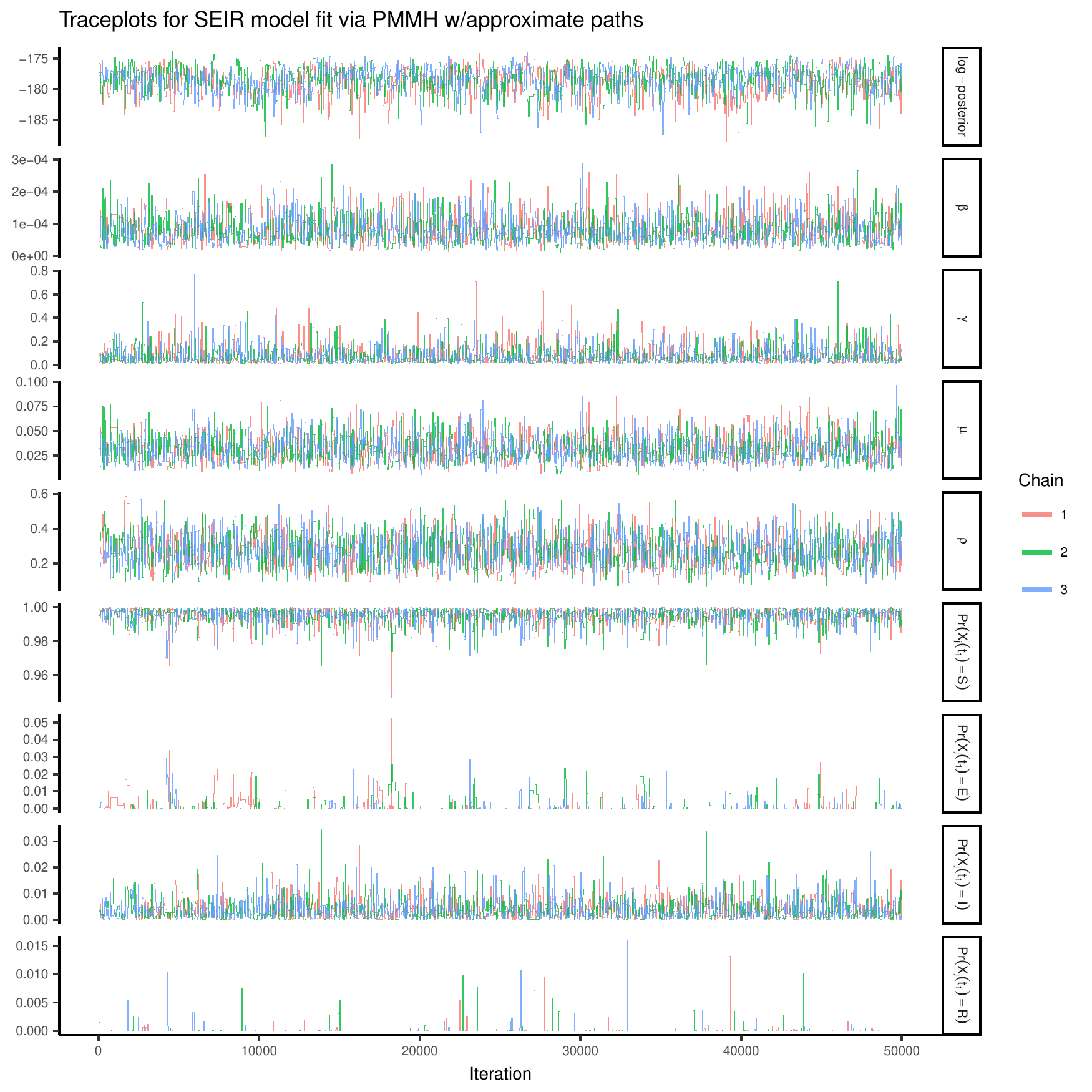}
	\caption{\added{Traceplots of the log--posterior and model parameters for the SIR model fit using PMMH with 200 particles and a time step of 8 hours, following a tuning run of 5,000 iterations used to estimate the covariance matrix for the RWMH and an initial burn--in of 100 iterations. $ \beta $ denotes the per--contact infectivity rate, $ \gamma $ is the rate at which exposed individuals become infectious, $ \mu $ is the recovery rate, $ \rho $ is the binomial sampling probability. Traceplots are thinned to display every 50\textsuperscript{th} iteration.}}
	\label{fig:seirpompapproxtraceplots}
\end{figure}

\begin{figure}
	\centering
	\includegraphics[width=0.9\linewidth]{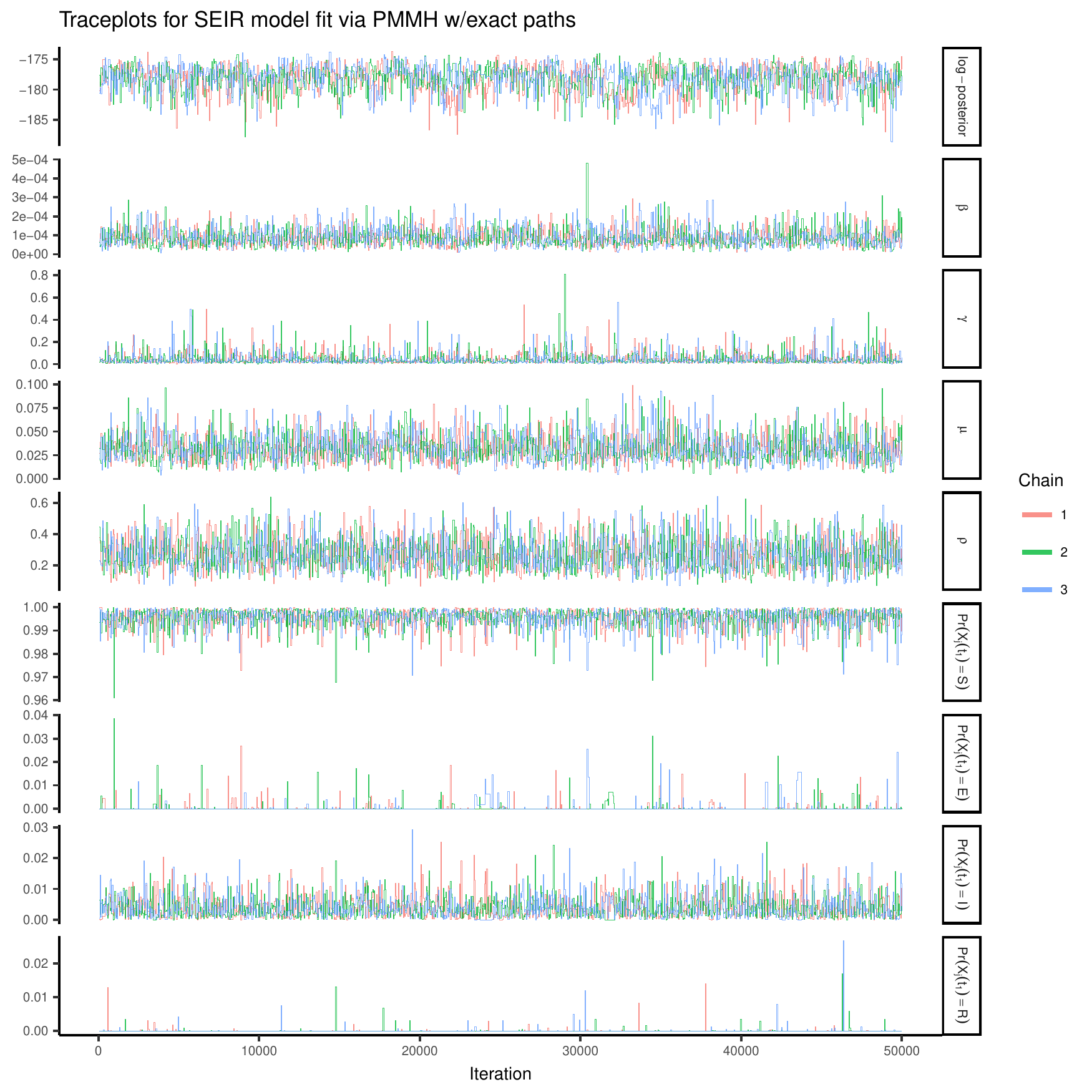}
	\caption{\added{Traceplots of the log--posterior and model parameters for the SIR model fit using PMMH with 200 particles, following a tuning run of 5,000 iterations used to estimate the covariance matrix for the RWMH and an initial burn--in of 100 iterations. $ \beta $ denotes the per--contact infectivity rate, $ \gamma $ is the rate at which exposed individuals become infectious, $ \mu $ is the recovery rate, $ \rho $ is the binomial sampling probability. Traceplots are thinned to display every 50\textsuperscript{th} iteration.}}
	\label{fig:seirpompexacttraceplots}
\end{figure}

\newpage
\subsection{\added{Simulation details for the SIRS model}}
\added{The final outbreak was simulated under SIRS dynamics in a population of 200 individuals, in which $ R_0 = \beta N / \mu = 2.52 $, the mean infectious period was $ 1/\mu = 14 $ days, and the mean time until loss of immunity was $ 1/\gamma = 150 $ days. One percent of of the population was initially at the time of the first observation and the rest of the individuals were susceptible. Prevalence was observed weekly, with detection probability $ \rho = 0.95 $, over a one year period that spanned the initial wave of the epidemic as well as most of the second wave of the epidemic.} 

\added{We ran three chains for 100,000 iterations each, sampling the paths for 100 subjects, chosen uniformly at random, per MCMC iteration. We discarded the first 2,000 iterations from each chain as burn-in. Priors for the rate parameters (summarized in Table \ref{tab:sim1_sirs_priors}) were scaled so that the prior mass spanned a reasonable range of values, but were otherwise mild. Similarly, the prior for the binomial sampling probability reflected a general prior belief that more than 60\% of cases were detected, but was not otherwise particularly informative. The prior for the initial distribution parameters was informative.}

\begin{table}[ht]
	\centering
	\begin{tabular}{lll}
		\hline
		Param. & True Value & Prior distribution \\ 
		\hline
		$ R_0 $ & 2.52 & Beta$ ^\prime $(0.1, 1.5, 1, 28) \\
		$ \beta $ & 0.1 & Gamma$ (0.1, 100) $ \\
		$ \mu $ & 0.036 & Gamma(1.8, 14)  \\ 
		$ \gamma $ & 0.071 & Gamma$ (0.0625, 10) $\\ 
		$ \bp_{t_1} $ & (0.99, 0.01, 0) & Dirichlet(90, 1.5, 0.01)  \\ 
		$ \rho $ & 0.95 & Beta(5, 1) \\
		\hline
	\end{tabular}
	\caption{\added{Prior distributions for SIRS model and measurement process parameters. The prior for $ R_0 $ is the induced prior implied by $ \beta $ and $ \mu $. The per--contact infectivity rate is $ \beta $, the recovery rate is $ \mu $, the rate at which immunity is lost is $ \gamma $, the binomial sampling probability is $ \rho $, and the initial state probabilities are $ \bp_{t_1} $.}}
	\label{tab:sim1_sirs_priors}
\end{table}

\added{We also fit the SIRS model to the data using PMMH. We ran three MCMC chains with the PMMH algorithm for 50,000 iterations each with 500 particles per chain, and discarded the first 100 iterations as burn-in. We also ran a set of chains with 200 particles but mixing was poor and not all of the chains converged. We attempted to exactly simulate particle paths but ultimately failed due to degeneracies in the algorithm. The time step for the $ \tau $--leaping algorithm was 8 hours. Parameters were updated using random walk Metropolis--Hastings (RWMH) with a proposal covariance matrix estimated from an initial run of 5,000 iterations using an adaptive RWMH algorithm with a target acceptance rate of 23.4\%. We updated parameters on transformed scales in order to remove restrictions on the parameter space, applying a log transformation to $ \beta $, $ \mu $, and $ \gamma $, a logit transformation to $ \rho $, and a generalized logit transformation to $ \bp_{t_1} $.}

\subsection{Additional results and MCMC diagnostics for the SIRS model}

\begin{table}[ht]
	\centering
	\begin{tabular}{llrrrr}
		\hline
		Model & Method & Chain & Time & ESS & ESS per CPU time \\ 
		\hline
		SIRS & BDA &  1 & 14.2 & 167.7 & 11.8 \\ 
		SIRS & BDA &  2 & 10.9 & 194.8 & 17.8 \\ 
		SIRS & BDA &  3 & 10.8 & 243.0 & 22.6 \\ 
		SIRS & PMMH - A &  1 & 3.1 & 670.8 & 214.1 \\ 
		SIRS & PMMH - A &  2 & 3.0 & 799.5 & 267.3 \\ 
		SIRS & PMMH - A &  3 & 3.5 & 766.2 & 217.1 \\ 
		SIRS & PMMH - E &  1 & 50.2 & 570.9 & 11.4 \\ 
		SIRS & PMMH - E &  2 & 48.6 & 667.6 & 13.7 \\ 
		SIRS & PMMH - E &  3 & 48.8 & 592.6 & 12.1 \\ 
		\hline
	\end{tabular}
	\caption{\added{Log--posterior run times, effective sample sizes (ESSs), and effective sample sizes per CPU time measure in hours (ESS.per.CPU.time). BDA indicates our Bayesian data augmentation algorithm, PMMH--A indicates PMMH with paths simulated approximately via $ \tau $--leaping algorithm, and PMMH--E indicates PMMH with paths simulated exactly using Gillespie's direct algorithm. The BDA chains were run for 100,000 iterations each, while the PMMH chains were run for 50,000 iterations following a tuning run of 5,000 iterations.}}
	\label{tab:sim1_sirs_ess}
\end{table}

\begin{figure}
	\centering
	\includegraphics[width=0.9\linewidth]{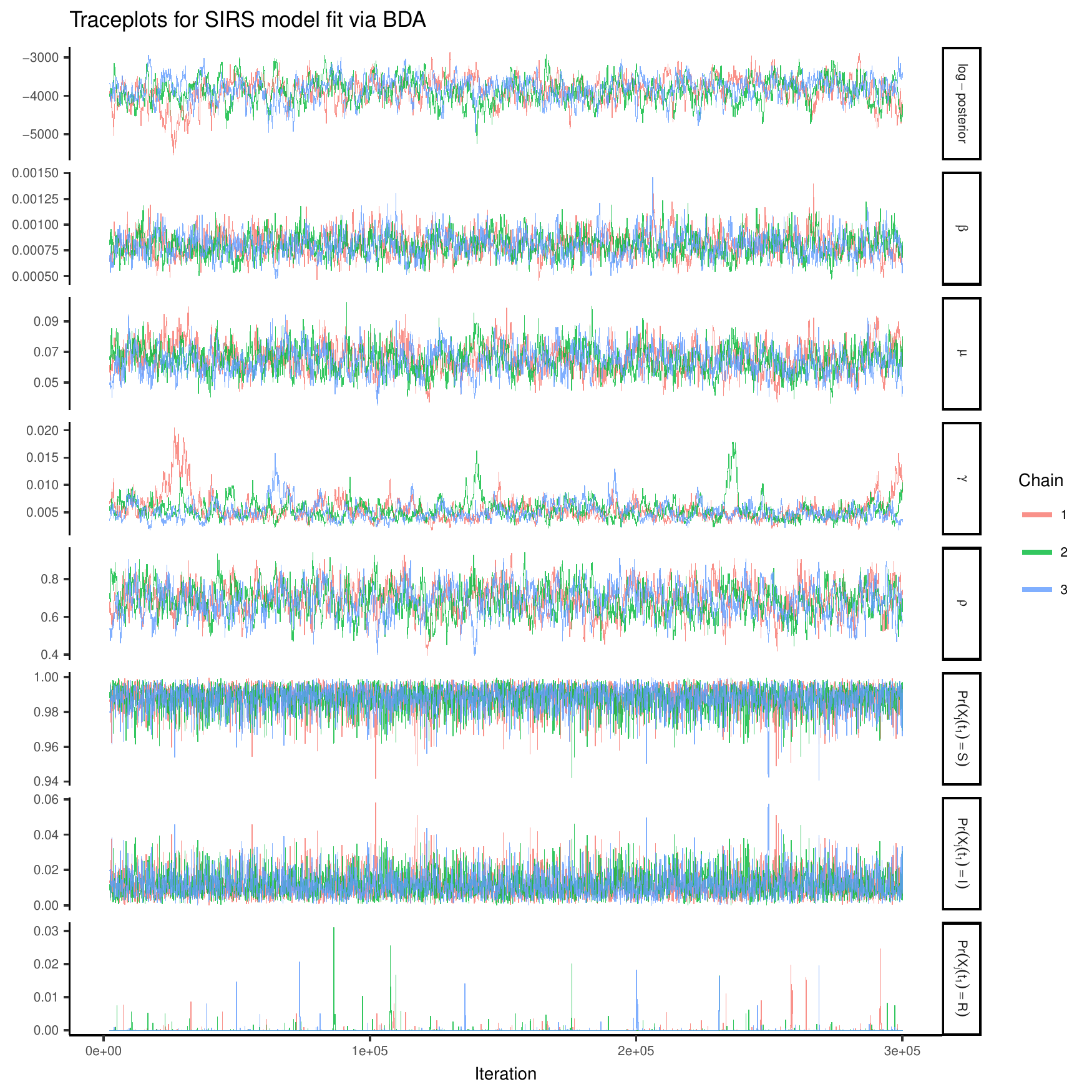}
	\caption{\added{Traceplots of the log--posterior and model parameters for the SIR model fit using Bayesian data augmentation following an initial burn--in of 2,000 iterations. $ \beta $ denotes the per--contact infectivity rate, $ \mu $ is the recovery rate, $ gamma $ is the rate at which immunity is lost, and $ \rho $ is the binomial sampling probability. Traceplots are thinned to display every 50\textsuperscript{th} iteration.}}
	\label{fig:sirsbdatraceplots}
\end{figure}

\begin{figure}
	\centering
	\includegraphics[width=0.9\linewidth]{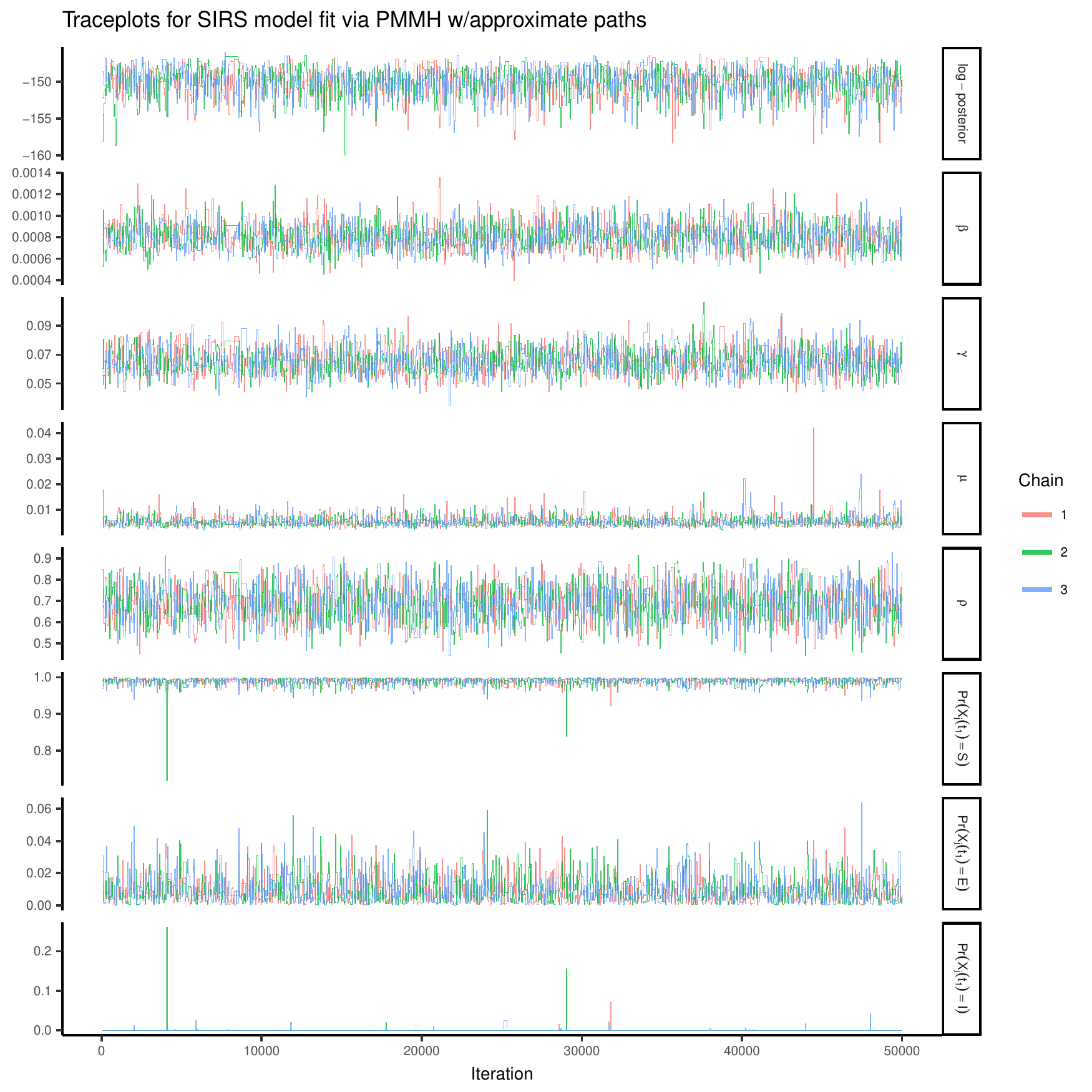}
	\caption{\added{Traceplots of the log--posterior and model parameters for the SIR model fit using PMMH with 500 particles per chain and a time--step of 8 hours in the approximate $ \tau $--leaping algorithm, following a tuning run of 5,000 iterations to estimate the RWMH covariance matrix and in initial burn--in of 100 iterations. $ \beta $ denotes the per--contact infectivity rate, $ \mu $ is the recovery rate, $ gamma $ is the rate at which immunity is lost, and $ \rho $ is the binomial sampling probability. Traceplots are thinned to display every 50\textsuperscript{th} iteration.}}
	\label{fig:sirspompapproxtraceplots}
\end{figure}

\begin{figure}
	\centering
	\includegraphics[width=0.9\linewidth]{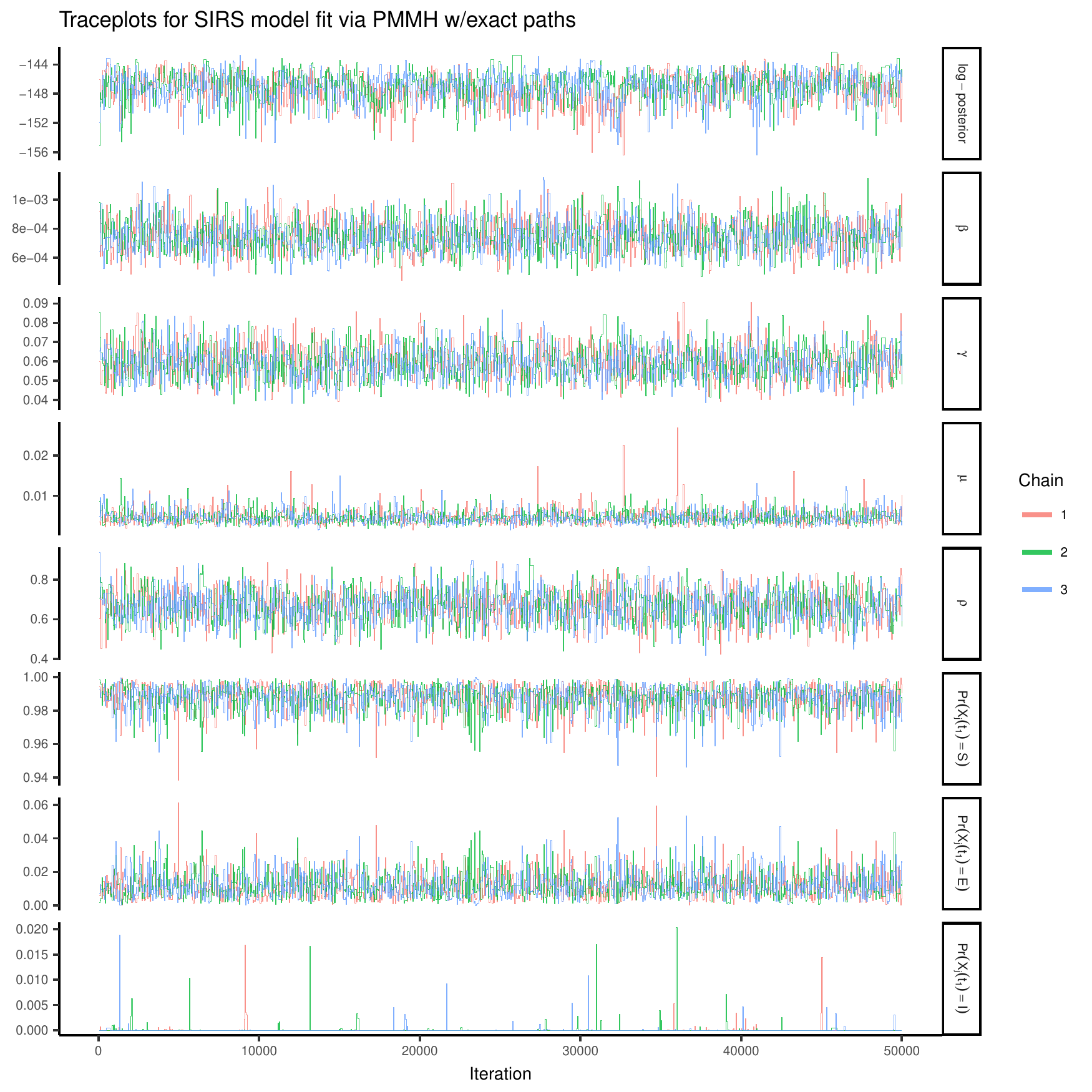}
	\caption{\added{Traceplots of the log--posterior and model parameters for the SIR model fit using PMMH with 500 particles per chain and particle paths simulated exactly via Gillespie's direct algorithm, following a tuning run of 5,000 iterations to estimate the RWMH covariance matrix and in initial burn--in of 100 iterations. $ \beta $ denotes the per--contact infectivity rate, $ \mu $ is the recovery rate, $ gamma $ is the rate at which immunity is lost, and $ \rho $ is the binomial sampling probability. Traceplots are thinned to display every 50\textsuperscript{th} iteration.}}
	\label{fig:sirspompexacttraceplots}
\end{figure}

\begin{figure}
	\centering
	\includegraphics[width=0.9\linewidth]{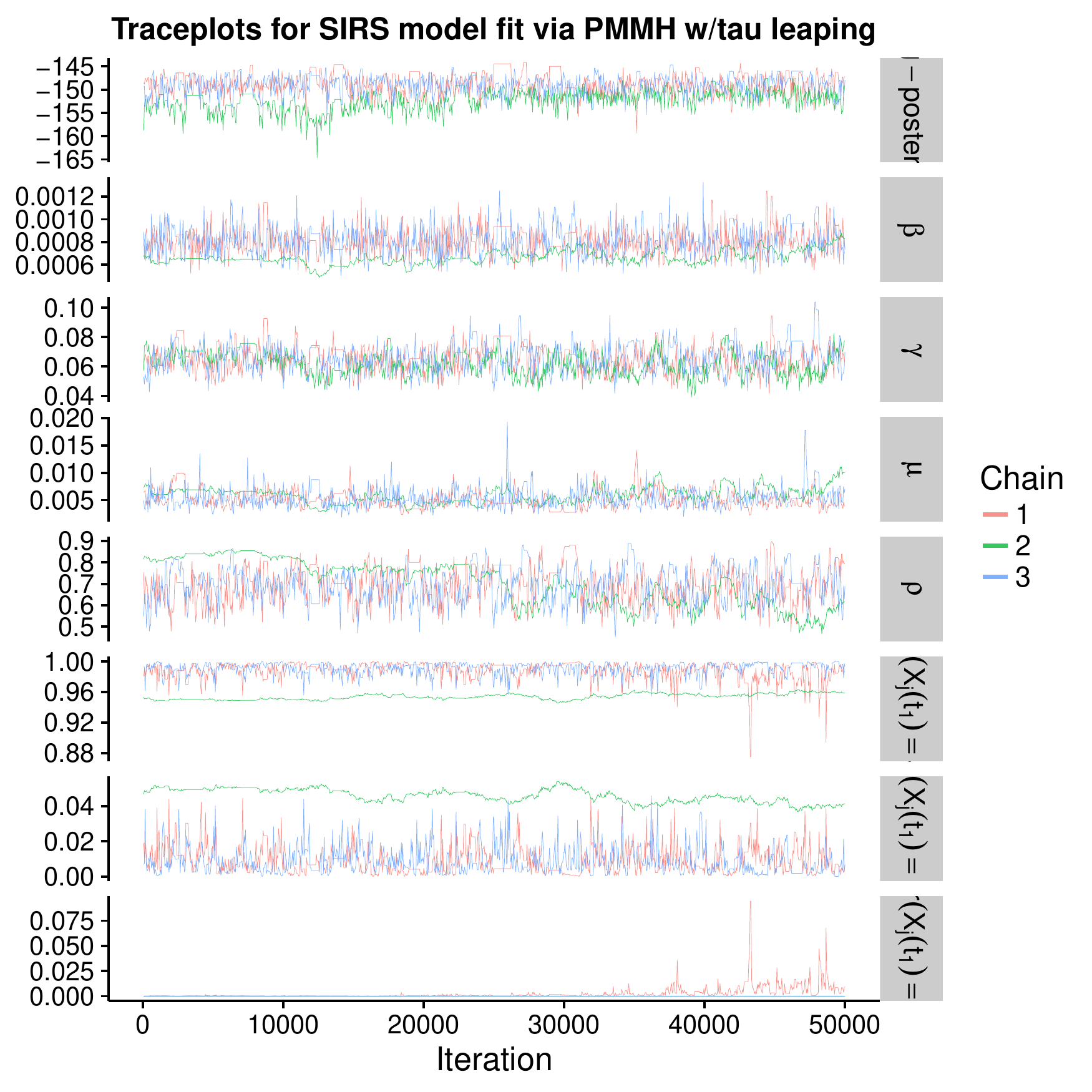}
	\caption{\added{Traceplots of the log--posterior and model parameters for the SIR model fit using PMMH with 200 particles per chain and a time--step of 8 hours in the approximate $ \tau $--leaping algorithm, following a tuning run of 5,000 iterations to estimate the RWMH covariance matrix and in initial burn--in of 100 iterations. $ \beta $ denotes the per--contact infectivity rate, $ \mu $ is the recovery rate, $ gamma $ is the rate at which immunity is lost, and $ \rho $ is the binomial sampling probability. Traceplots are thinned to display every 50\textsuperscript{th} iteration.}}
	\label{fig:sirspompapproxtraceplotsnp200}
\end{figure}
\newpage
\subsection{Estimated latent posterior distributions for all models}

\begin{figure}[!h]
	\centering
	\includegraphics[width=\linewidth]{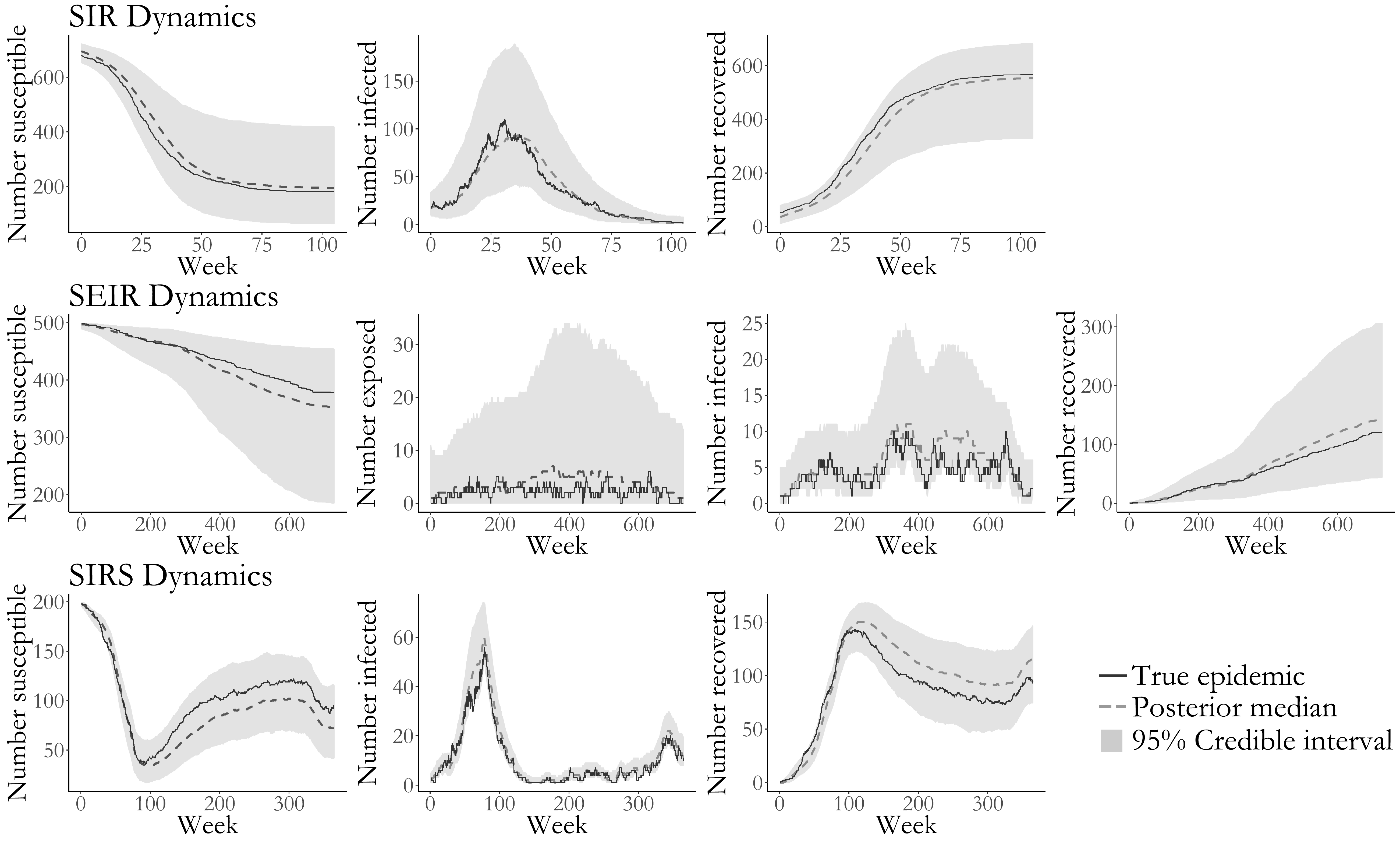}
	\caption{\added{Pointwise posterior medians (dashed lines) and pointwise 95\% credible intervals for the numbers of individuals in each disease state for the SIR, SEIR, and SIRS models. True compartment counts are shown as solid lines. Estimates are based on a thinned sample, retaining the collection of disease histories at the end of every 250$ ^{th} $ MCMC iteration.}}
	\label{fig:sim1_latent_post_all}
\end{figure}

\newpage

\section{\added{Simulation 2 --- Inference under Model\\ Misspecification --- Setup and Additional Results}}
\label{sec:SEM_misspec_details}
\subsection{Simulation setup}
\added{We simulated an epidemic in a population of size N=400 with time--varying dynamics using Gillespie's direct algorithm over a four year period. Weekly prevalence counts were binomially distributed with detection probability $ \rho = 0.95 $. The epidemic dynamics varied over four epochs, based on the parameters given in Table \ref{tab:misspec_sim_params}. We fit SIR and SEIR models to the data, running three MCMC chains per model, discarding the first 100 iterations as burn--in, and sampling the paths of 150 subjects, chosen uniformly at random, per MCMC iteration. After discarding the burn--in, the resulting samples were combined to form the final sample. We also attempted to fit the models using PMMH. We ran three chains per model, each using 2,500 particles, the paths for which were simulated approximately via $ \tau$--leaping with a one day time step. The PMMH chains were plagued by severe particle degeneracy and did not converge.}

\begin{table}[ht]
	\centering
	\begin{tabular}{ll}
		\textbf{Epoch 1: Weeks 0 -- 26} & \\
		\hline\hline
		Param. & True value \\ 
		\hline
		$ \beta $ &  0.00025 \\
		$ \gamma $ & 1/210\\
		$ \mu $ &  1/150 \\ 
		$ \rho $ &  0.95 \\
		$ \bX(t_0) $ &  $ S_0 = 397, E_0 = 2, I_0 = 1, R_0 =0 $  \\ 
		\hline	
		&\\
		\textbf{Epoch 2: Weeks 26--105} & \\
		\hline\hline
		$ \beta $ &  0.0001 \\
		$ \gamma $ & 1/210\\
		$ \mu $ &  1/330 \\ 
		$ \rho $ &  0.95 \\
		$ \bX(t_{26}) $ &  $ S_0 = 279, E_0 = 98, I_0 = 20, R_0 =3 $  \\ 
		\hline
		&\\
		\textbf{Epoch 3: Weeks 105--167} & \\
		\hline\hline
		$ \beta $ &  0.00035 \\
		$ \gamma $ & 1/90\\
		$ \mu $ &  1/300 \\ 
		$ \rho $ &  0.95 \\
		$ \bX(t_{105}) $ &  $ S_0 = 1, E_0 = 43, I_0 = 145, R_0 =211 $  \\ 
		\hline
		&\\
		\textbf{Epoch 4: Weeks 167 -- 209} & \\
		\hline\hline
		$ \beta $ &  0.0001 \\
		$ \gamma $ & 1/180\\
		$ \mu $ &  1/70 \\ 
		$ \rho $ &  0.95 \\
		$ \bX(t_{167}) $ &  $ S_0 = 0, E_0 = 1, I_0 = 52, R_0 = 347 $\\
		\hline
	\end{tabular}
	\caption{\added{Parameter values governing the time--varying SEIR dynamics and binomial emissions process. The epidemic was simulated using Gillespie's direct algorithm and the process was restarted with the new parameter values at the beginning of each epoch.}}
	\label{tab:misspec_sim_params}
\end{table}

\begin{table}[!ht]
	\centering
	\begin{tabular}{ll}
		\textbf{SIR model} & \\
		\hline\hline
		Parameter & Prior distribution \\ 
		\hline
		$ R_0 $ &  Beta$ ^\prime $(0.6, 0.7, 1, 4) \\
		$ \beta $ &  Gamma$ (0.6, 10000) $ \\
		$ \mu $ &  Gamma(0.7, 100) \\ 
		$ \bp_{t_1} $ &  Dirichlet(90, 0.5, 0.01)  \\ 
		$ \rho $ &  Beta(10, 1) \\
		\hline	
		&\\
		\textbf{SEIR model} & \\
		\hline\hline
		$ R_0 $ &  Beta$ ^\prime $(0.6, 0.7, 1, 4) \\
		$ \beta $ &  Gamma$ (0.6, 10000) $ \\
		$ \gamma $ &  Gamma$ (0.5, 100) $\\ 
		$ \mu $ &  Gamma(0.7, 100) \\ 
		$ \bp_{t_1} $ &  Dirichlet(90, 0.5, 0.5, 0.01)  \\ 
		$ \rho $ &  Beta(10, 1) \\
		\hline
	\end{tabular}
	\caption{\added{Prior distributions for the SIR and SEIR model and measurement process parameters for the models fit to the dataset simulated under time--varying SEIR dynamics. The prior for $ R_0 $ is the induced prior implied by $ \beta $ and $ \mu $. The per--contact infectivity rate is $ \beta $, the rate at which an exposed individual becomes infectious is $ \gamma $, the recovery rate is $ \mu $, the binomial sampling probability is $ \rho $, and the initial state probabilities are $ \bp_{t_1} $.}}
	\label{tab:misspec_priors}
\end{table}

\newpage

\subsection{Additional results}
\label{sec:misspec_additional_results}
\begin{table}[!ht]
	\centering
	\begin{tabular}{ll}
		\textbf{SIR model} & \\
		\hline\hline
		Parameter & Posterior median (95\% Credible interval) \\ 
		\hline
		$ R_0 $ & 4.05 (3.40, 4.81) \\ 
		$\beta $ & 0.000035 (0.000030, 0.000040) \\ 
		$ \mu $ & 0.0034 (0.0031, 0.0038) \\ 
		$ \rho $ & 0.95 (0.93, 0.97) \\
		\hline & \\
		\textbf{SEIR model} & \\
		\hline\hline
		Parameter & Posterior median (95\% Credible interval) \\ 
		\hline 
		$ R_0 $ & 23.80 (15.10, 36.98) \\ 
		$ \beta $ & 0.00021 (0.00013, 0.00032) \\ 
		$ \gamma $ & 0.0047 (0.0038, 0.0061) \\ 
		$ \mu $ & 0.0035 (0.0032, 0.0038) \\ 
		$ \rho $ & 0.95 (0.94, 0.97) \\ 
		\hline
	\end{tabular}
	\caption{\added{Posterior median estimates and 95\% credible intervals for SIR and SEIR model parameters fit under a binomial emission distribution to the epidemic simulated with time--varying SEIR dynamics.}}
\end{table}

\begin{figure}
	\centering
	\includegraphics[width=0.9\linewidth]{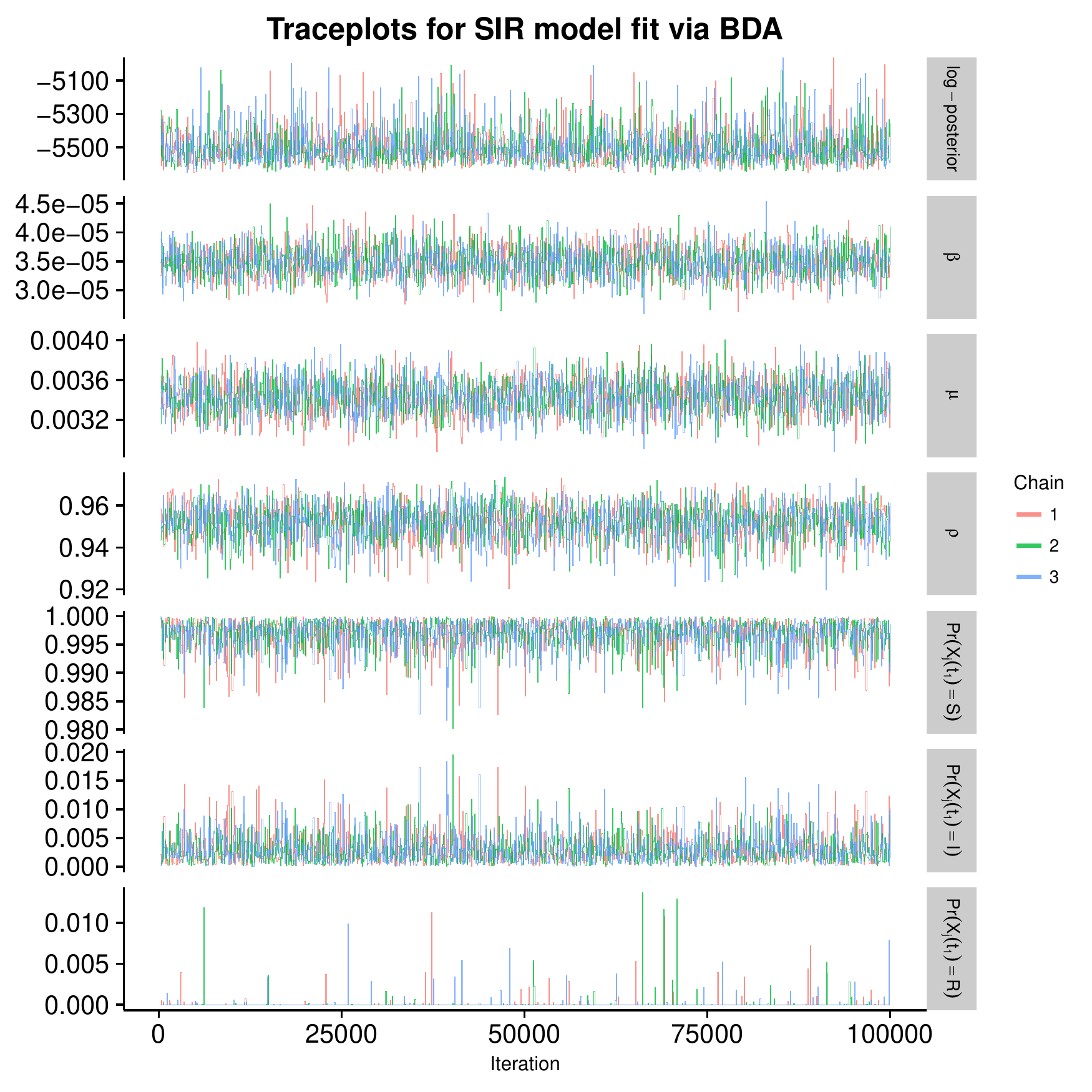}
	\caption{\added{Traceplots of the log--posterior and model parameters for the SIR model fit using BDA following an initial burn--in of 100 iterations. $ \beta $ denotes the per--contact infectivity rate, $ \mu $ is the recovery rate, $ \rho $ is the binomial sampling probability. Traceplots are thinned to display every 50\textsuperscript{th} iteration.}}
	\label{fig:misspec_sir_bda_traceplots}
\end{figure}

\begin{figure}
	\centering
	\includegraphics[width=0.9\linewidth]{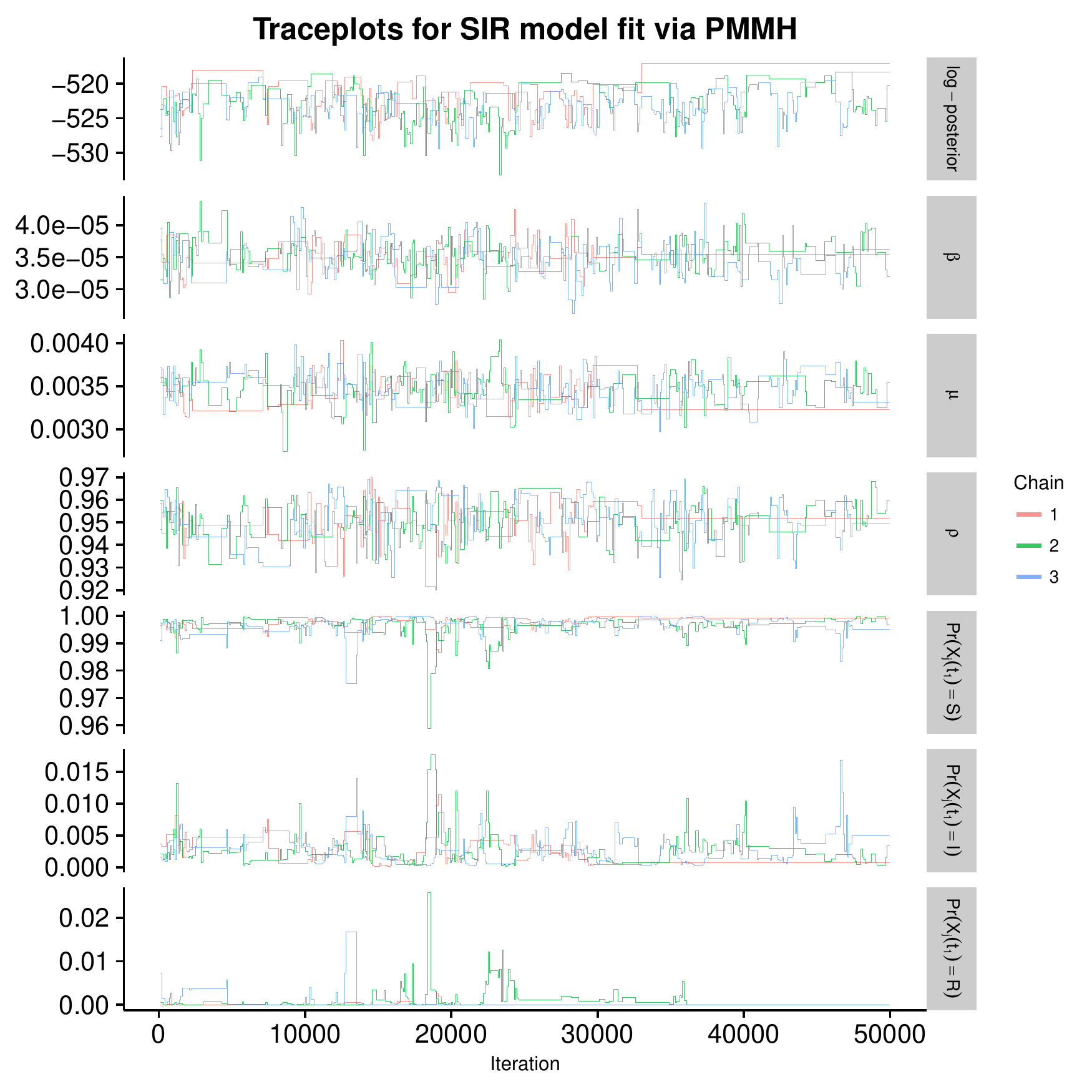}
	\caption{\added{Traceplots of the log--posterior and model parameters for the SIR model fit using PMMH with 2,500 particles, following a tuning run of 5,000 iterations used to estimate the covariance matrix for the RWMH. $ \beta $ denotes the per--contact infectivity rate, $ \mu $ is the recovery rate, $ \rho $ is the binomial sampling probability. Traceplots are thinned to display every 50\textsuperscript{th} iteration.}}
	\label{fig:misspec_sir_pmmh_traceplots}
\end{figure}

\begin{figure}
	\centering
	\includegraphics[width=0.9\linewidth]{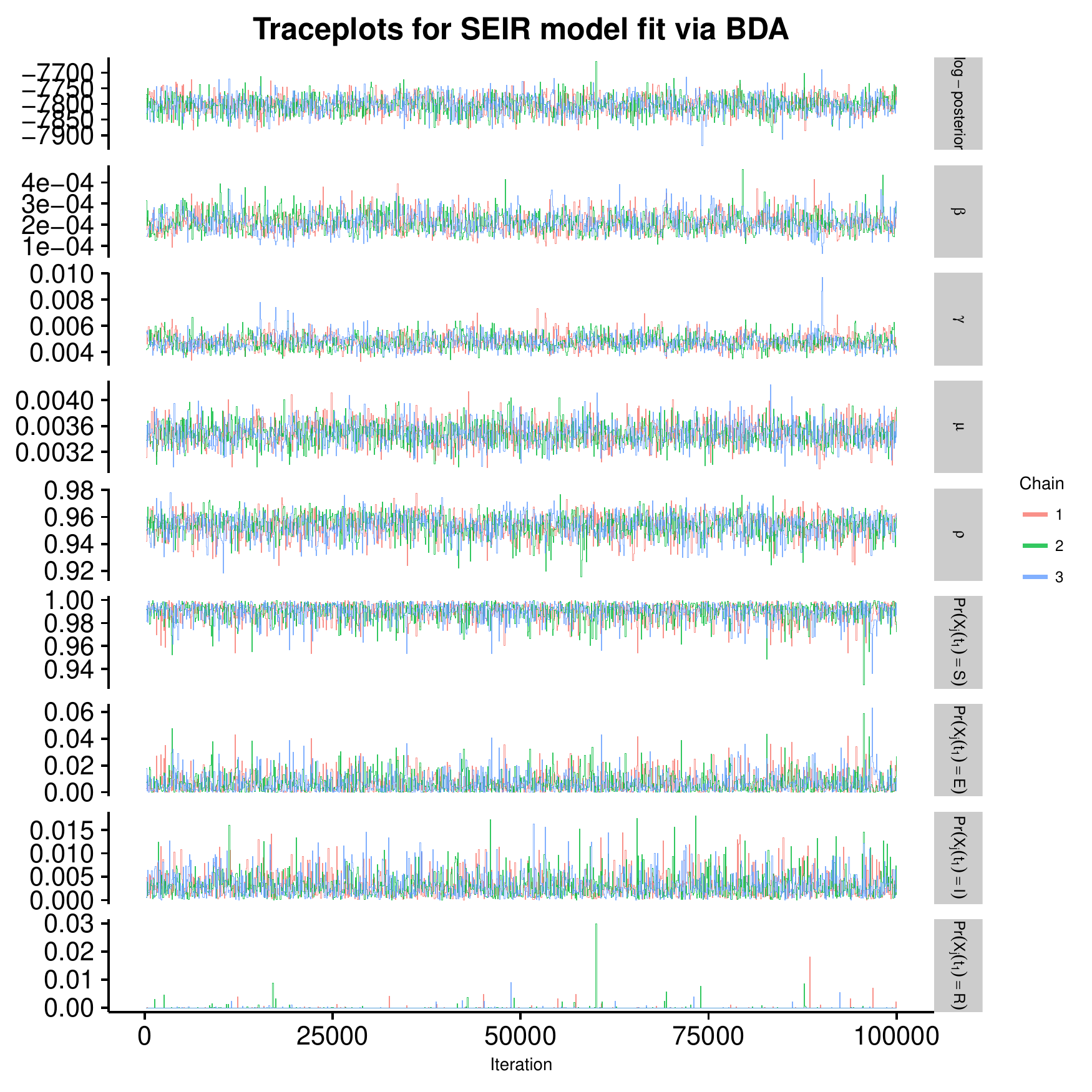}
	\caption{\added{Traceplots of the log--posterior and model parameters for the SEIR model fit using BDA following an initial burn--in of 100 iterations. $ \beta $ denotes the per--contact infectivity rate, $ \gamma $ is the rate at which an exposed individual becomes infectious, $ \mu $ is the recovery rate, $ \rho $ is the binomial sampling probability. Traceplots are thinned to display every 50\textsuperscript{th} iteration.}}
	\label{fig:misspec_seir_bda_traceplots}
\end{figure}

\begin{figure}
	\centering
	\includegraphics[width=0.9\linewidth]{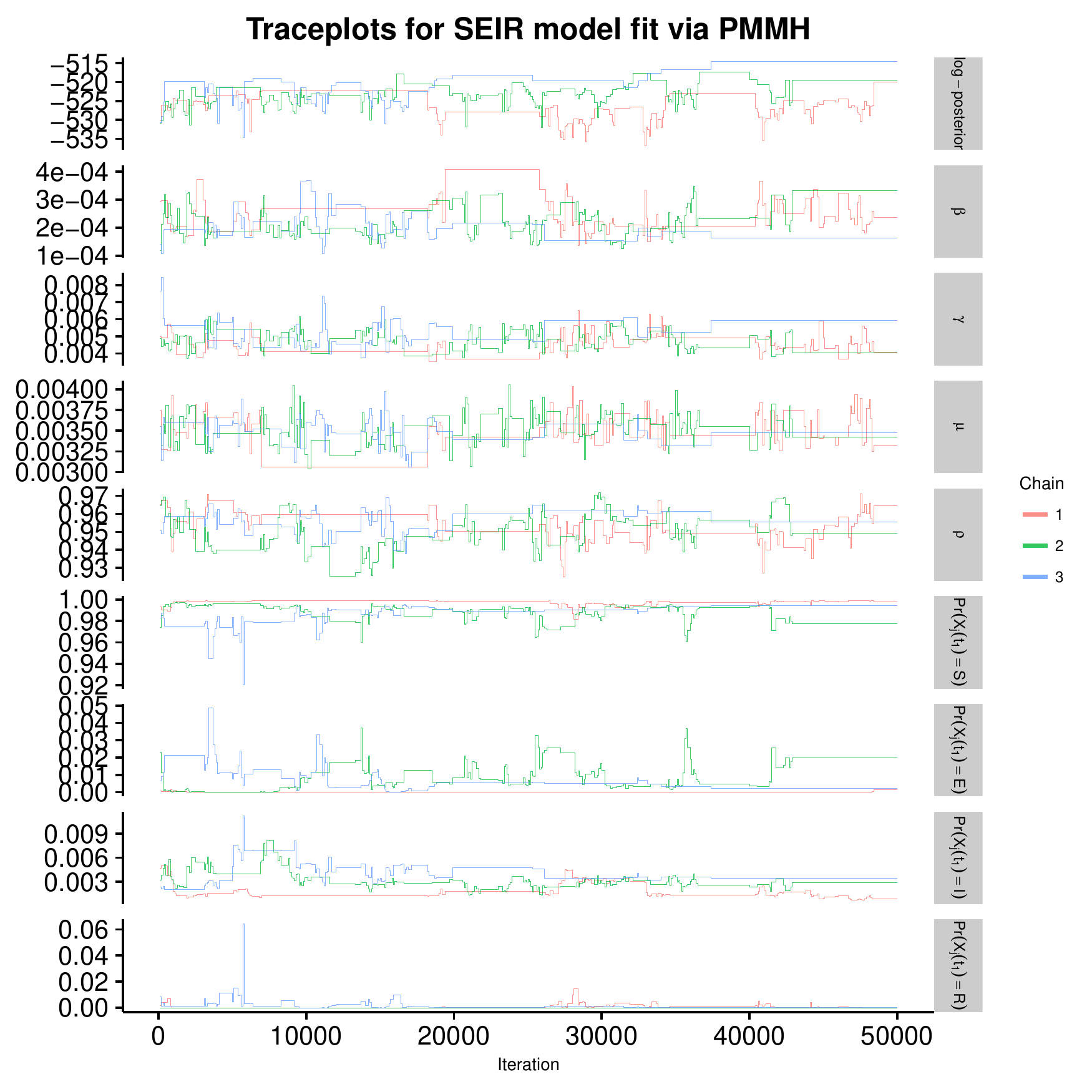}
	\caption{\added{Traceplots of the log--posterior and model parameters for the SEIR model fit using PMMH with 2,500 particles, following a tuning run of 5,000 iterations used to estimate the covariance matrix for the RWMH. $ \beta $ denotes the per--contact infectivity rate,  $ \gamma $ is the rate at which an exposed individual becomes infectious, $ \mu $ is the recovery rate, $ \rho $ is the binomial sampling probability. Traceplots are thinned to display every 50\textsuperscript{th} iteration.}}
	\label{fig:misspec_seir_pmmh_traceplots}
\end{figure}

\newpage

\section{\added{Simulation 3 --- Inference under Population Size\\ Misspecification --- Details}}
\label{sec:popsize_misspec_details}
\added{We simulated an outbreak under SIR dynamics, with $ R_0 = \beta N / \mu =3.5 $, in a population of 1,250 individuals. Roughly 0.2\% of the population was initially infected, and 95\% were initially susceptible. The mean infectious period was $ 1/\mu = 7 $ days. Prevalence was observed at weekly intervals, with detection probability $ \rho = 0.3 $, over a one year period.} 

\added{We ran three chains for 100,000 iterations each under the following assumed population sized: 150, 300, 500, 900, 1100, 1200, 1250, 1300, 1400. We sampled the paths for 10\% of the subjects, chosen uniformly at random, per MCMC iteration. We discarded the first 500 iterations from each chain as burn-in. Diffuse priors were specified for all model parameters, with the prior for the per--contact infectivity rate depending on the assumed population size (summarized in Table \ref{tab:popsize_misspec_priors}).}

\begin{table}[!ht]
	\centering
	\begin{tabular}{ll}
		\hline
		Param. & Prior distribution \\ 
		\hline
		$ R_0 $ &  Beta$ ^\prime $($ 0.00042 \times \frac{1250}{N} $, 0.35, 1, 2 / N) \\
		$ \beta $ &  Gamma$ (0.00042 \times \frac{1250}{N}, 1) $ \\ 
		$ \mu $ & Gamma(0.35, 2)  \\ 
		$ \bp_{t_1} $ & Dirichlet(100, 1, 5)  \\ 
		$ \rho $ & Beta(1,1) \\
		\hline
	\end{tabular}
	\caption{\added{Prior distributions for SIR model and measurement process parameters. The prior for $ R_0 $ is the induced prior implied by $ \beta $ and $ \mu $. The per--contact infectivity rate is $ \beta $, the recovery rate is $ \mu $, the binomial sampling probability is $ \rho $, and the initial state probabilities are $ \bp_{t_1} $. The prior for $ \beta $ was scaled in accordance with the assumed population size.}}
	\label{tab:popsize_misspec_priors}
\end{table}

\newpage

\section{\added{Simulation 4 --- Effect of Prior Specification on\\ Inference --- Setup and Additional Results}}
\label{sec:prior_effect_details}
\subsection{\added{Simulation details}}
\added{We ran three MCMC chains for each of the SIR models fit under the prior regimes that are specified in Table \ref{tab:prior_effect_priors} along with the true parameter values under which the data were simulated. Each chain was run for 100,000 MCMC iterations  with 75 subject--paths per iteration. The first 100 iterations of each were discarded as burn--in, after which the samples from all three chains for each model were combined to form the posterior sample. }

\begin{table}[!ht]
	\centering	
	\footnotesize
	\begin{tabular}{lcccc}
		\hline & \multicolumn{4}{c}{Prior Distribution} \\
		\hline
		Parameter & Regime 1 & Regime 2 & Regime 3 & Regime 4\\
		\hline
		$R_0 = 1.84$ & Beta$ ^\prime $(3, 3, 1, 1.526) & Beta$ ^\prime $(0.3, 0.1, 1, 0.6) & Beta$ ^\prime $(3, 3, 1, 1.526) & Beta$ ^\prime $(0.3, 0.1, 1, 0.6) \\
		$\beta = 0.00035$ & Gamma(3, 10000) & Gamma(0.3, 1000) & Gamma(3, 10000)& Gamma(0.3, 1000) \\
		$\mu = 0.14$ & Gamma(3, 20) & Gamma(0.1, 0.8) & Gamma(3, 20)& Gamma(0.1, 0.8)  \\
		$\rho = 0.2$ & Beta(21, 75) & Beta(21, 75) & Beta(1,1) & Beta(1,1)\\
		\hline 
	\end{tabular} 
	\caption{\added{True parameter values and prior distributions under four different prior regimes. The prior for $ R_0 $ is the implied prior induced by the priors for $ \beta $ and $ \mu $. In regimes one and three, the central 80\% of the prior mass for $ R_0 $ lay between 1.25 and 4.56, while in regimes two and four, 80\% of the prior mass lay between $ 3.8\times10^{-4} $ and $ 2.7\times 10^4 $. In regimes one and two, 80\% of the prior mass for $ \rho $ lay between 0.17 and 0.27, while in regimes three and four the prior mass for $ \rho $ was uniformly distributed between 0 and 1. We used the same mildly informative Dirichlet(9, 0.2, 0.5) prior for $ \bp_{t_1} $ in all prior regimes.}}
	\label{tab:prior_effect_priors}
\end{table}

\subsection{Convergence diagnostics}
\begin{figure}
	\centering
	\includegraphics[width=0.9\linewidth]{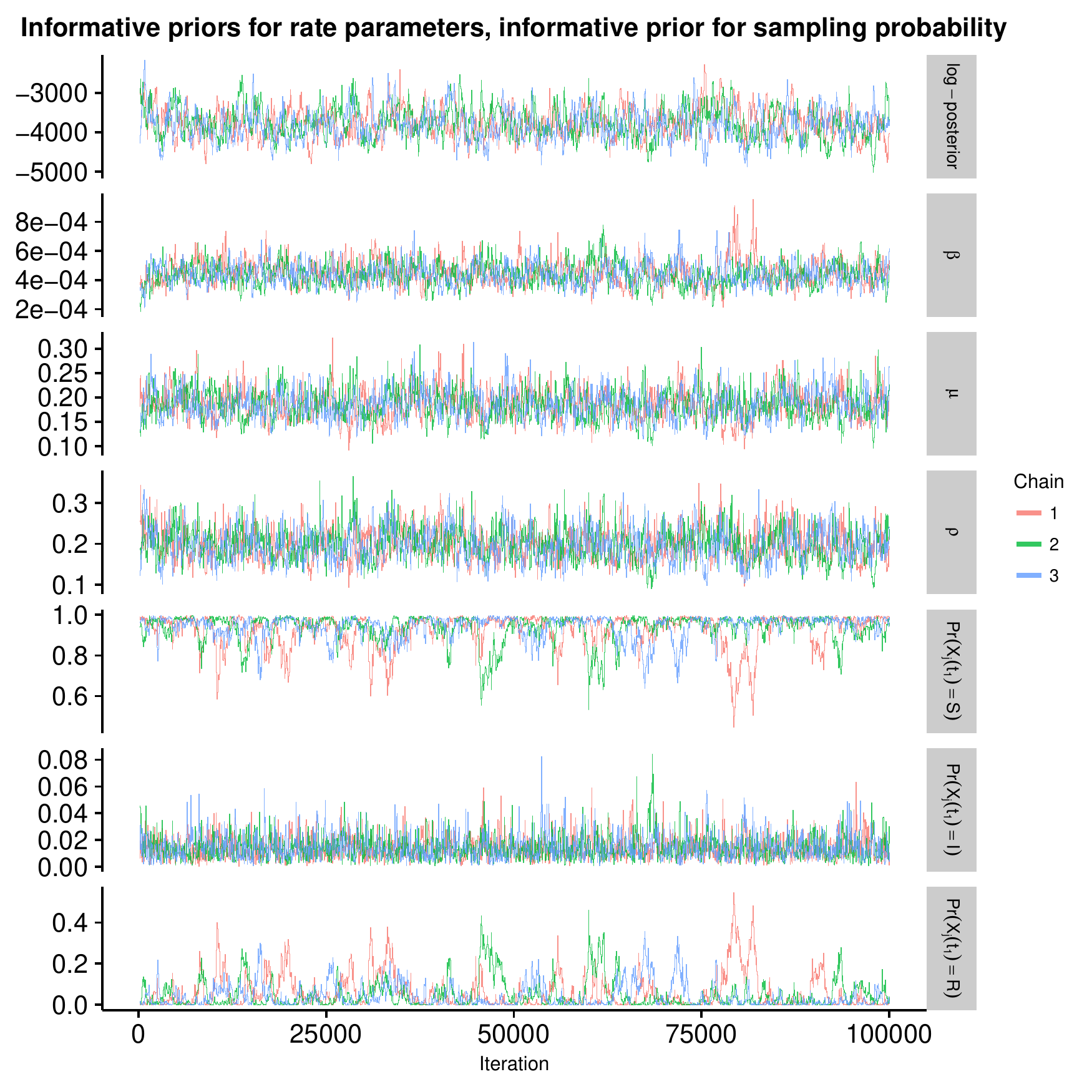}
	\caption{\added{Traceplots of the log--posterior and model parameters for the SIR model fit under informative priors for all model parameters. $ \beta $ denotes the per--contact infectivity rate, $ \mu $ is the recovery rate, $ \rho $ is the binomial sampling probability. Traceplots are thinned to display every 50\textsuperscript{th} iteration.}}
	\label{fig:inform_inform_traces}
\end{figure}

\begin{figure}
	\centering
	\includegraphics[width=0.9\linewidth]{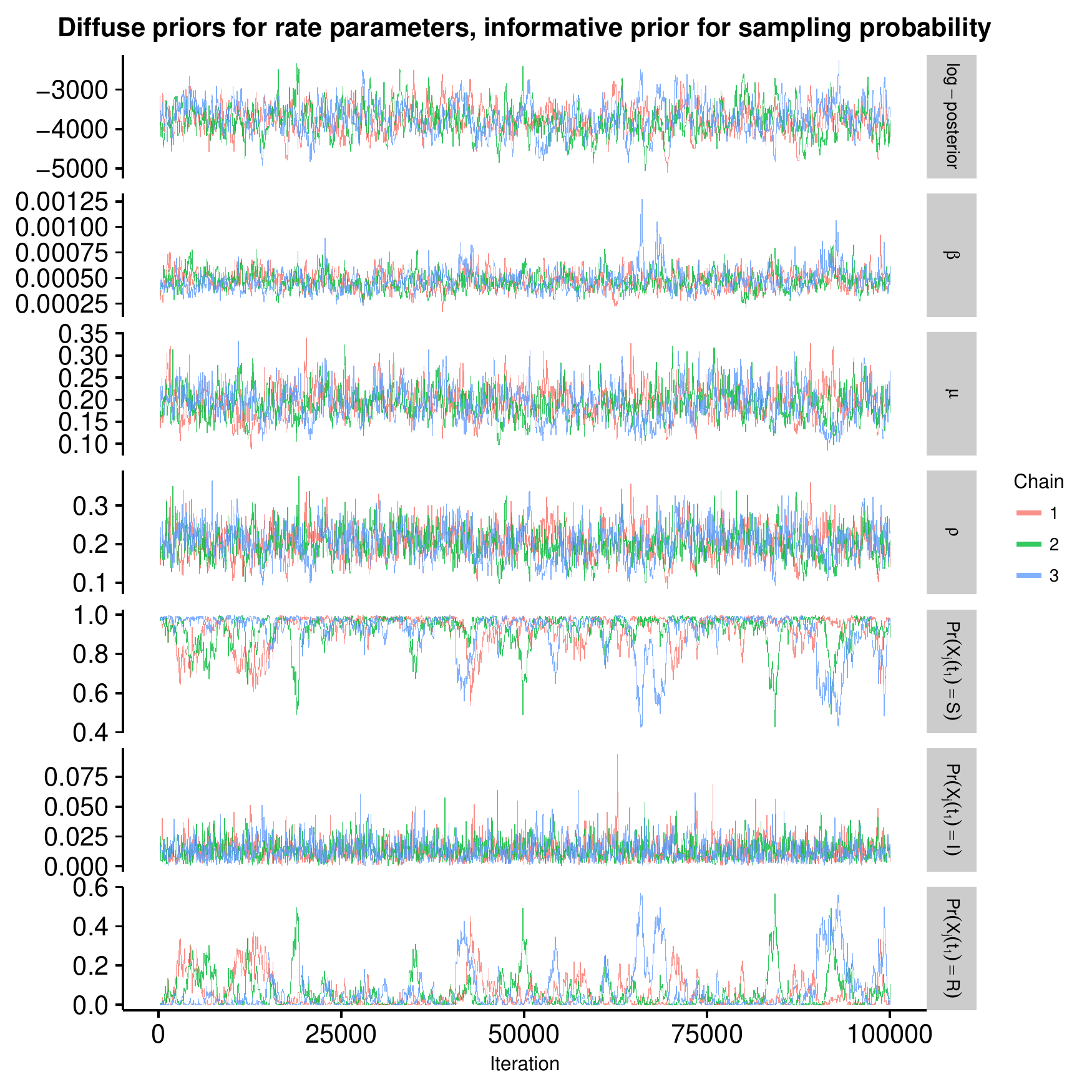}
	\caption{\added{Traceplots of the log--posterior and model parameters for the SIR model fit under diffuse priors for the rate parameters and an informative prior for the binomial sampling probability. $ \beta $ denotes the per--contact infectivity rate, $ \mu $ is the recovery rate, $ \rho $ is the binomial sampling probability. Traceplots are thinned to display every 50\textsuperscript{th} iteration.}}
	\label{fig:diffuse_inform_traces}
\end{figure}

\begin{figure}
	\centering
	\includegraphics[width=0.9\linewidth]{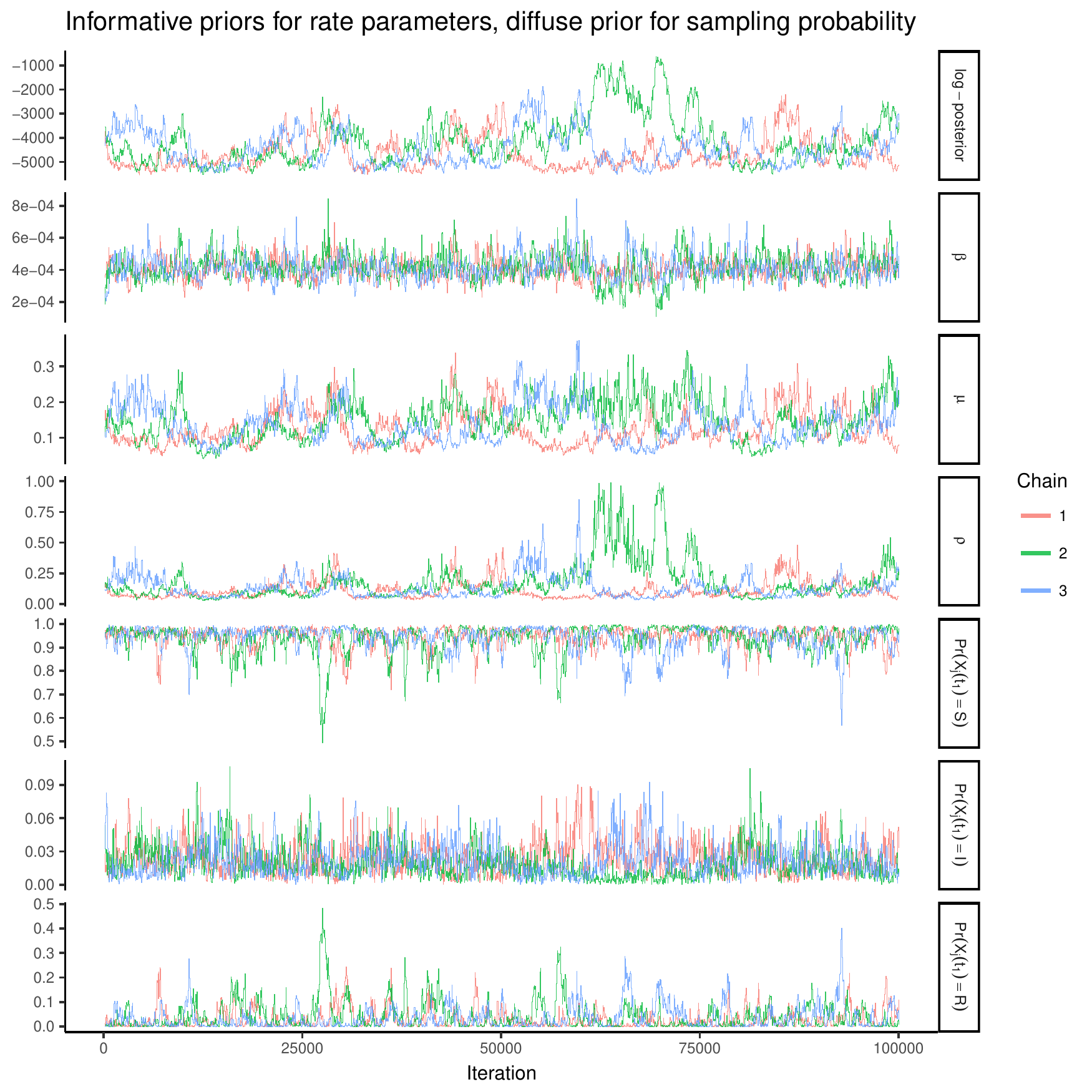}
	\caption{\added{Traceplots of the log--posterior and model parameters for the SIR model fit under informative priors for the rate parameters and a diffuse prior for the binomial sampling probability. $ \beta $ denotes the per--contact infectivity rate, $ \mu $ is the recovery rate, $ \rho $ is the binomial sampling probability. Traceplots are thinned to display every 50\textsuperscript{th} iteration.}}
	\label{fig:inform_diffuse_traces}
\end{figure}

\begin{figure}
	\centering
	\includegraphics[width=0.9\linewidth]{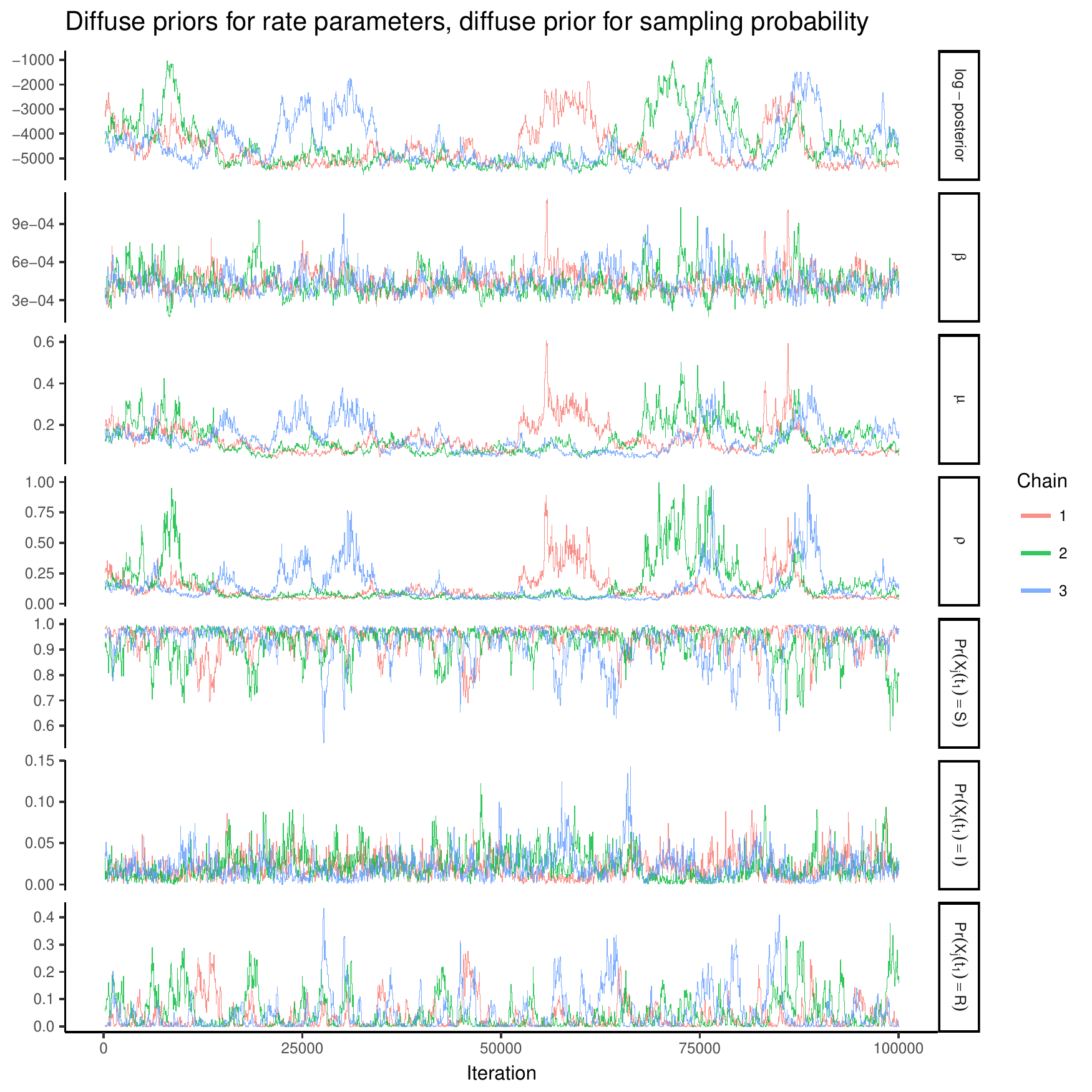}
	\caption{\added{Traceplots of the log--posterior and model parameters for the SIR model fit under diffuse priors for all model parameters. $ \beta $ denotes the per--contact infectivity rate, $ \mu $ is the recovery rate, $ \rho $ is the binomial sampling probability. Traceplots are thinned to display every 50\textsuperscript{th} iteration.}}
	\label{fig:diffuse_diffuse_traces}
\end{figure}

\newpage
 \section{\replaced{Setup, additional results,}{MCMC output} and MCMC diagnostics for British boarding school example}
 \label{sec:bbs_supp}
 
 \added{We ran three MCMC chains per model to fit the SIR and SEIR models to the British boarding school dataset, for 100,000 iterations per chain. We sampled the paths for 100 subjects, chosen uniformly at random, per MCMC iteration, and discarded, as burn-in, the first 100 iterations of each chain for the SIR model, and the first 5,000 iterations of each chain for the SEIR model. Prior distributions, along with posterior medians and credible intervals are given in tables \ref{tab:bbs_SIR_prior_binom} and \ref{tab:bbs_SEIR_prior_binom}. The induced prior for $ R_0 $ is highly diffuse due to the diffuse prior on the per--contact infectivity rate. The prior distribution for the recovery rate and the rate at which exposed individuals became infectious reflected prior knowledge of the natural history of influenza. The prior for the detection probability has roughly 90\% of its mass above 0.3, but is arguably quite diffuse given that it is known that over 90\% of the boys were eventually infected.}  
 
 \added{We also fit the SIR and SEIR models using PMMH with paths for 5,000 particles simulated approximately via a multinomial modification of $ \tau $--leaping over two hour increments. The same priors were used as for the chains fit using BDA. Parameters were updated via random walk Metropolis--Hastings on transformed scales with a proposal covariance matrix that was estimated from an initial run of 2,000 MCMC iterations. We applied a log transformation to the rate parameters, a logit transformation to the binomial sampling probability, and a generalized logit transformation to the initial state probabilities. Results for PMMH are not reported since the MCMC never converged (see traceplots below).} 
 
 \begin{table}[ht!]
 	\begin{center}
 		\begin{tabular}{ccc}
 			\hline Parameter &  Prior Distribution & Posterior Median (95\% BCI)  \\ 
 			\hline
 			\hline $R_0$ & Beta$ ^\prime $(0.001, 1, 1, 1526) & 3.89 (3.40, 4.47) \\
 			\hline $\beta$ & Gamma(0.001, 1) & 0.0024 (0.0021, 0.0026) \\ 
 			\hline $\mu$ & Gamma(1,2) & 0.46 (0.42, 0.50) \\ 
 			\hline $\rho $ & Beta(1,2) & 0.98 (0.92, 1.00)\\
 			\hline $\Pr(X_j(t_1) = S)$& \multirow{3}{*}{Dirichlet(900,3,9)} & 0.99 (0.98, 0.99) \\
 			$\Pr(X_j(t_1) = I)$& & 0.003 (0.001, 0.007) \\
 			$\Pr(X_j(t_1) = R)$&  & 0.009 (0.004, 0.017)\\
 			\hline 
 		\end{tabular} 
 		\caption{\added{Prior distributions and posterior estimates for parameters of the SIR model with binomial emissions fit to the British boarding school outbreak data. The per--contact infectivity rate is $ \beta $, the recovery rate is $ \mu $, and the binomial sampling probability is $ \rho $. The prior for $ R_0 $ is the implied prior induced by the priors for $ \beta $ and $ \mu $. Effective sample size were $\beta$: 11,304; $\mu$: 16,238; $\rho$: 3,920; $p_{S_{t_1}}$: 26,989; $p_{I_{t_1}}$: 284,431; $p_{R_{t_1}}$: 22,761.}}
	 	\label{tab:bbs_SIR_prior_binom}
 	\end{center}
 \end{table}

\begin{table}[ht!]
	\begin{center}
			\begin{tabular}{ccc}
				\hline Parameter &  Prior Distribution & Posterior Median (95\% BCI)  \\ 
				\hline
				\hline $R_0$ & Beta$ ^\prime $(0.001, 1, 1, 1526) & 3.89 (3.40, 4.47) \\
				\hline $\beta$ & Gamma(0.001, 1) & 0.0064 (0.0046, 0.0086) \\ 
				\hline $ \gamma $ & Gamma(0.001, 1) & 0.84 (0.66, 1.19) \\
				\hline $\mu$ & Gamma(1,2) & 0.47 (0.43, 0.51) \\ 
				\hline $\rho $ & Beta(1,2) & 0.98 (0.91, 1.00)\\
				\hline $\Pr(X_j(t_1) = S)$& \multirow{4}{*}{Dirichlet(900, 6,3,9)} & 0.98 (0.97, 0.99) \\
				\hline $ \Pr(X_j(t_1) = E) $ & & 0.006 (0.002, 0.01)\\
				$\Pr(X_j(t_1) = I)$& & 0.003 (0.001, 0.007) \\
				$\Pr(X_j(t_1) = R)$&  & 0.009 (0.004, 0.016)\\
				\hline 
		\end{tabular}
		\caption{\added{Prior distributions and posterior estimates for parameters of the SEIR model with binomial emissions fit to the British boarding school outbreak data. The per--contact infectivity rate is $ \beta $, the rate at which an exposed individual becomes infectious is $ \gamma $, the recovery rate is $ \mu $, and the binomial sampling probability is $ \rho $. The prior for $ R_0 $ is the implied prior induced by the priors for $ \beta $ and $ \mu $. Effective sample size were $\beta$: 679; $\gamma$: 658; $\mu$: 10,069; $ \rho $: 3,244; $p_{S_{t_1}}$: 26,868; $p_{I_{t_1}}$: 26,168; $p_{R_{t_1}}$: 273,613.}}
	 	\label{tab:bbs_SEIR_prior_binom}
	\end{center}
\end{table}
 
 \begin{table}[ht!]
\begin{center}
\caption{\deleted{Prior distributions and posterior estimates for SIR model parameters from three chains with 75 subject path updates per parameter update. The prior for $ R_0 $ is the implied prior induced by the priors for $ \beta $ and $ \mu $. Effective sample size were $\beta$: 20,958; $\mu$: 37,266; $\rho$: 9,336; $p_{S_{t_1}}$: 49,605; $p_{I_{t_1}}$: 715,284; $p_{R_{t_1}}$: 41,806.}}
\end{center}
\end{table}

\subsection{\added{Boarding school example --- MCMC diagnostics}}

\begin{figure}
	\centering
	\includegraphics[width=\linewidth]{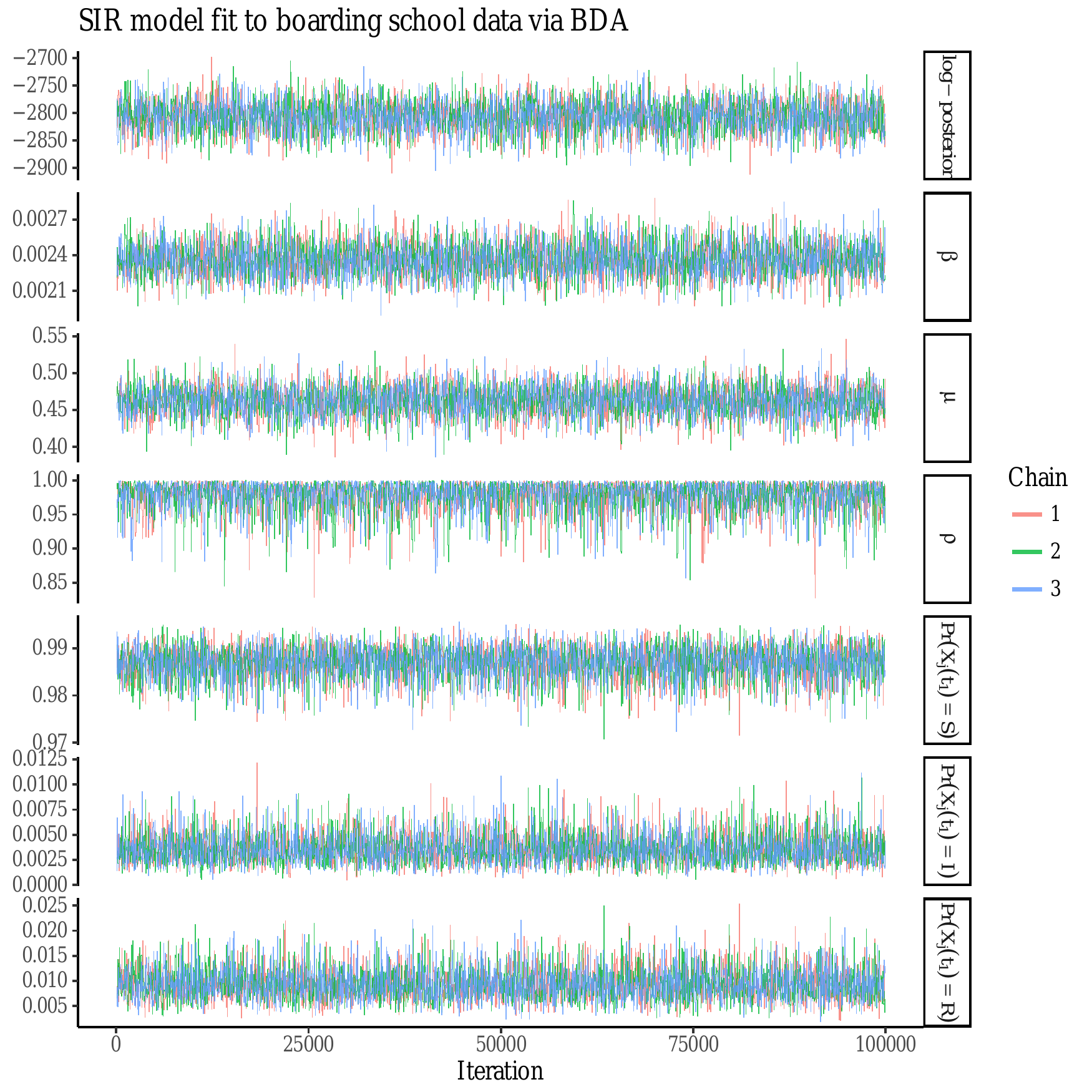}
	\caption{\added{Traceplots of the log--posterior and model parameters for the SIR model fit under binomial emissions using BDA following an initial burn--in of 100 iterations. $ \beta $ denotes the per--contact infectivity rate, $ \mu $ is the recovery rate, and $ \rho $ is the binomial sampling probability. Traceplots are thinned to display every 50\textsuperscript{th} iteration.}}
	\label{fig:bbs_sir_bda_traceplots}
\end{figure}

\begin{figure}
	\centering
	\includegraphics[width=\linewidth]{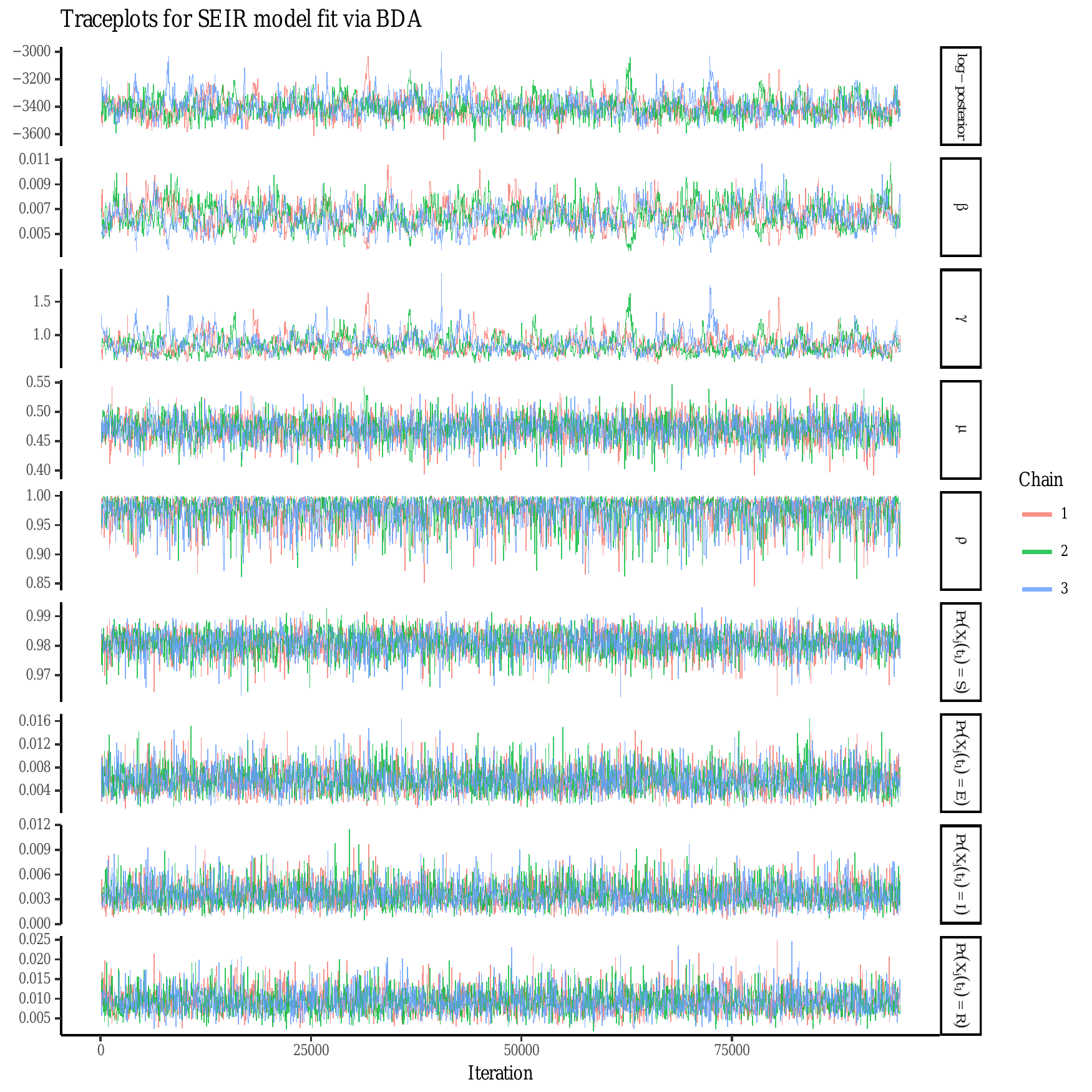}
	\caption{\added{Traceplots of the log--posterior and model parameters for the SEIR model fit under binomial emissions via BDA following an initial burn--in of 5,000 iterations. $ \beta $ denotes the per--contact infectivity rate, $ \mu $ is the recovery rate, $ \gamma $ is the rate at which immunity is lost, and $ \rho $ is the binomial sampling probability. Traceplots are thinned to display every 50\textsuperscript{th} iteration.}}
	\label{fig:bbs_seir_bda_traceplots}
\end{figure}

\begin{figure}
	\centering
	\includegraphics[width=\linewidth]{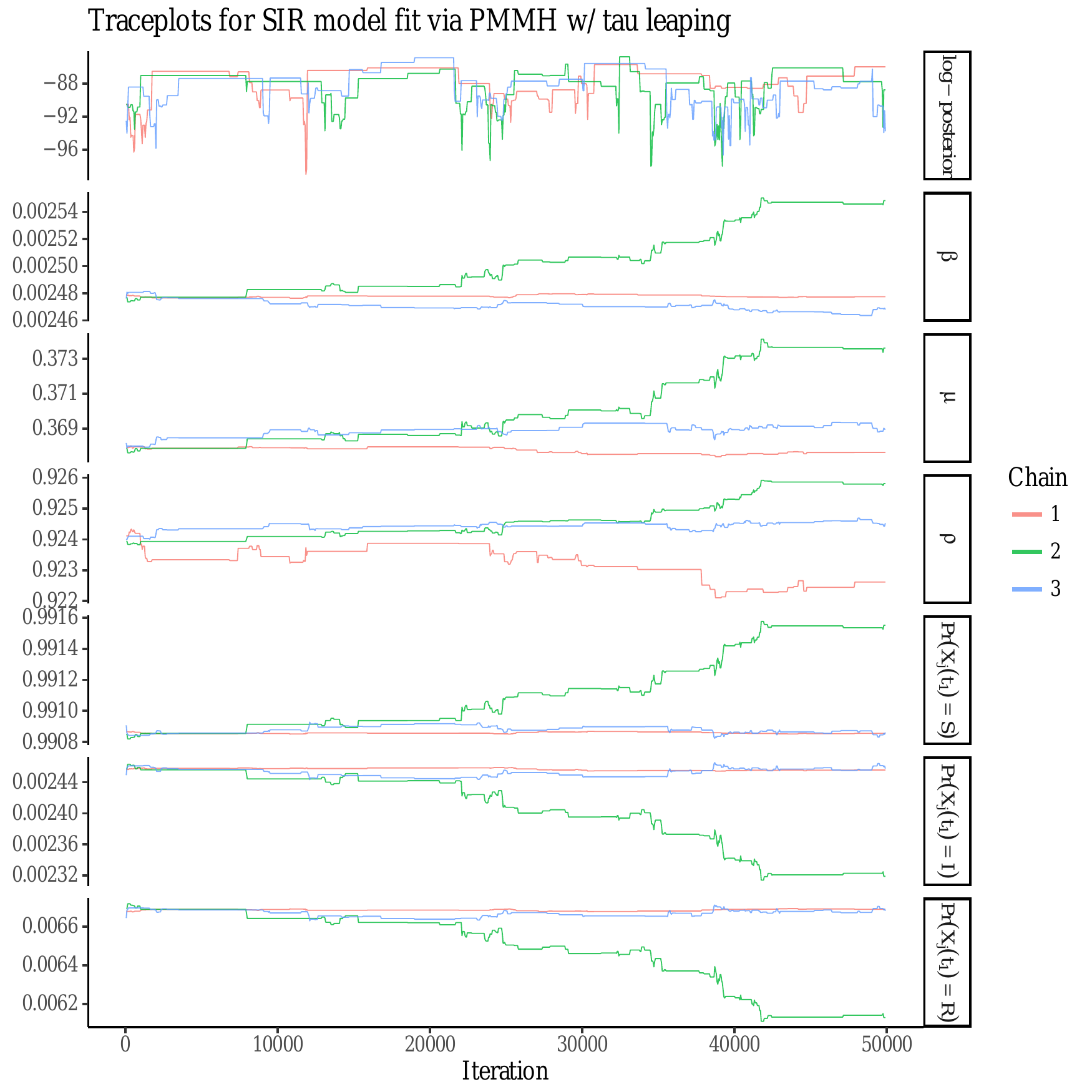}
	\caption{\added{Traceplots of the log--posterior and model parameters for the SIR model fit under binomial emissions using PMMH with 5,000 particles per chain and a time--step of 2 hours in the approximate $ \tau $--leaping algorithm, following a tuning run of 2,000 iterations to estimate the RWMH covariance matrix and in initial burn--in of 100 iterations. $ \beta $ denotes the per--contact infectivity rate, $ \mu $ is the recovery rate, and $ \rho $ is the binomial sampling probability. Traceplots are thinned to display every 50\textsuperscript{th} iteration.}}
	\label{fig:bbs_sir_pmmh_traceplots}
\end{figure}

\begin{figure}
	\centering
	\includegraphics[width=\linewidth]{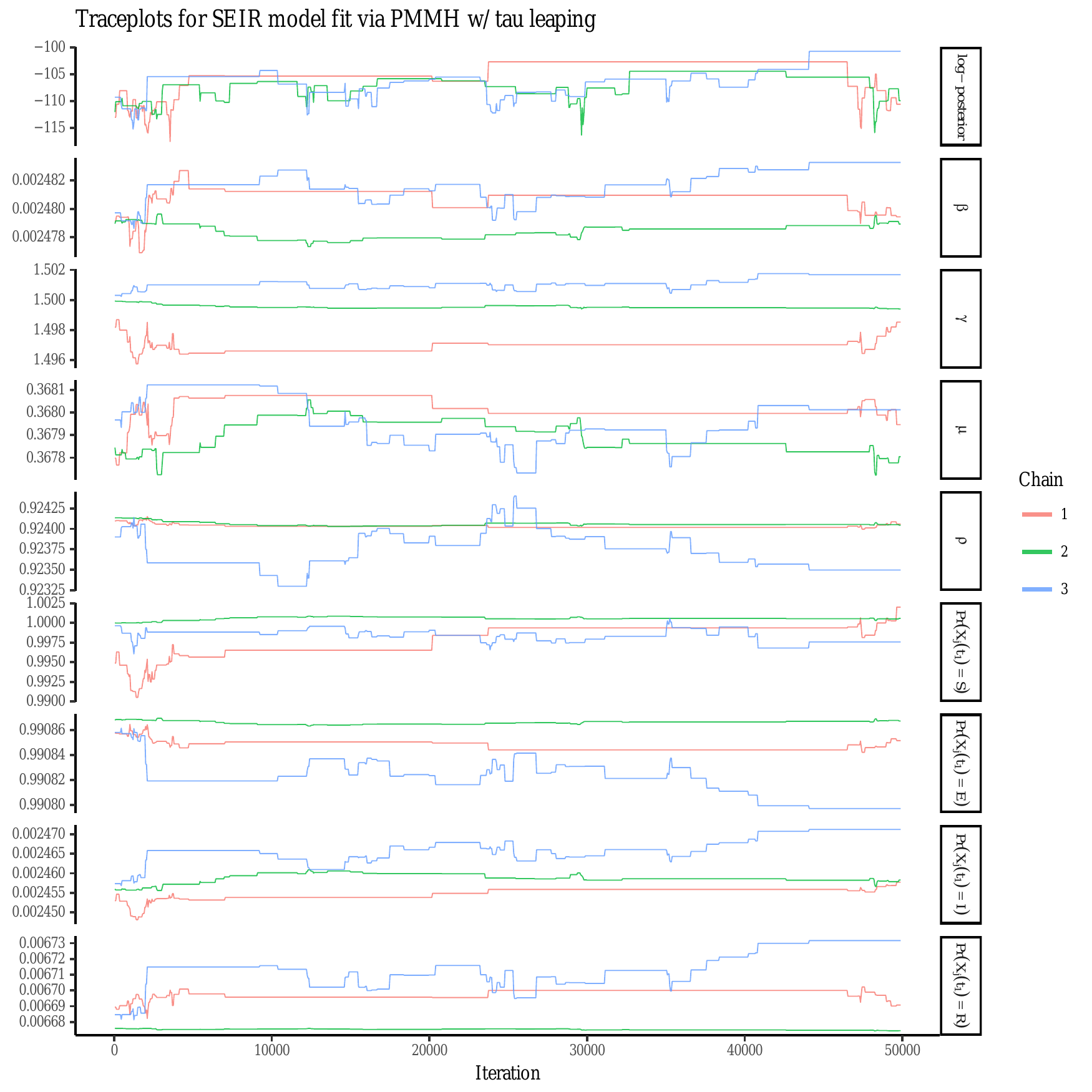}
	\caption{\added{Traceplots of the log--posterior and model parameters for the SEIR model fit under binomial emissions using PMMH with 5,000 particles per chain and a time--step of 2 hours in the approximate $ \tau $--leaping algorithm, following a tuning run of 2,000 iterations to estimate the RWMH covariance matrix and in initial burn--in of 100 iterations. $ \beta $ denotes the per--contact infectivity rate, $ \mu $ is the recovery rate, $ \gamma $ is the rate at which immunity is lost, and $ \rho $ is the binomial sampling probability. Traceplots are thinned to display every 50\textsuperscript{th} iteration.}}
	\label{fig:bbs_seir_pmmh_traceplots}
\end{figure}

\newpage 

\subsection{\added{Supplementary analysis of the British boarding school example under negative binomial emissions}}
\label{sec:bbs_neg_binom}

\begin{figure}
	\centering
	\includegraphics[width=0.6\linewidth]{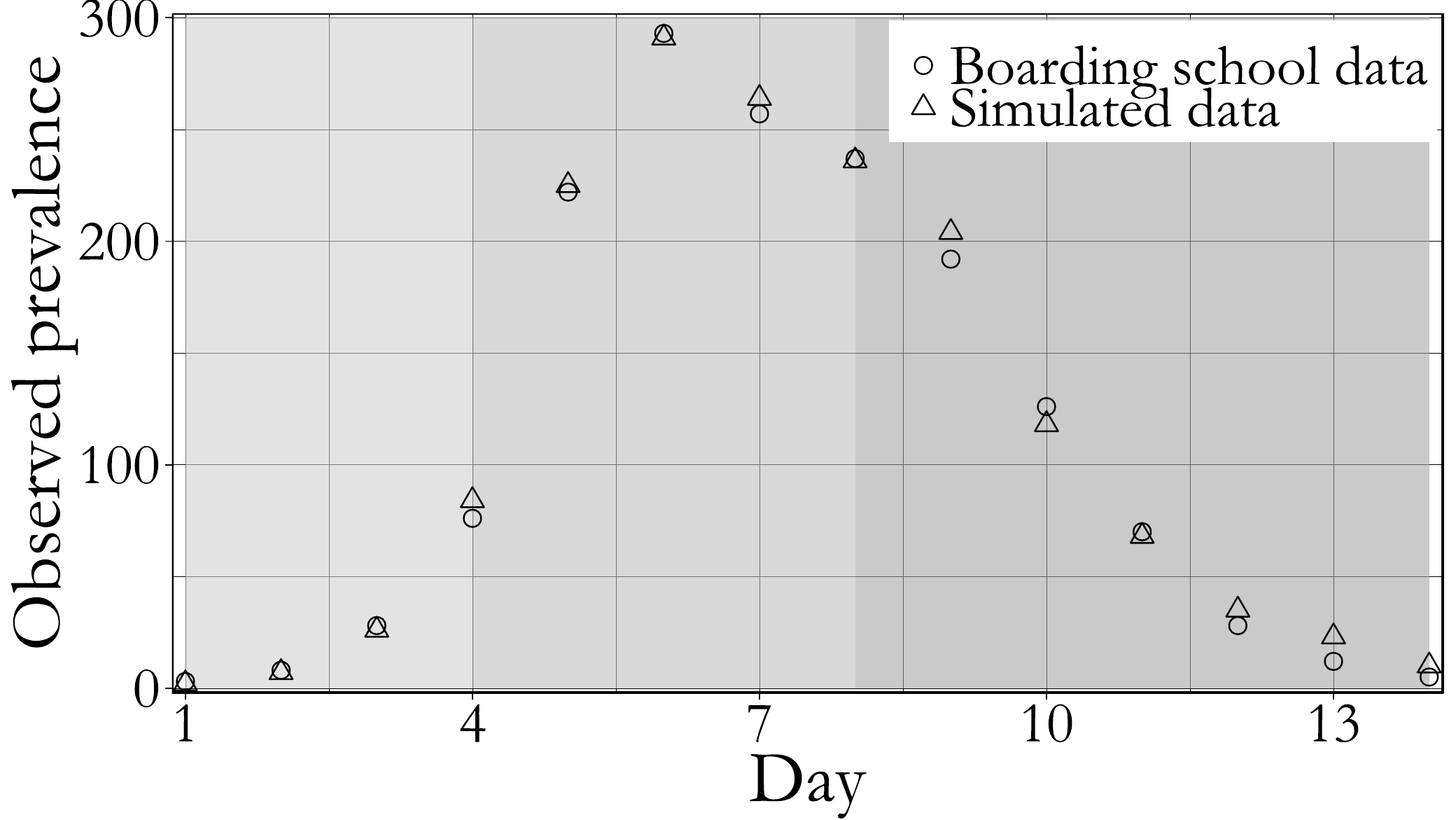}
	\caption{\added{British boardings school data and data simulated under SEIR dynamics with time--varying dynamics over three epochs (indicated by different shaded regions). The simulated dataset was generated using the following parameters: in the first epoch (days 1-4), $ \beta = 0.0035,\ \gamma = 1.25,\  \mu = 0.3$. In the second epoch (days 4-8), $ \beta = 0.065,\ \gamma = 0.51,\ \mu = 0.41 $. In the third epoch (days 8-14), $ \beta = 0.06,\ \gamma = 2.5,\ \mu=0.54 $. The data were a binomial sample of the true prevalence with detection probability $ \rho = 0.98 $. There were three exposed individuals and two infected individuals at the beginning of day 1.}}
	\label{fig:bbs_dat_sim}
\end{figure}

\added{The PMMH MCMC runs in which SIR and SEIR models were fit to the boarding school data under a binomial emission distribution were plagued by severe particle degeneracy (Figures \ref{fig:bbs_sir_pmmh_traceplots} and \ref{fig:bbs_seir_pmmh_traceplots}). The binomial emission distribution requires that the latent prevalence always be at least as great as the observed prevalence. However, this seemed to be a very stringent criterion with such a high case detection rate. That this criterion was so stringent is suggestive of non--trivial model misspecification. We attempted to confirm this by simulating a dataset that resembled the boarding school data. One possible data generating mechanism that yielded to similar prevalence counts resulted from an outbreak evolving under SEIR dynamics that varied over three epochs (Figure \ref{fig:bbs_dat_sim}). That even a model with this simple set of time--varying dynamics would undoubtedly still be misspecified with respect to the real world circumstances in the boarding school is suggestive of a non--trivial level of model misspecification for both the simple SIR and SEIR models that we attempted to fit. Still, the inability of PMMH to fit simple, easily interpretable, SEMs to this data under binomial emissions is a severe limitation.} 

\added{We fit an alternative set of SIR and SEIR models to the data using BDA and PMMH in which the observed prevalence was modeled as a negative binomial sample of the true prevalence, parameterized by its mean and overdispersion. This is a somewhat unrealistic emission distribution because it allows for the observed prevalence to be greater than the true prevalence. That this tended to occur more often in the later parts of the epidemic when boys were being discharged from the infirmary was particularly odd. However, the negative binomial emission distribution allows us to avoid degeneracy in the collection of PMMH particles by doing away with the constraint that the latent prevalence be no smaller than the observed prevalence. Parameters were assigned the same priors given in Tables \ref{tab:bbs_SIR_prior_binom} and \ref{tab:bbs_SEIR_prior_binom}, and the negative binomial overdispersion parameter, $ \phi $, was assigned a Gamma(1, 0.1) prior parameterized by rate. When fitting the model with BDA, we sampled new values for the rate parameters and initial state probabilities from their univariate full conditional distributions via Gibbs sampling. New values for the negative binomial sampling probability and the overdispersion parameter were sampled using multivariate random walk Metropolis--Hastings on the logit scale for $ \rho $ and on the log scale for $ \phi $. An empirical covariance matrix for the RWMH was estimated from an initial run of 10,000 iterations and scaled until the acceptance rate was between 15\%--50\%. We ran three chains per model for 100,000 iterations each, updating the paths of 100 subjects per MCMC iteration, and discarding the first 10,000 iterations as burn--in. We also ran three chains for 50,000 iterations each using PMMH for each of the models, with 500 particles per chain for the SIR model and 5,000 particles per chain for the SEIR model. Particle paths were simulated approximately using $ \tau $--leaping over a time step of 2 hours. Parameters were updated via multivariate RWMH whose covariance matrix was estimated from an initial tuning run of 2,000 iterations. Rate parameters and the overdispersion parameter were updated on the log scale, the negative binomial sampling probability was updated on the logit scale, and the initial state probabilities were updated on the generalized logit scale. We discarded the first 1,000 of each PMMH chain as burn--in.}

\begin{figure}[ht!]
	\centering
	\includegraphics[width=\linewidth]{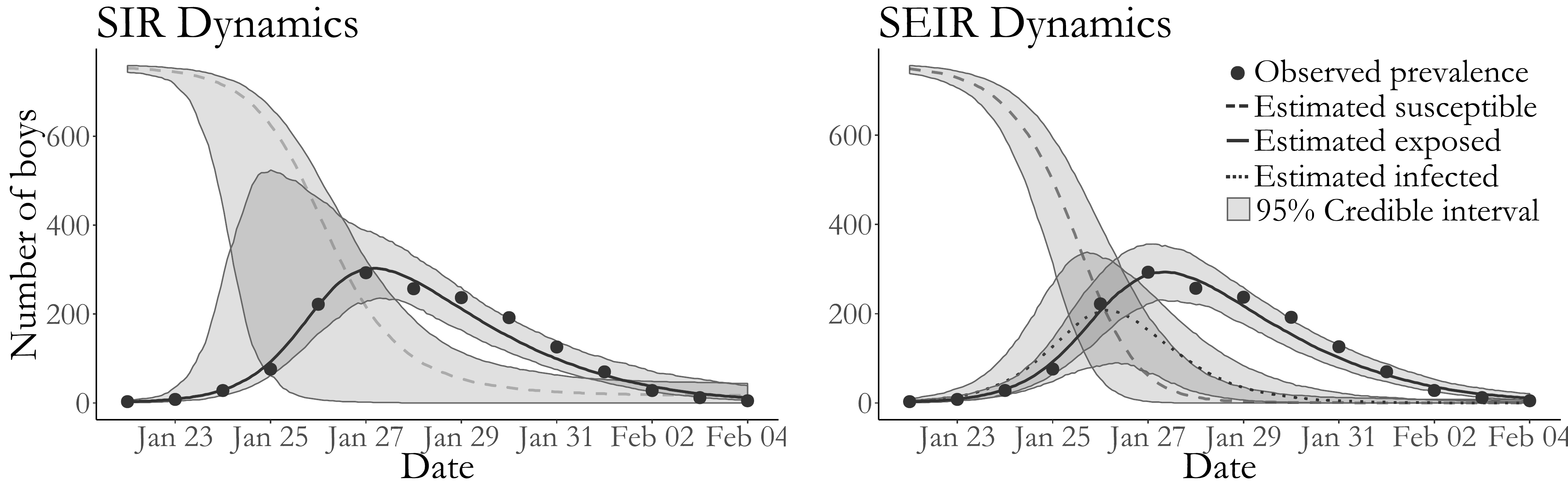}
	\caption{\added{Boarding school data, pointwise posterior median estimates and pointwise 95\% credible intervals under negative binomial emissions (grey shaded areas) for the numbers of infected boys (solid line) and susceptible boys (dashed line). Posterior estimates based on a thinned sample, with every 250$ ^{th} $ configuration retained.}}
	\label{fig:bbs_dat_negbinom}
\end{figure}

\begin{figure}[ht!]
	\centering
	\includegraphics[width=\linewidth]{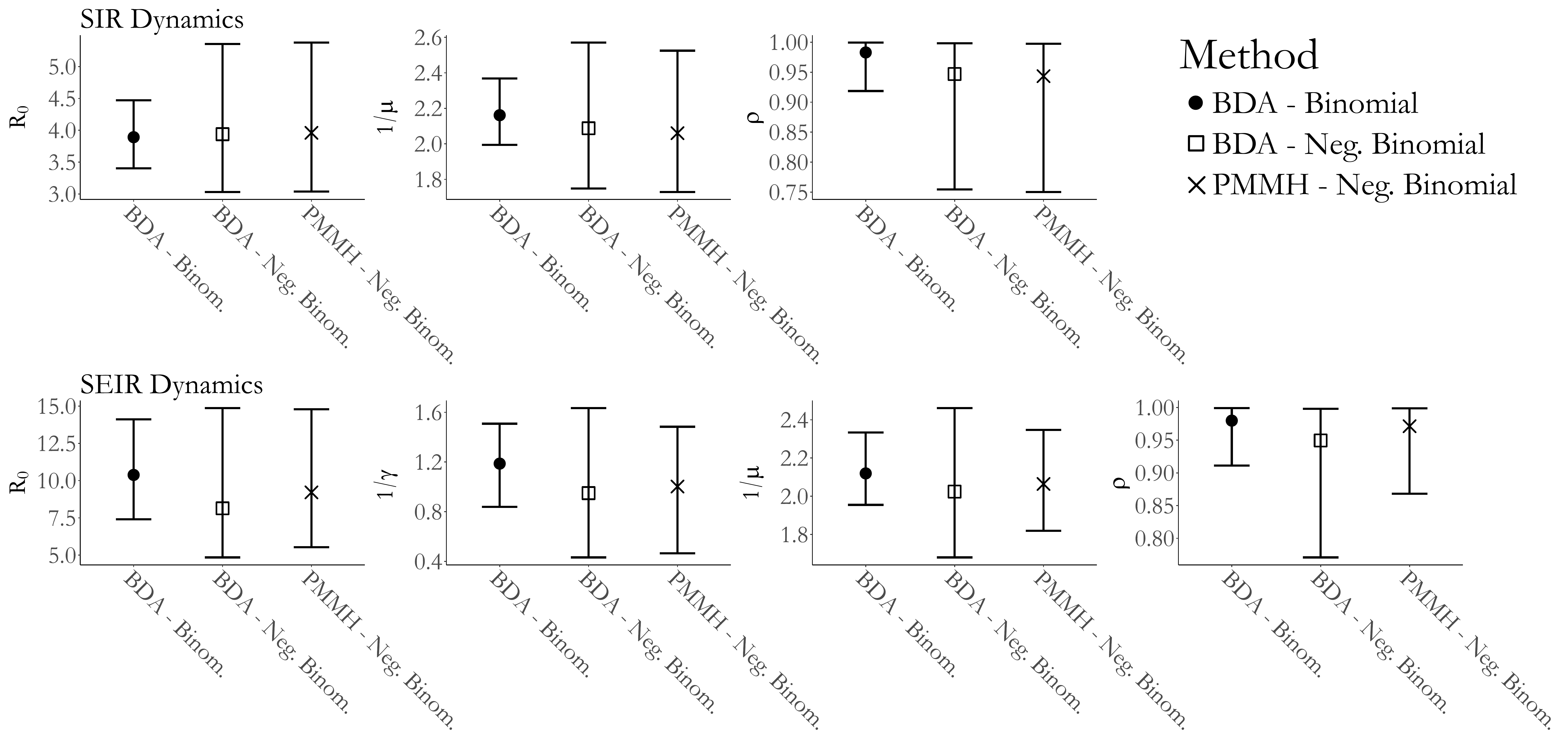}
	\caption{\added{Posterior medians and 95\% credible intervals for SIR and SEIR models fit with BDA and PMMH to the British boarding school data under binomial and negative binomial emission distributions.}}
	\label{fig:bbs_negbinom_credint_comp}
\end{figure}

\begin{figure}
	\centering
	\includegraphics[width=\linewidth]{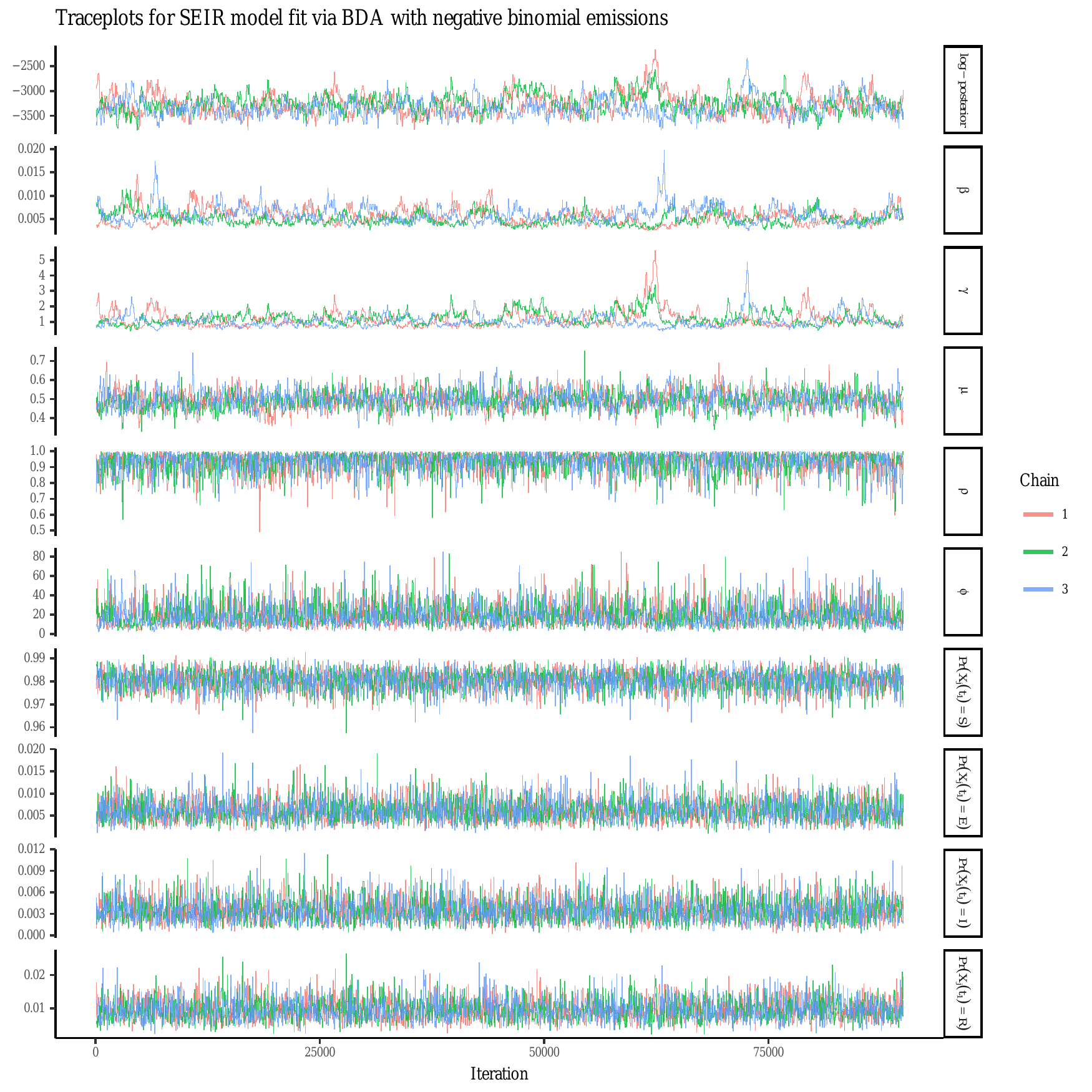}
	\caption{\added{Traceplots of the log--posterior and model parameters for the SEIR model fit under negative binomial emissions via BDA following an initial burn--in of 5,000 iterations. $ \beta $ denotes the per--contact infectivity rate, $ \mu $ is the recovery rate, $ \gamma $ is the rate at which immunity is lost, and $ \rho $ is the binomial sampling probability. Traceplots are thinned to display every 50\textsuperscript{th} iteration.}}
	\label{fig:bbs_seir_bda_negbinom_traceplots}
\end{figure}

\added{Although the posterior median estimates under binomial and negative binomial emissions for the SIR and SEIR dynamics and detection rate are generally quite similar, the posterior credible intervals are considerably wider when the data modeled as a negative binomial sample of the true prevalence. This manifests both in the widths of the credible intervals for the latent process (Figure \ref{fig:bbs_dat_negbinom}), and the credible intervals for the model parameters (Figure \ref{fig:bbs_negbinom_credint_comp}). This is not unexpected given that the negative binomial distribution is substantially more flexible than the binomial distribution. In comparing the posterior estimates obtained using BDA and PMMH under negative binomial emissions, we find that the estimates are essentially identical for the SIR model. For the SEIR model, estimates of the dynamics are generally similar, though not to the same degree as those for the SIR model. We notice that the credible intervals for the mean infectious period and the negative binomial detection probability obtained using PMMH are substantially wider than those obtained using BDA. Upon closer inspection of the traceplots of the model parameters, it is clear that the negative binomial overdispersion parameter in the PMMH chains did not converge (Figure \ref{fig:bbs_seir_bda_negbinom_traceplots}).}

\begin{figure}
	\centering
	\includegraphics[width=\linewidth]{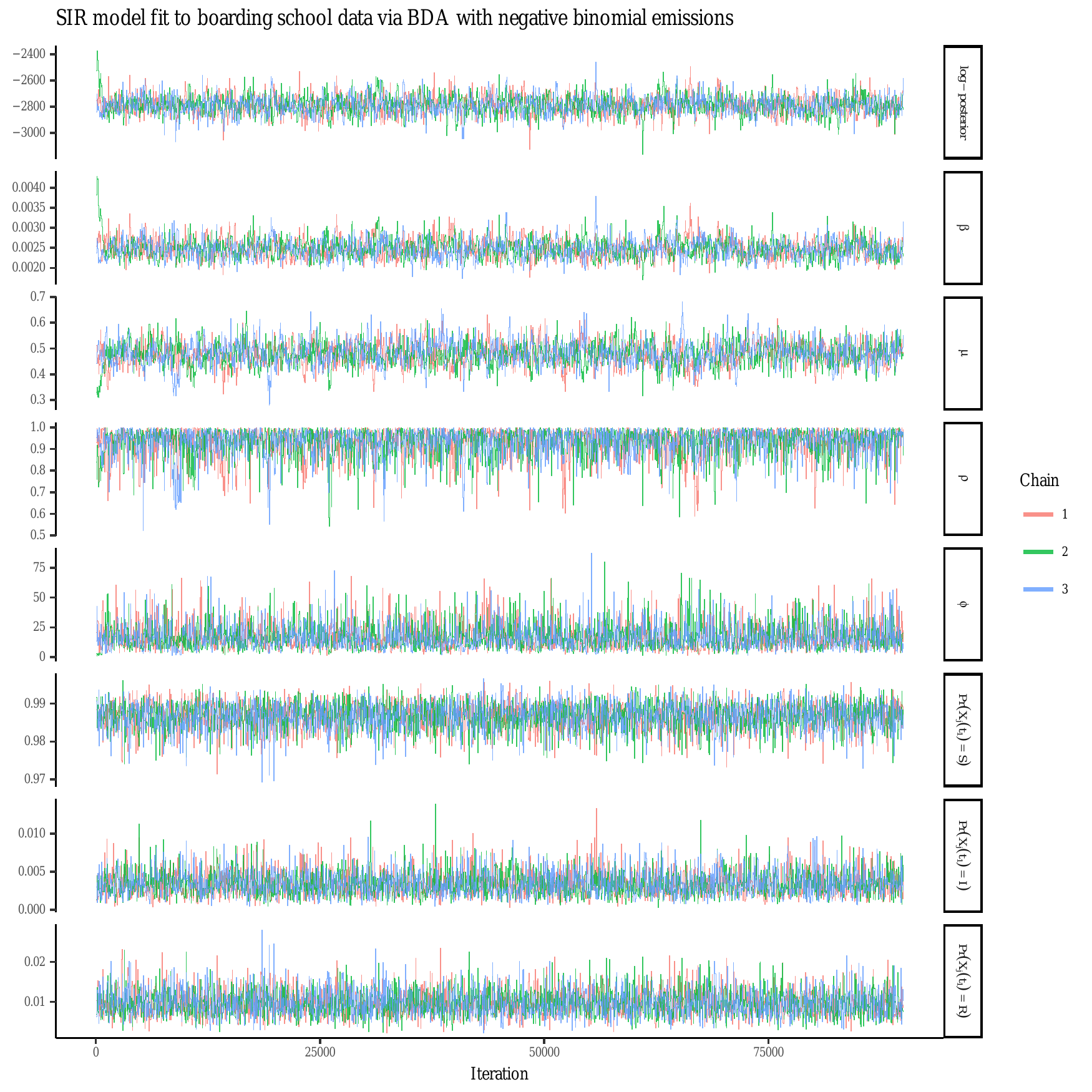}
	\caption{\added{Traceplots of the log--posterior and model parameters for the SIR model fit under negative binomial emissions using BDA following an initial burn--in of 100 iterations. $ \beta $ denotes the per--contact infectivity rate, $ \mu $ is the recovery rate, and $ \rho $ is the binomial sampling probability. Traceplots are thinned to display every 50\textsuperscript{th} iteration.}}
	\label{fig:bbs_sir_bda_negbinom_traceplots}
\end{figure}

\begin{figure}
	\centering
	\includegraphics[width=\linewidth]{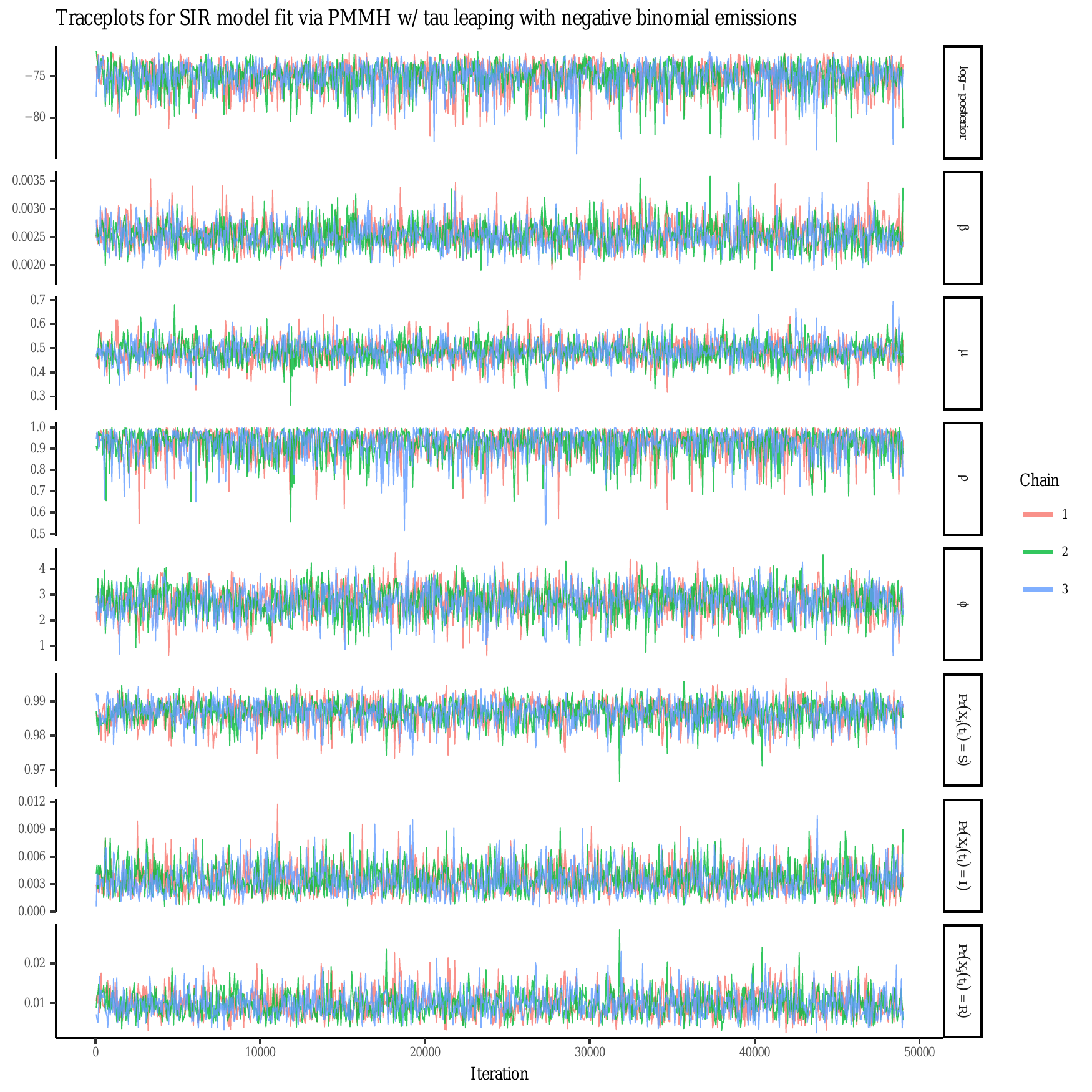}
	\caption{\added{Traceplots of the log--posterior and model parameters for the SIR model fit under negative binomial emissions using PMMH with 5,000 particles per chain and a time--step of 2 hours in the approximate $ \tau $--leaping algorithm, following a tuning run of 2,000 iterations to estimate the RWMH covariance matrix and in initial burn--in of 100 iterations. $ \beta $ denotes the per--contact infectivity rate, $ \mu $ is the recovery rate, and $ \rho $ is the binomial sampling probability. Traceplots are thinned to display every 50\textsuperscript{th} iteration.}}
	\label{fig:bbs_sir_pmmh_negbinom_traceplots}
\end{figure}

\begin{figure}
	\centering
	\includegraphics[width=\linewidth]{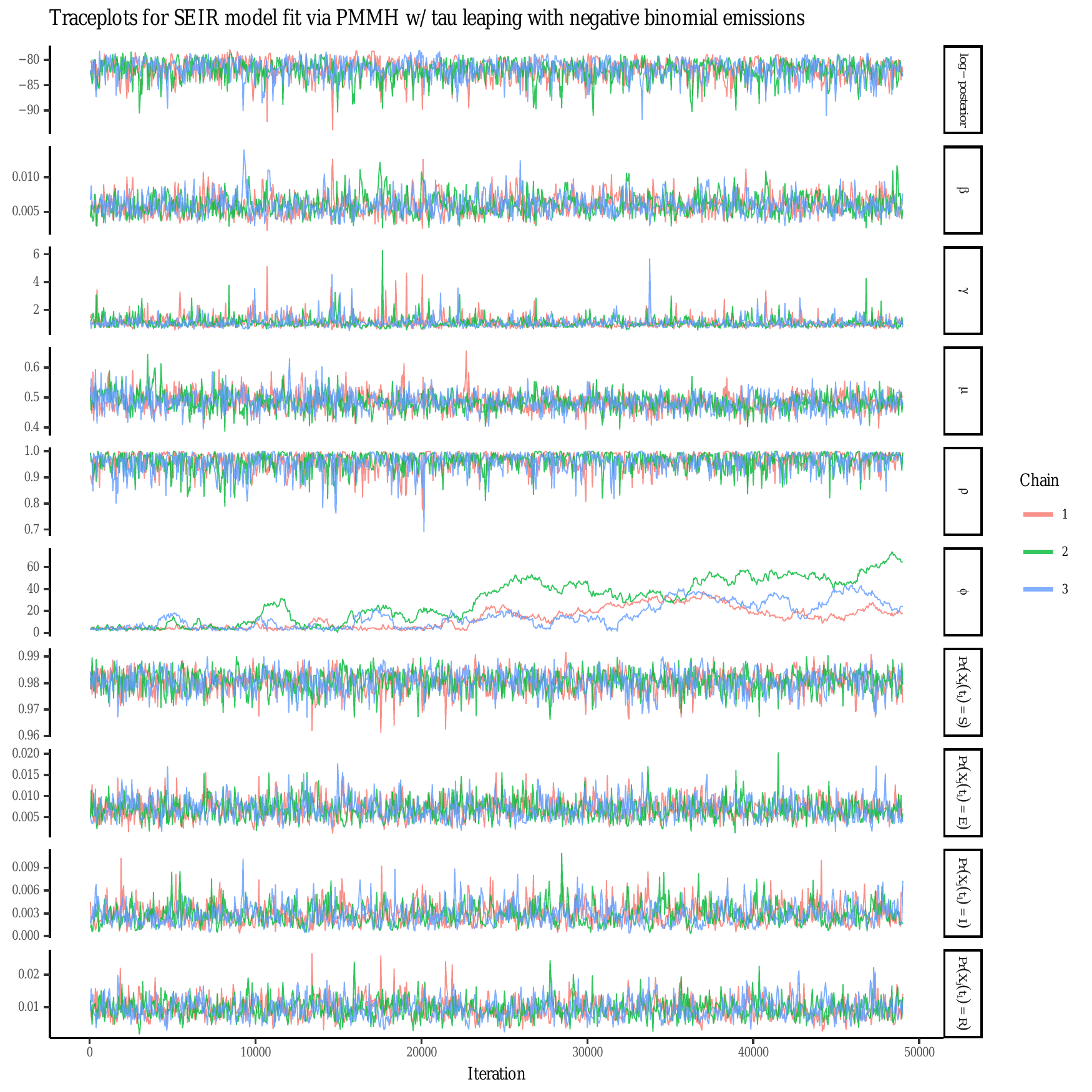}
	\caption{\added{Traceplots of the log--posterior and model parameters for the SEIR model fit under negative binomial emissions using PMMH with 5,000 particles per chain and a time--step of 2 hours in the approximate $ \tau $--leaping algorithm, following a tuning run of 2,000 iterations to estimate the RWMH covariance matrix and in initial burn--in of 100 iterations. $ \beta $ denotes the per--contact infectivity rate, $ \mu $ is the recovery rate, $ \gamma $ is the rate at which immunity is lost, and $ \rho $ is the binomial sampling probability. Traceplots are thinned to display every 50\textsuperscript{th} iteration.}}
	\label{fig:bbs_seir_pmmh_negbinom_traceplots}
\end{figure}